\documentclass[10pt,english,floatfix,superscriptaddress,aps,prd,preprint,showkeys,nofootinbib]{revtex4}
\usepackage{amsmath}
\usepackage{amssymb}
\usepackage{amsbsy}
\usepackage{amsfonts}
\usepackage{amsopn}
\usepackage{amstext}
\usepackage{graphicx}
\usepackage[english]{babel}
\usepackage{color}
\usepackage{slashed}
\usepackage{esint}
\usepackage[dvips]{epsfig}
\usepackage[dvips]{graphicx}
\usepackage{float}
\usepackage{units}
\usepackage{textcomp}
\usepackage{wasysym}
\usepackage{hyperref}
\usepackage{slashed}

\begin{document}

\title{ModMax-black hole surrounded by cloud of strings in Bumblebee gravity}

\author{Fernando M. Belchior}
\email{fernandobelcks7@gmail.com}
\affiliation{Departamento de Física, Universidade Federal da Paraíba, Centro de Ciências Exatas e da Natureza, 58051-970, João Pessoa, Paraíba, Brazil}

\author{Allan R. P. Moreira}
\email{allan.moreira@fisica.ufc.br}
\affiliation{Secretaria da Educação do Ceará (SEDUC), Coordenadoria Regional de Desenvolvimento da Educação (CREDE
9), 62880-384, Horizonte, Ceará, Brazil}

\author{Abdelmalek Bouzenada}
\email{abdelmalekbouzenada@gmail.com}
\affiliation{Laboratory of Theoretical and Applied Physics, Echahid Cheikh Larbi Tebessi University, 12001, Algeria}
\affiliation{Research Center of Astrophysics and Cosmology, Khazar University, Baku, AZ1096, 41 Mehseti Street, Azerbaijan}

\author{Faizuddin Ahmed}
\email{faizuddinahmed15@gmail.com}
\affiliation{Department of Physics, The Assam Royal Global University, 781035, Guwahati, Assam, India}

\begin{abstract}
In this article, we investigate the optical, thermodynamic, and scattering properties of a ModMax black hole surrounded by a cloud of strings within the framework of Einstein-bumblebee gravity. We then analyze in detail the thermodynamic properties of this black hole, including the Hawking temperature, entropy, and other relevant thermodynamic quantities, and examine the outcomes. Furthermore, we study the greybody factors (GFs) associated with the emission of various perturbative fields propagating in this black hole background. In particular, we consider spin-0 scalar fields, spin-1 electromagnetic fields, and spin-2 graviton fields, and evaluate the corresponding absorption probabilities and energy emission rates. Our analysis demonstrates how the optical features, thermodynamics and GFs depend on the underlying parameters of the system, such as the Lorentz symmetry violation parameter, the cloud of strings parameter, the ModMax parameter, the electric charge, and the black hole mass, thereby providing a comprehensive understanding of the physical effects of these parameters on the radiation and scattering processes around the black hole.
\end{abstract}

\keywords{\bf Einstein-Bumblebee gravity; black hole; ModMax theory; topological defects; geodesic structure; thermodynamics}

\maketitle

\section{Introduction}

Black holes (BHs) have evolved from theoretical model constructs into fundamental objects of modern astrophysical observational information, although their direct observation remains among the most difficult tasks because of the extreme conditions in their vicinity \cite{BHM}. A central method in this direction is the analysis of the shadow produced by a BH, which represents an optical signature of its gravitational field in the region near the event horizon. In this context, a large portion of information is obtained from the detection of gravitational wave (GW) signals generated by BH mergers observed by the LIGO collaboration \cite{AbbottPRL2016,AbbottPRD2020,AbbottPRL2020,LiuNature20219}. In addition, these findings are supported by radial velocity observations that identified a distant binary system composed of a star and a BH, providing independent observational confirmation for the existence of such compact objects. In investigations of shadow measurements, the Event Horizon Telescope (EHT) has successfully imaged the BHs located at the centers of $M87^{*}$ and Sgr $A^{*}$, providing the first direct observational evidence of BH shadows \cite{AkiyamaL12019, AkiyamaL52019,AkiyamaL62019, AkiyamaL122022, AkiyamaL172022}, where these images display a central dark region, identified as the shadow, surrounded by a bright photon ring formed by photons undergoing strong gravitational lensing due to spacetime curvature. Gravitational lensing \cite{39,40,41}, defined as the deflection of light trajectories in the presence of a massive object such as a BH, not only confirms the existence of the BH but also allows a detailed reconstruction of its surrounding geometry \cite{42,43,44,45,46,47}. The observed attenuation of radiation caused by the trapping nature of the BH gravitational field constitutes the physical origin of the shadow and provides essential information about radiation propagation in curved spacetime. The increasing accuracy and amount of observational data have motivated extensive theoretical and numerical analyses aimed at understanding how the structure of the BH shadow encodes the fundamental properties of BHs and can be used to discriminate between different gravitational theories \cite{TsupkoPRD2017, BroderickApJ2022, MizunoNatAst2018, AtamurotovPRD2013, AbdujabbarovSS2016, AbdikamalovPRD2019, AtamurotovPRD2015, AtamurotovCPC2023, BelhajPLB2021, BelhajCQG2021,CunhaPLB2017,WeiJCAP2019,LingPRD2021,TsukamotoPRD2018,AraujoFilhoCQG2024,RayimbaevPoDU2022,PerlickPRD2018,VagnozziCQG2023,KocherlakotaPRD2021,gy1,gy2}. In this framework, variations in the shadow geometry and intensity distribution can originate from several parameters, including charge ($q$), spin, environmental matter distributions, or deviations from general relativity \cite{q,qq,qqq,qqqq,qqqqq,qqqqqq,qqqqqqq,BZQ}. Furthermore, these developments refine the current understanding of BH physics, indicating that BH observations may provide constraints on new physical effects and offer tests for the underlying structure of spacetime and gravitational interactions \cite{WuPoDU2024, PantigJCAP2022, CapozzielloJCAP2023, Capozziello2023, XuJCAP2018, KonoplyaApJ2022}.

The investigation of BH solutions within the framework of Bumblebee gravity has attracted considerable attention because it consistently incorporates spontaneous Lorentz symmetry breaking and allows the study of its measurable consequences in strong gravitational regimes. In particular, gravitational lensing in these backgrounds has been examined in detail, demonstrating that the presence of the Bumblebee field modifies the deflection angle of light and consequently provides a potential observational channel to probe Lorentz-violating effects through precise astrophysical measurements \cite{BG1}. Furthermore, the motion of particles in this spacetime has been systematically analyzed, indicating that the Lorentz-violating parameter affects the perihelion precession of bounded orbits while preserving the orbital period for circular motion, which reflects a nontrivial modification of geodesic dynamics \cite{BG2}. Beyond these foundational results, several physical properties of Bumblebee BHs have been thoroughly explored, testing the computation of greybody factors, where the modification of the effective potential induced by Lorentz symmetry breaking produces observable deviations in Hawking radiation spectra \cite{BG3, BG4}. Also, the analysis of BH shadows has established that the photon sphere structure and the apparent angular size of the shadow are significantly influenced by the Bumblebee parameter, providing additional observational signatures \cite{BG5, BG6}. The influence on gravitational waves has also been investigated, showing that Lorentz violation can generate detectable imprints on the waveform and energy emission mechanisms \cite{BG7}. Also, from a thermodynamic viewpoint, the inclusion of the Bumblebee field modifies fundamental quantities such as temperature, entropy, and heat capacity, thereby affecting the stability conditions and phase structure of BHs \cite{BG8, BG9, BG10}. In this case, the quasinormal mode spectrum, which describes the response of BHs under external perturbations, has been analyzed and shows that Lorentz-violating effects alter both oscillation frequencies and damping rates, thus modifying the ringdown phase \cite{BG11, BG12}. Also, slowly rotating and charged BH solutions in Bumblebee gravity have been derived, and their shadow characteristics have been studied, revealing the combined influence of rotation, charge, and Lorentz symmetry breaking on photon trajectories \cite{BG13}. Subsequent analyses have further extended these findings by examining particle motion in such generalized geometries \cite{BG14, BG15}, testing scalar quasinormal modes and their stability properties \cite{BG16}, and evaluating greybody factors together with Hawking radiation spectra and strong gravitational lensing in these extended configurations, thereby establishing a comprehensive description of both classical and quantum aspects of Bumblebee BHs \cite{BG17}.

ModMax electrodynamics, introduced as a self-consistent nonlinear generalization of Maxwell theory, defines a framework that continuously connects the standard linear regime with nonlinear corrections that arise in the low-energy sector of the Born-Infeld construction. In this formulation, a dimensionless control parameter $\gamma$ is incorporated to quantify the departure from Maxwell electrodynamics, while guaranteeing that the conventional Maxwell equations are exactly recovered in the limit $\gamma \to 0$ \cite{MAX1}. In this setting, BH configurations have been explicitly derived within the ModMax model, where it is shown that nonlinear electromagnetic contributions induce nontrivial modifications of the spacetime metric as well as of the associated conserved charges \cite{MAX2}. Subsequent analyses have generalized these solutions to a variety of physical processes, including greybody factors, neutrino propagation near dyonic ModMax BHs, shadow structure, gravitational lensing observables, and quasinormal mode spectra, thereby establishing the relevance of ModMax electrodynamics in strong gravitational fields \cite{MAX3}. A central feature of these configurations is the $SO(2)$ duality symmetry between electric and magnetic components, which maintains rotational invariance in the electromagnetic field space, while the parameter $\gamma$ regulates the regularity properties and stability of physical charges. The internal consistency, formal structure, and physical implications of ModMax electrodynamics have been systematically tetsed across multiple studies, confirming its robustness as a nonlinear extension of classical electrodynamics in curved backgrounds \cite{MAX4, MAX5, MAX6, MAX7, MAX8, MAX9, MAX10, MAX11, MAX12, MAX13, MAX14, MAX15, MAX16, MAX17, MAX18, MAX19, MAX20, MAX21}. 

In parallel, nonlinear electrodynamics (NED) frameworks have been extensively employed to analyze astrophysical mechanisms in the vicinity of regular BH geometries. In particular, the properties of thin accretion disks in NED spacetimes have been examined using circular geodesic motion, allowing the determination of the innermost stable circular orbit (ISCO), together with the corresponding energy flux, temperature distribution, and luminosity emission \cite{MAX22}. These investigations indicate that the ISCO radius is displaced inward with respect to the Schwarzschild limit, resulting in higher disk temperature, increased luminosity, and improved radiative efficiency. Further studies have explored both strong-field and weak-field regimes through the evaluation of effective potentials, stable circular trajectories, photon spheres, marginally bound orbits, and epicyclic oscillations, combined with accretion observables such as radiation flux, temperature behavior, and mass accretion rate \cite{MAX23}. Also, the results show that a reduction of the NED parameter shifts characteristic radii closer to the BH, whereas larger values enhance the energy flux, decrease the disk temperature, and increase the accretion rate. In addition, joint analyses of weak gravitational lensing, BH shadow formation, and accretion disk radiation demonstrate that the NED parameter produces measurable effects on the photon sphere radius, shadow angular size, and light deflection, which can be constrained using observational data from the Event Horizon Telescope for M87$^*$ and Sgr A$^*$ \cite{MAX24}. Within this approach, the Novikov-Thorne thin disk model has been applied to consistently determine ISCO characteristics, luminosity output, and thermal properties, showing that increasing the NED parameter shifts the ISCO to larger radii, reduces the energy flux, lowers the disk temperature, and decreases luminosity, while enhancing the weak deflection angle and modifying the shadow radius. More recently, detailed investigations of circular geodesics and accretion processes around dyonic ModMax BHs have been reported \cite{MAX25}, where the ISCO radius, photon sphere, and orbital dynamics of test particles were computed, and the dependence of radiative flux, temperature profile, and accretion efficiency on the ModMax parameter $\gamma$ was determined. It is found that increasing $\gamma$ shifts the ISCO outward and leads to a decrease in radiative flux, disk temperature, and efficiency. Moreover, the behavior of accreting matter has been analyzed through radial velocity, density distribution, and mass accretion rate in isothermal fluid configurations, together with epicyclic motion analysis. The results indicate that larger values of $\gamma$ reduce the radial velocity, density, and mass accretion rate, while the radial epicyclic frequency increases near the BH and decreases at large distances, and the vertical epicyclic frequency exhibits an increasing trend. These measurmment demonstrate that nonlinear electromagnetic effects play a crucial role in shaping both the geometric structure and astrophysical signatures of BH spacetimes, In this context, the behavior of the mass accretion rate obtained here differs from that presented in Ref. \cite{MAX25}, testing that its dependence on the ModMax parameter is sensitive to the choice of fluid description, assumptions, and boundary conditions. Lorentz-violation gravity, specifically within the framework of Bumblebee models, incorporates a vector field $B_\mu$ acquiring a nonzero vacuum expectation value (VEV) and a nonminimal gravitational coupling $\xi B^\mu B^\nu R_{\mu\nu}$, resulting in Schwarzschild or RN-like BH solutions deformed by the Lorentz-violating parameter $\ell=\xi b^2$, which modifies the geodesic structure, gravitational lensing, quasinormal mode spectrum, and thermodynamic quantities. For a purely radial configuration $B_\mu=(0,,B_r(r),,0,,0)$ with vanishing field strength $B_{\mu\nu}=0$ and zero potential $V=V_X=0$, the energy-momentum contribution satisfies $T^B_{\mu\nu}=0$, while the spacetime metric acquires corrections through the tensor $\mathcal K_{\mu\nu}$ and the nonminimal coupling term $B^\mu B^\nu R_{\mu\nu}$, giving \cite{MAX25}:
\begin{align}
A(r)=1-\frac{2M}{r},\qquad B(r)=\frac{1+\ell}{A(r)},
\end{align}
with the event horizon located at $r_h=2M$. When matter fields are included, for instance Maxwell fields, the metric function generalizes to \cite{MAX25}:
\begin{align}
F(r)=1-\frac{2M}{r}+\frac{(2+\ell)\mathcal Q^2}{2(1+\ell),r^2}-\frac{\Lambda_{\rm eff}}{3}r^2,
\end{align}
where the effective cosmological constant $\Lambda_{\rm eff}$ accounts for potential background contributions. Linearized perturbations around flat spacetime, $g_{\mu\nu}=\eta_{\mu\nu}+h_{\mu\nu}$ with $B_\mu=b_\mu+\beta_\mu$, induce anisotropic corrections to the gravitational field and alter the propagation of gravitational waves. ModMax black holes, characterized by the nonlinear electromagnetic parameter $\lambda$, modify the electric field contribution and horizon geometry, recovering the standard RN solution in the limit $\lambda\to 0$. Observable quantities are sensitive to the parameters $\alpha$, $\ell$, $Q$, and $\lambda$: the charge $Q$ and Lorentz-violation $\ell$ reduce both absorption and Hawking radiation flux, thereby enhancing sparsity; $\alpha$ adjusts the asymptotic form of the effective potential; and the nonlinearity parameter $\lambda$ diminishes barrier suppression through a factor $e^{-\lambda}$, partially restoring mode transmission and reducing sparsity.

This paper is organized as follows: in Sec. (\ref{s2}), we present the theoretical framework of bumblebee gravity and derive the ModMax BH solution in the presence of a cloud of strings, specifying the associated geometrical structure and physical characteristics of the spacetime. Also, in Sec. (\ref{s3}), we examine the geodesic configuration, including null trajectories, and determine the photon sphere together with the corresponding BH shadow through the analysis of light propagation. Sec. (\ref{s4}) illustrates the thermodynamic properties of the system, where fundamental quantities such as temperature, entropy, and stability conditions are systematically defined. In this case, in Sec. (\ref{s5}), we investigate the sparsity of Hawking radiation and analyze its dependence on the relevant parameters governing the model. In Sec. (\ref{s6}), we study the greybody factors and absorption probabilities by considering different field perturbations, defined scalar (spin 0) in Sec. (\ref{s6-1}), vector (spin 1) in Sec. (\ref{s6-2}), and tensor (spin 2) in Sec. (\ref{s6-3}). In this context, in Sec. (\ref{S7}), we explain the principal results and provide concluding remarks.

\section{Bumblebee gravity and ModMax black hole with cloud strings}\label{s2}

In this section, we discuss some aspects of Bumblebee gravity and derive a ModMax black hole solution with cloud strings in Bumblebee gravity. In this gravitational framework, local Lorentz symmetry is spontaneously broken by a vector field $B_\mu$ acquiring a nonzero vacuum expectation value (VEV). We will use a nonminimal gravitational action given by \cite{Maluf:2020kgf,Liu:2024axg,Li:2026uwx}
\begin{align}
S=\int d^4x\,\sqrt{-g}\bigg[
\frac{1}{2\kappa}\Big(R-2\Lambda+\xi\,B^\mu B^\nu R_{\mu\nu}\Big)
-\frac14 B_{\mu\nu}B^{\mu\nu}
-V\!\left(X\right)
+\mathcal L_{\rm m}+\mathcal L_{\rm CS}
\bigg],
\label{eq:bb_action}
\end{align}
where $\kappa\equiv 8\pi G$, $\xi$ is a nonminimal curvature coupling,
$B_{\mu\nu}\equiv \nabla_\mu B_\nu-\nabla_\nu B_\mu$ is the field strength associated to field $B_\mu$, $X\equiv B_\mu B^\mu\mp b^2$ is the argument of the potential, and $b^2>0$ sets the symmetry-breaking scale. In this context, the potential $V(X)$ plays an essential role since it enforces a vacuum manifold $X=0$, so that $\langle B_\mu\rangle \equiv b_\mu$, $b_\mu b^\mu=\pm b^2$, where the upper (lower) sign corresponds to a timelike (spacelike) VEV. A common sigma-model choice $V=\lambda X$ with a Lagrange multiplier $\lambda$ imposes $X=0$ as a constraint, while a smooth Mexican-hat
potential, e.g. $V=\frac{\lambda}{2}X^2$, generates massive excitations orthogonal to the vacuum manifold.
The spontaneous nature of the breaking implies that diffeomorphism invariance is preserved in the action,
but the vacuum selects a preferred direction in local frames.

Variation of \eqref{eq:bb_action} with respect to $g^{\mu\nu}$ yields modified Einstein equations,
\begin{align}
G_{\mu\nu}+\Lambda g_{\mu\nu}
=\kappa\Big(T_{\mu\nu}^{\rm m}+T_{\mu\nu}^{B}\Big)
+\mathcal K_{\mu\nu}+T^{\rm CS}_{\mu\nu},
\label{eq:bb_einstein}
\end{align}
where the bumblebee stress tensor (for the Maxwell-type kinetic term and potential) is
\begin{align}
T_{\mu\nu}^{B}
&=
B_{\mu\alpha}B_{\nu}{}^{\alpha}
-\frac14 g_{\mu\nu}B_{\alpha\beta}B^{\alpha\beta}
-g_{\mu\nu}V(X)+2V_X\,B_\mu B_\nu,
\label{eq:bb_TB}\\
V_X&\equiv \frac{dV}{dX},
\end{align}
and the curvature-coupling contribution can be expressed as
\begin{align}
\mathcal K_{\mu\nu}=&\xi\bigg[\frac{1}{2}\,g_{\mu\nu}\,B^\alpha B^\beta R_{\alpha\beta}-
\frac12\nabla_\alpha\nabla_\mu\!\left(B^\alpha B_\nu\right)
-\frac12\nabla_\alpha\nabla_\nu\!\left(B^\alpha B_\mu\right)+\frac12\nabla^2\!\left(B_\mu B_\nu\right)
+\frac12 g_{\mu\nu}\nabla_\alpha\nabla_\beta\!\left(B^\alpha B^\beta\right)\bigg].
\label{eq:bb_Kmunu}
\end{align}
The variation of \eqref{eq:bb_action} with respect to $B_\mu$ provides the bumblebee equation of motion, namely \cite{Maluf:2020kgf}
\begin{align}
\nabla_\mu B^{\mu\nu}=2V_X\,B^\nu-\frac{\xi}{\kappa}\,B_\mu R^{\mu\nu}.
\label{eq:bb_Beom}
\end{align}
For the Lagrange-multiplier potential $V=\lambda X$, one has $V_X=\lambda$ and the constraint $X=0$ is enforced.
On the vacuum manifold with $X=0$ and $V=0$ (and, in many solutions, also $V_X=0$), the vector field can be covariantly frozen to a background configuration $B_\mu=b_\mu(x)$, but the nonminimal coupling $\xi\,B^\mu B^\nu R_{\mu\nu}$ generically still feeds back into the metric equations through \eqref{eq:bb_einstein}.

It is important to highlight two simplifying branches that are frequently employed. The first one is the Potential-minimized branch, where one considers $X=0$ and $V_X=0$ (e.g. in a smooth potential at its minimum). Then $V=0$ and the source term $2V_X B^\nu$ vanishes in \eqref{eq:bb_Beom}. The second one is Constraint branch (sigma model) $X=0$ imposed by $\lambda$, so that $V=0$ but $V_X=\lambda$
need not vanish; $\lambda$ is determined by the coupled system and can act as an auxiliary source. In many black-hole applications, one adopts the potential-minimized branch to avoid additional matter sources and to highlight purely geometric consequences of spontaneous Lorentz breaking.

The energy-momentum tensor $T^{\rm CS}_{\mu\nu}$ for a static, spherically symmetric cloud of radial strings is
\begin{align}
&T^{t}{}_{t} = T^{r}{}_{r} = -\rho(r),\\
&T^{\theta}{}_{\theta} = T^{\phi}{}_{\phi} = 0,
\label{eq:Tmunu}
\end{align}
where the proper energy density is
\begin{equation}
\rho(r) = \frac{\alpha}{r^{2}}.
\end{equation}
Here \(\alpha\) (\(0 \le \alpha < 1\)) is a positive dimensionless parameter that quantifies the strength of the string cloud. This source obeys the null energy condition and behaves as an anisotropic fluid with vanishing tangential pressure.

In addition to cloud of strings, we will consider ModMax electrodynamics, which is a nonlinear, conformally invariant deformation of Maxwell theory controlled by a real parameter $\gamma\ge 0$. It is conveniently written in terms of the two electromagnetic invariants
\begin{align}
S \equiv -\frac14\,F_{\mu\nu}F^{\mu\nu},\\
P \equiv -\frac14\,F_{\mu\nu}\tilde F^{\mu\nu},
\end{align}
where $\tilde F^{\mu\nu}\equiv \frac12\,\varepsilon^{\mu\nu\alpha\beta}F_{\alpha\beta}$. With this, Maxwell theory corresponds to $\mathcal L_{\rm Mxw}=S$. The ModMax Lagrangian density may be expressed as
\begin{align}
\mathcal L_{\rm MM}(S,P;\gamma)=
S\,\cosh\gamma+\sqrt{S^2+P^2}\,\sinh\gamma,
\label{eq:modmax_lagrangian}
\end{align}
which is analytic for $(S,P)\neq (0,0)$, reduces to Maxwell at $\gamma=0$, and preserves electric-magnetic duality with a deformation of the constitutive relations. In the static and spherically symmetric case, the line element is commonly written as
\begin{align}
ds^2=-f(r)\,dt^2+\frac{dr^2}{f(r)}+r^2(d\theta^2+\sin^2\theta\,d\phi^2),
\end{align}
where the metric function $f(r)$ contains the gravitational mass contribution and the effective electric term generated by the ModMax sector. In Einstein gravity, the metric function for the ModMax black hole is given by
\begin{align}
    f(r)=1-\frac{2M}{r}+\frac{Q^2\,e^{-\lambda}}{r^2}
\end{align}

In this case, the ModMax black hole preserves the basic Reissner-Nordstr\"om-type structure, but the charge sector is effectively dressed by nonlinear electrodynamic effects, making the horizon structure, thermodynamics, and wave absorption explicitly dependent on the parameter $\lambda$. 

Now, let us obtain a ModMax black hole solution with a cloud of strings. Black holes surrounded by a cloud of strings constitute one of the simplest and most tractable departures from vacuum general relativity that still admit exact, asymptotically flat solutions. These solutions provide us with an interesting theoretical testing ground for exploring the relationship between topological defects, black hole physics, and observational tests of strong-field gravity. To obtain a consistent ModMax black hole solution in Bumblebee, we need to consider the modified Lagrangian
\begin{align}
    S=-\frac{1}{4}F_{\mu\nu}F^{\mu\nu}-\frac{\rho}{4}B_{\lambda}B^{\lambda}\,F_{\mu\nu}F^{\mu\nu}
\end{align}

Thus, the geometry is described by the following element of line
\begin{align}
ds^2=-A(r)\,dt^2+B(r)\,dr^2+r^2(d\theta^2+\sin^2\theta\,d\phi^2),
\label{eq:sss_metric}
\end{align}
where
\begin{align}
A(r)=1-\alpha-\frac{2M}{r}+\frac{2(1+\ell)\,Q^2\, e^{-\lambda}}{(2+\ell)\,r^2},
\label{eq:ModMax-A-bumblebee}
\end{align}
and
\begin{align}
B(r)=\frac{1+\ell}{A(r)},
\label{eq:ModMax-B-bumblebee}
\end{align}
From these expressions, one sees that the ModMax parameter $\lambda$ modifies the electric contribution through the factor $e^{-\lambda}$, while the Lorentz-violating parameter $\ell$ and the cloud-of-strings parameter $\alpha$ further deform the geometry.

\begin{figure}[ht!]
\begin{center}
\begin{tabular}{ccc}
\includegraphics[height=5cm]{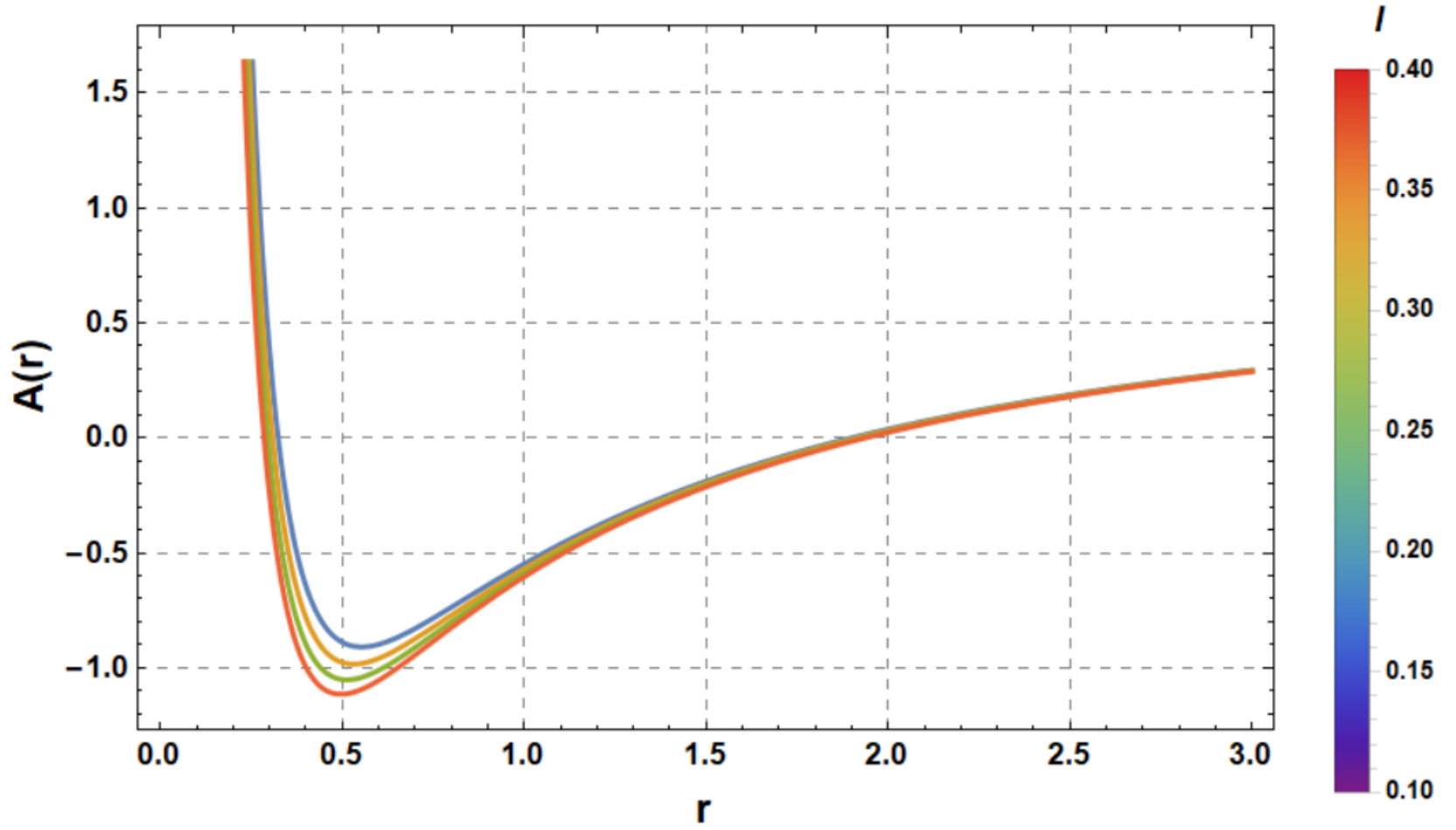} 
\includegraphics[height=5cm]{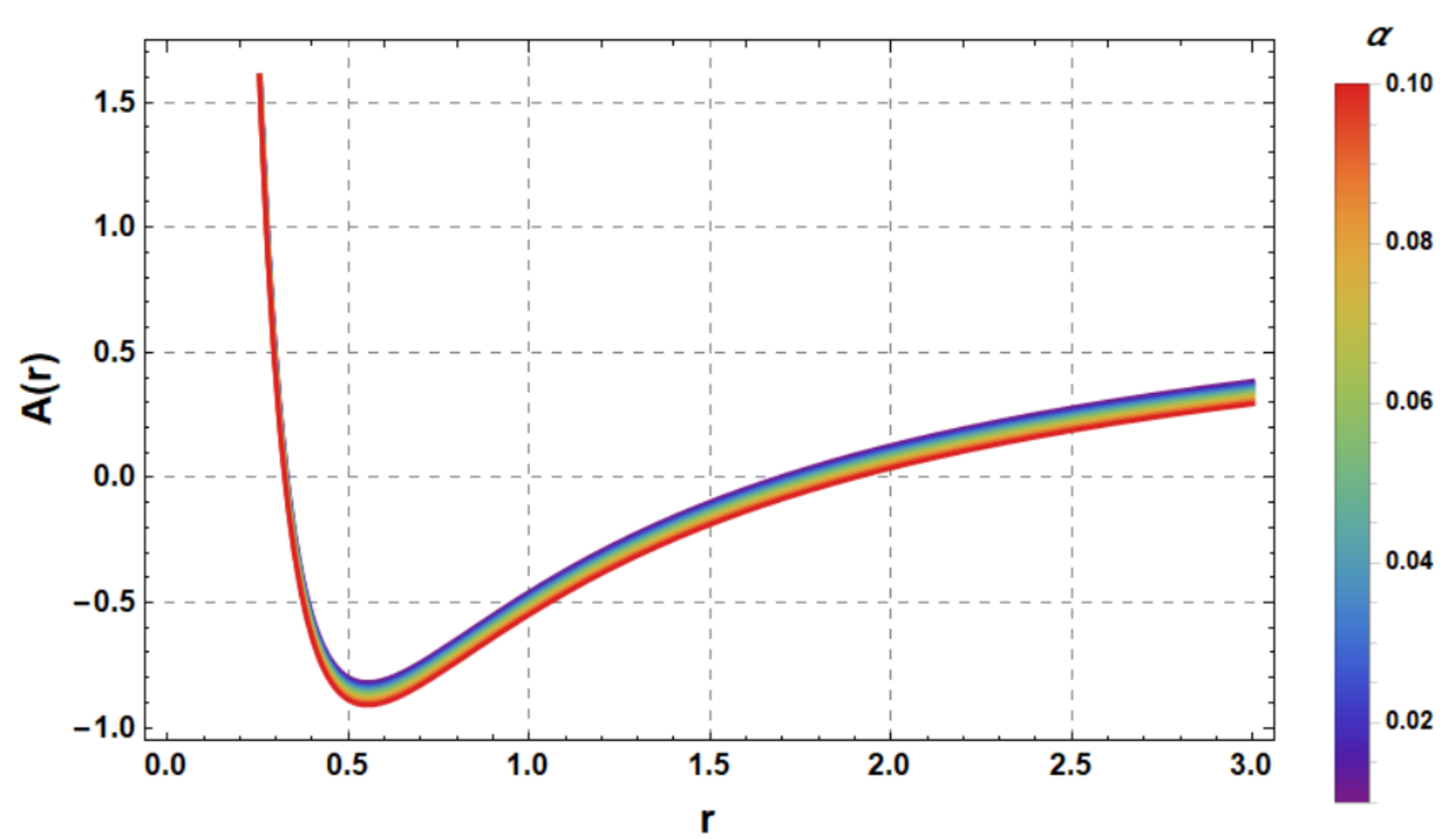}\\
(a) $Q=\lambda=0.1$ and $\alpha=0.01$\hspace{5cm}(b)$\ell=Q=\lambda=0.1$\\
\includegraphics[height=5cm]{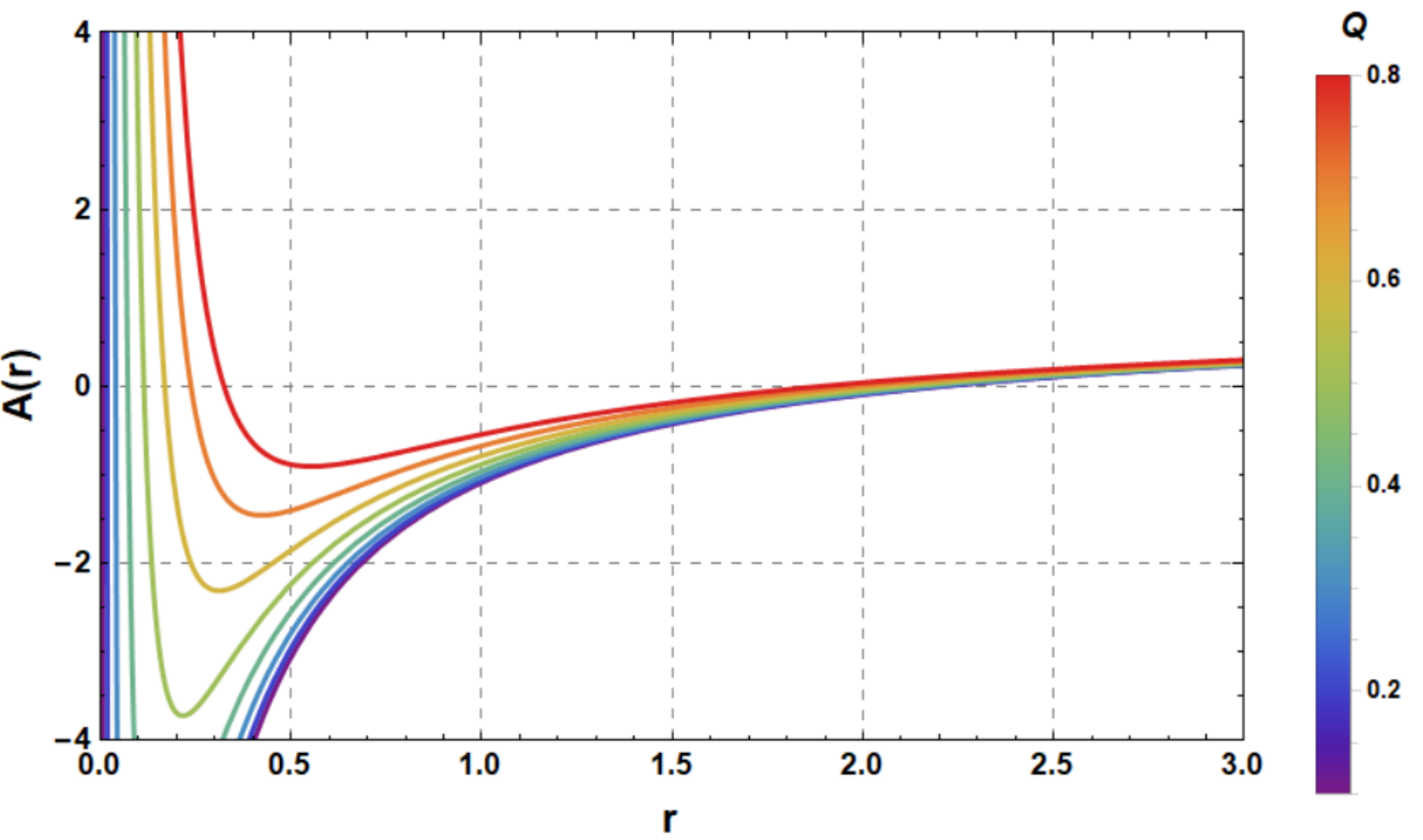}
\includegraphics[height=5cm]{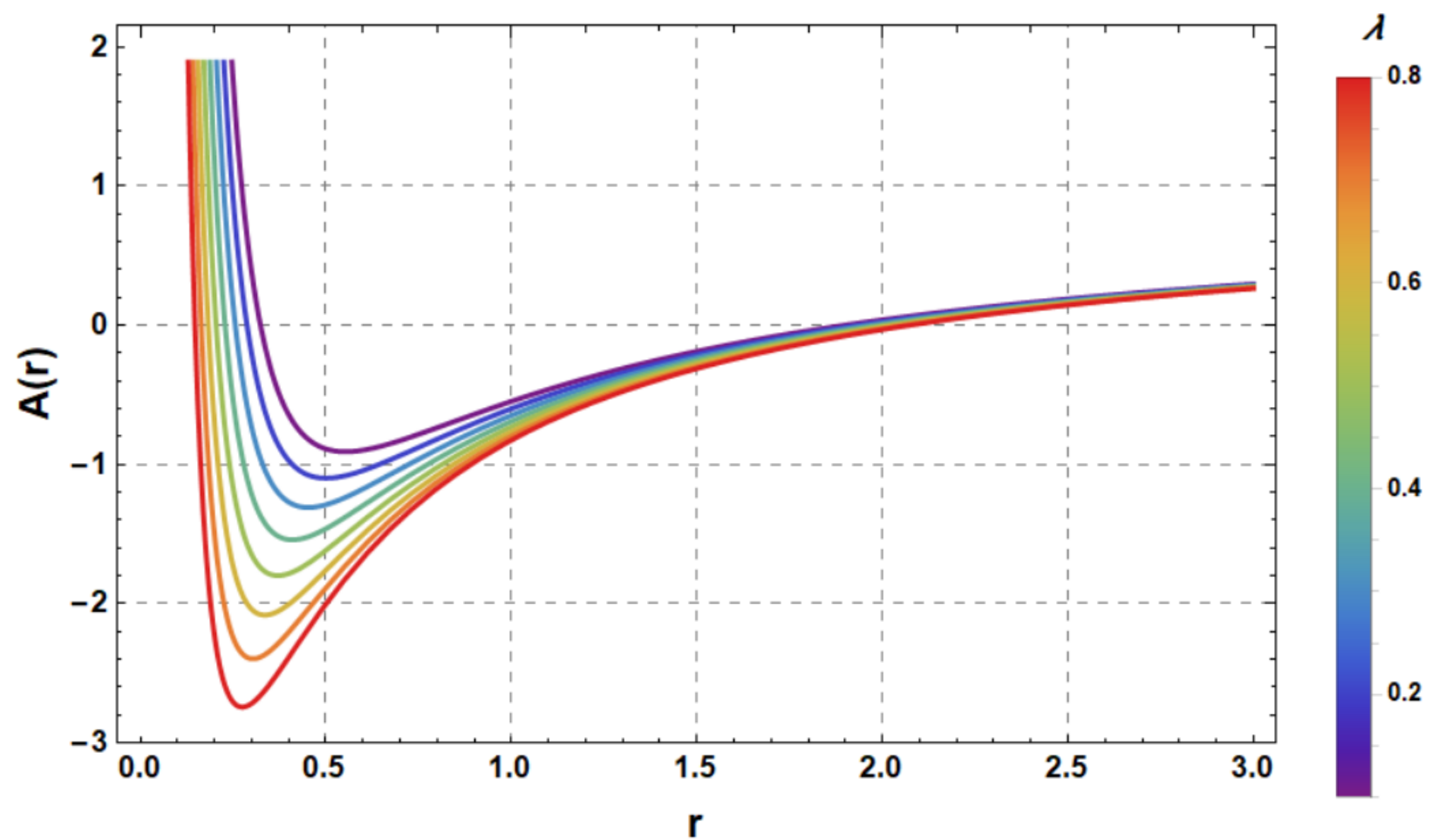}
\\
(c)$\ell=\lambda=0.1$ and $\alpha=0.01$\hspace{5cm}(d)$\ell=Q=0.1$ and $\alpha=0.01$\\ 
\end{tabular}
\end{center}
\vspace{-0.5cm}
\caption{Behavior of function $A(r)$ varying the parameters $(\alpha,\ell,Q,\lambda)$.
\label{A}}
\end{figure}

The behavior of the metric function \eqref{eq:ModMax-A-bumblebee} with relation to the black hole parameters $( \alpha,\ell,Q,\lambda)$ is illustrated in Fig.\ref{A}. This function plays a central role in determining the causal structure of the black hole, and its graphical behavior reveals how the different parameters affect the geometry. For very large radial distances, the function approaches the constant value $A(r)\to 1-\alpha$, showing that the cloud-of-strings parameter $\alpha$ shifts the asymptotic level of the curve downward. On the other hand, near the origin, the term proportional to $1/r^2$ dominates, so that the electric sector, modulated by the LV parameter $\ell$ and the ModMax parameter $\lambda$, strongly affects the behavior of the function in the small-$r$ region. The mass term $-2M/r$ contributes negatively and tends to pull the curve downward at intermediate distances, favoring the appearance of horizons. Graphically, the zeros of $A(r)$ correspond to the horizons of the black hole, and depending on the values of $\alpha$, $Q$, $\ell$, and $\lambda$, the curve may exhibit two distinct roots, a degenerate root in the extremal case, or no real root at all. In particular, increasing $Q$ enhances the positive contribution near the origin and tends to separate the inner and outer horizons, while increasing $\lambda$ suppresses the charge contribution through the factor $e^{-\lambda}$, making the curve resemble the uncharged case more closely. Likewise, the LV parameter $\ell$ changes the strength of the charge term through the factor $(1+\ell)/(2+\ell)$, thereby modifying the position of the roots and the overall shape of the curve. Therefore, the graph of $A(r)$ provides a clear and direct visualization of how the competition between the mass, charge, cloud-of-strings background, and LV effects controls the horizon structure and the global properties of the spacetime.

\section{Geodesic, photon sphere and BH shadow}\label{s3}

After obtaining the black hole solution, let us examine the geodesic associated with this geometry. To do so, we begin with the Lagrangian describing the motion of test particles in a curved spacetime, namely
\begin{equation}
L=\frac{1}{2}g_{\mu\nu}\frac{dx^{\mu}}{d\lambda}\frac{dx^{\nu}}{d\lambda},
\label{eq13}
\end{equation}
where $\lambda$ denotes an affine parameter along the geodesics. For the spherically symmetric metric~(6), restricting the motion to the equatorial plane $(\theta=\pi/2)$, the Lagrangian reduces to
\begin{equation}
L=\frac{1}{2}\left[
-f(r)\left(\frac{dt}{d\lambda}\right)^{2}
+\frac{1+\ell}{f(r)}\left(\frac{dr}{d\lambda}\right)^{2}
+r^{2}\left(\frac{d\phi}{d\lambda}\right)^{2}
\right].
\label{eq14}
\end{equation}

Because $t$ and $\phi$ are cyclic coordinates, the system admits two conserved quantities: the total energy $E$ and the angular momentum $L$ of the particle, given by
\begin{equation}
E=A(r)\frac{dt}{d\lambda},
\qquad
L=r^{2}\frac{d\phi}{d\lambda}.
\label{eq15}
\end{equation}

For null geodesics $(ds^{2}=0)$, the metric condition leads to
\begin{equation}
-A(r)\left(\frac{dt}{d\lambda}\right)^{2}
+\frac{1+\ell}{A(r)}\left(\frac{dr}{d\lambda}\right)^{2}
+r^{2}\left(\frac{d\phi}{d\lambda}\right)^{2}=0.
\label{eq16}
\end{equation}

Substituting the conserved quantities from Eq.~\eqref{eq15} into Eq.~\eqref{eq16}, the equation motion for the radial coordinate takes the form
\begin{equation}
\left(\frac{dr}{d\lambda}\right)^{2}=V_{\mathrm{eff}}(r),
\label{eq17}
\end{equation}
with the effective potential expressed as
\begin{equation}
V_{\mathrm{eff}}(r)=\frac{1}{1+\ell}\left[E^2-\frac{L^{2}}{r^{2}}\left\{1-\alpha-\frac{2M}{r}+\frac{2(1+\ell)\,Q^2\, e^{-\lambda}}{(2+\ell)\,r^2}\right\}\right].
\label{eq18}
\end{equation}

Considering the effective potential for null geodesics derived from the metric function~(19), we obtain the following expression for the effective radial force:

The photon sphere represents a spherical surface around a black hole where photons can propagate along unstable circular trajectories due to the strong curvature of spacetime. Any small perturbation will inevitably cause the photon either to plunge into the black hole horizon or to escape to infinity. This region is fundamental in gravitational lensing phenomena and directly determines the size of the black hole shadow observed by distant observers.

For circular null orbits located at $r=r_c$, the conditions
\begin{equation}
\dot{r}=0,
\qquad
\ddot{r}=0
\end{equation}
must hold. Using Eqs.~\eqref{eq17} and \eqref{eq18}, one obtains
\begin{equation}
E^{2}=\frac{L^{2}}{r^{2}}\left[1-\alpha-\frac{2M}{r}+\frac{2(1+\ell)\,Q^2\, e^{-\lambda}}{(2+\ell)\,r^2}\right]\Bigg|_{r=r_c},
\label{eq22}
\end{equation}
which leads to the definition of the critical impact parameter
\begin{equation}
\beta_c=\frac{L_{\mathrm{(ph)}}}{E_{\mathrm{(ph)}}}
=\frac{r}{\sqrt{1-\alpha-\frac{2M}{r}+\frac{2(1+\ell)\,Q^2\, e^{-\lambda}}{(2+\ell)\,r^2}}}\Bigg|_{r=r_c}.
\label{eq23}
\end{equation}

The photon sphere is determined from the null circular orbit condition
\begin{equation}
\frac{dV_{\rm eff}}{dr}=0.\label{eq24}
\end{equation}
The following quadratic equation will arrive
\begin{equation}
    (1-\alpha)r^2-3 M r+\frac{1+\ell}{2+\ell}\,e^{-\lambda} Q^2=0.
\end{equation}
Its real valued analytical solution will give us the photon sphere radius as,
\begin{equation}
    r_s=\frac{3 M}{2(1-\alpha)}\left[1+\sqrt{1-\frac{1+\ell}{2+\ell}\,\frac{8 e^{-\lambda} Q^2 (1-\alpha)}{9 M^2}}\right].
\end{equation}
One can see that the hoton sphere exists provided we have the following constraint
\begin{equation}
    Q^2 <\frac{2+\ell}{1+\ell}\frac{9 M^2 e^{\lambda}}{8 (1-\alpha)}.
\end{equation}

As the selected spacetime is not asymptotically flat, we followed the procedure in \cite{Perlick2022} to determine the shadow radius. For a static observation located at position $r_O$, the shadow radius is given by
\begin{equation}
    R_{\rm sh}=r_s \sqrt{\frac{1-\alpha-\frac{2M}{r_O}+\frac{2(1+\ell)\,Q^2\, e^{-\lambda}}{(2+\ell)\,r^2_O}}{1-\alpha-\frac{2M}{r_s}+\frac{2(1+\ell)\,Q^2\, e^{-\lambda}}{(2+\ell)\,r^2_s}}}
\end{equation}

For a distant observer, the shadow radius simplifies as
\begin{equation}
    R_{\rm sh}=\frac{r_s (1-\alpha)^{1/2}}{\sqrt{1-\alpha-\frac{2M}{r_s}+\frac{2(1+\ell)\,Q^2\, e^{-\lambda}}{(2+\ell)\,r^2_s}}}=(1-\alpha)^{1/2} \beta_c
\end{equation}

Finally, we focus on photon trajectories and analyze how the LV, string, and ModMax parameters influence the trajectories of light. The equation of the orbit is given by
\begin{equation}
    \left(\frac{dr}{d\phi}\right)^2=\frac{r^4}{1+\ell}\left[\frac{1}{\beta^2}-\frac{1-\alpha}{r^2}+\frac{2 M}{r^3}-\frac{2(1+\ell)\,Q^2\, e^{-\lambda}}{(2+\ell) r^4}\right]
\end{equation}
Transforming to a new variable via $u(\phi)=\frac{1}{r(\phi)}$ and after simplification results
\begin{equation}
    (1+\ell)\left(\frac{du}{d\phi}\right)^2+(1-\alpha)u^2=\frac{1}{\beta^2}+2 M u^3-\frac{2(1+\ell)\,Q^2\, e^{-\lambda}}{(2+\ell) } u^4. 
\end{equation}
Differentiating both sides w. to $\phi$ and after simplification results
\begin{equation}
(1+\ell)\,\frac{d^2u}{d\phi^2}+(1-\alpha) u=3 M u^2-\frac{(1+\ell)\,Q^2\, e^{-\lambda}}{(2+\ell) } u^3
\end{equation}

The perturbative solution of the above second-order differential equation by considering $u \simeq u_0+\epsilon u_1$, where $\epsilon \ll 1$ is given by

\begin{align}
u(\phi) &= \frac{\cos(k\phi)}{\beta} + \epsilon \Bigg[ 
C \cos(k\phi) + D \sin(k\phi) 
+ \frac{3M}{2 k^2 (1+\ell) \beta^2} 
- \frac{M}{2 k^2 (1+\ell) \beta^2} \cos(2 k \phi)- \frac{6 Q^2 e^{-\lambda}}{4 k (2+\ell)^2 \beta^3} \, \phi \sin(k\phi) \notag \\
&\quad - \frac{ Q^2 e^{-\lambda}}{8 k^2 (2+\ell)^2 \beta^3} \cos(3 k \phi)
\Bigg],
\end{align}
where $C, D$ are arbitrary constants and $k^2 = \frac{1-\alpha}{1+\ell}$.

In that scenario, the deflection angle of light is obtained as
\begin{equation}
\hat{\alpha} \approx 
\frac{4 M (1+\ell)^2}{(1-\alpha)^3\, \beta} 
- \frac{3 \pi (2+\ell)(1+\ell) Q^2 e^{-\lambda}}{8  (1-\alpha)^3 \, \beta^2},
\end{equation}
where $\beta$ is the impact parameter for light.

The behavior of $\hat{\alpha}$ in various limits can be discussed as follows:

\begin{itemize}
    \item \textbf{Schwarzschild limit:} Setting $\ell = 0$, $\alpha = 0$, and $Q = 0$ reduces the expression to
    \begin{equation}
        \hat{\alpha} \approx \frac{4 M}{\beta},
    \end{equation}
    which reproduces the classical general relativity result for light deflection by a Schwarzschild black hole.

    \item \textbf{Reissner-Nordstr\"{o}m limit:} For $Q \neq 0$ but $\ell = 0 = \alpha$, we obtain
    \begin{equation}
        \hat{\alpha} \approx \frac{4 M}{\beta} - \frac{3 \pi Q^2}{8 \beta^2},
    \end{equation}
    showing the reduction of the deflection angle due to the black hole charge, consistent with the Reissner-Nordstr\"{o}m solution.
\end{itemize}

\section{Thermodynamics}\label{s4}

In this section, we will study the thermodynamic properties of the static and spherically symmetric black hole described by \eqref{eq:sss_metric}. This study is relevant in this context since it establishes a deep connection between gravity, quantum theory, and statistical physics. In particular, thermodynamic quantities, such as the Hawking temperature, entropy, heat capacity, and free energy, provide essential information about the physical viability and stability of a given black hole solution. Through these quantities, one can determine whether the geometry admits extremal states, whether it undergoes phase transitions, and whether it is locally or globally stable under thermal fluctuations. 

Since we deal with black hole solutions in modified gravity scenarios and nonlinear electrodynamics, thermodynamic analysis becomes even more relevant, since additional parameters may meaningfully alter the horizon structure and the thermal behavior of the solution. In the present context, the parameters associated with Lorentz violation, the cloud of strings, electric charge, and the ModMax sector not only modify the metric functions but also leave direct imprints on the temperature, entropy, and stability properties of the black hole, as we will see ahead. In this sense, this analysis will be an important step toward understanding the physical consistency of the black hole solution discussed previously. Besides, it will help us clarify how each parameter affects its horizon physics and possible phase structure.

For convenience, let us initially define
\begin{align}
\beta\equiv \frac{2(1+\ell)\,Q^2\,e^{-\lambda}}{(2+\ell)},
\label{eq:def_beta}
\end{align}
so that the lapse function becomes
\begin{align}
A(r)=1-\alpha-\frac{2M}{r}+\frac{\beta}{r^2}.
\label{eq:A_beta}
\end{align}
Besides, the event horizon is determined by the largest root of $A(r)=0$, namely
\begin{align}
(1-\alpha)r^2-2Mr+\beta=0.
\end{align}
Thus, the two horizons are
\begin{align}
r_{\pm}=\frac{M\pm \sqrt{M^2-(1-\alpha)\beta}}{1-\alpha},
\label{eq:rpm}
\end{align}
provided that
\begin{align}
M^2\geq (1-\alpha)\beta
=\frac{2(1-\alpha)(1+\ell)}{(2+\ell)}\,Q^2 e^{-\lambda}.
\label{eq:extremality_condition}
\end{align}
The extremal configuration is obtained when the discriminant vanishes,
$M^2=(1-\alpha)\beta$, for which the two horizons coincide:
\begin{align}
r_{+}=r_{-}=r_{\rm ext}=\frac{M}{1-\alpha}
=\sqrt{\frac{\beta}{1-\alpha}}.
\label{eq:rext}
\end{align}

Equation \eqref{eq:extremality_condition} shows that the cloud-of-strings parameter $\alpha$ weakens the effective asymptotic gravitational potential through the factor $(1-\alpha)$, while the combination $\beta$ encodes the joint effect of the electric sector, ModMax nonlinearity, and Lorentz violation. In particular, we observe that
\begin{itemize}
\item increasing $Q$ increases $\beta$, making extremality easier to reach;
\item increasing $\lambda$ decreases $e^{-\lambda}$ and therefore decreases $\beta$, pushing the system away from extremality;
\item increasing $\ell$ modifies $\beta$ through the factor $(2+\ell)/(1+\ell)$ and also affects the radial sector through $B(r)$.
\end{itemize}

From the horizon condition $A(r_h)=0$, where $r_h\equiv r_+$ is the event-horizon radius, one finds
\begin{align}
2M=(1-\alpha)r_h+\frac{\beta}{r_h},
\end{align}
or equivalently,
\begin{align}
M(r_h,Q)=\frac{1-\alpha}{2}\,r_h+\frac{\beta}{2r_h}.
\label{eq:mass_rh}
\end{align}
Restoring the original parameters, we have
\begin{align}
M(r_h,Q)=\frac{1-\alpha}{2}\,r_h
+\frac{(1+\ell)\,Q^2 e^{-\lambda}}{(2+\ell)\,r_h}.
\label{eq:mass_rh_explicit}
\end{align}
This expression clearly separates two contributions: a geometric term proportional to $(1-\alpha)r_h$, and an effective charge contribution proportional to $Q^2 e^{-\lambda}/r_h$. 

\begin{figure}[ht!]
\begin{center}
\begin{tabular}{ccc}
\includegraphics[height=5cm]{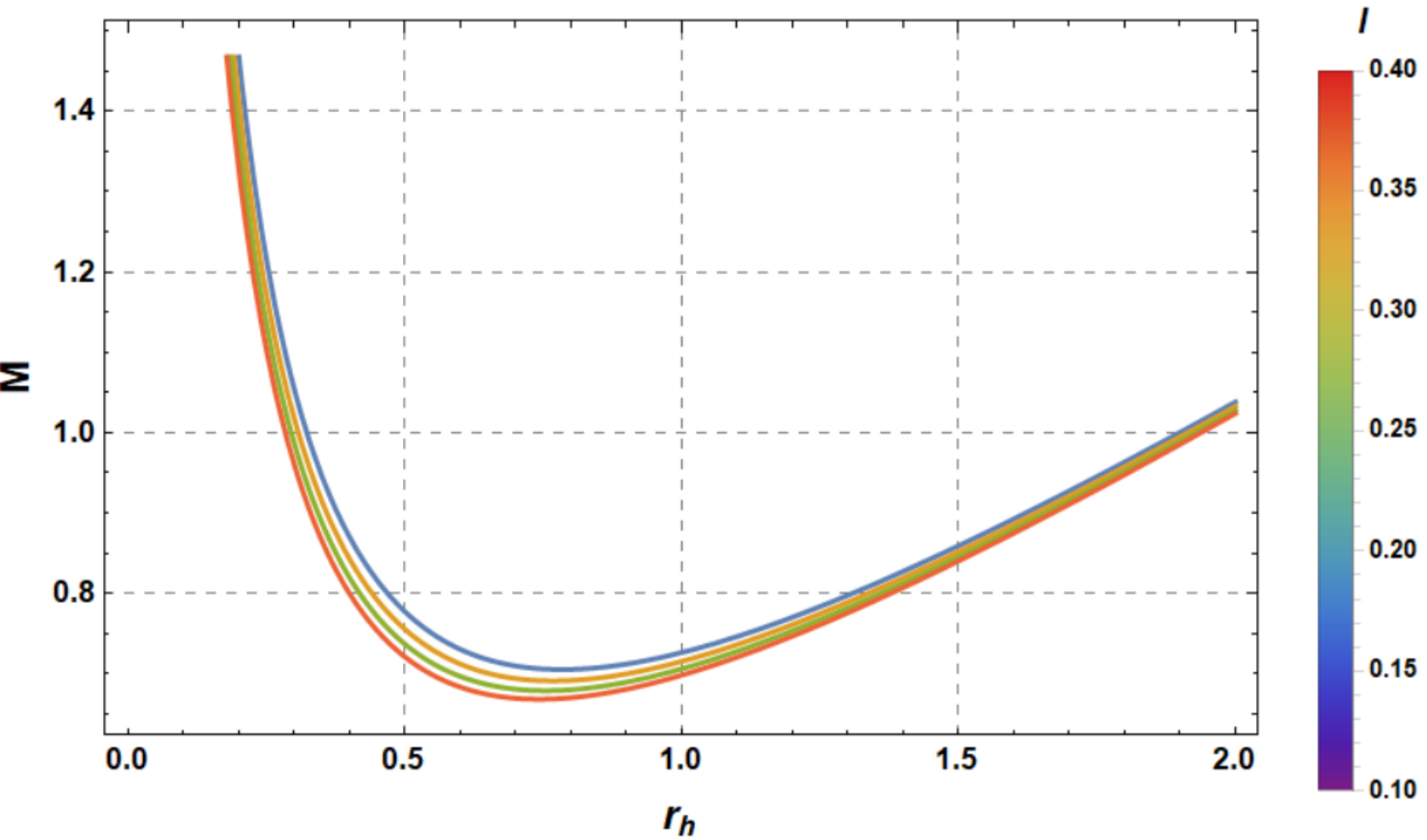} 
\includegraphics[height=5cm]{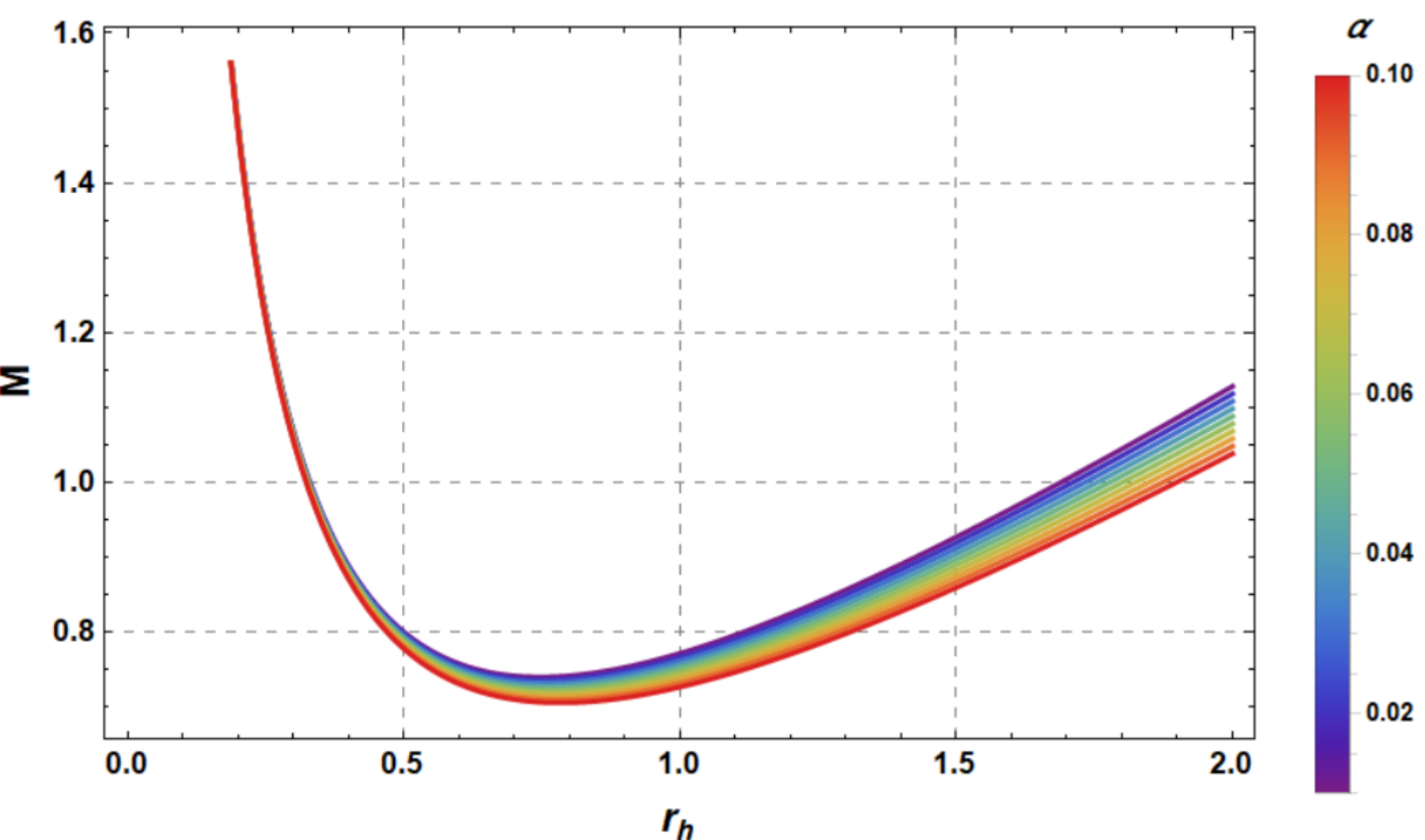}\\
(a) $Q=\lambda=0.1$ and $\alpha=0.01$\hspace{5cm}(b)$\ell=Q=\lambda=0.1$\\
\includegraphics[height=5cm]{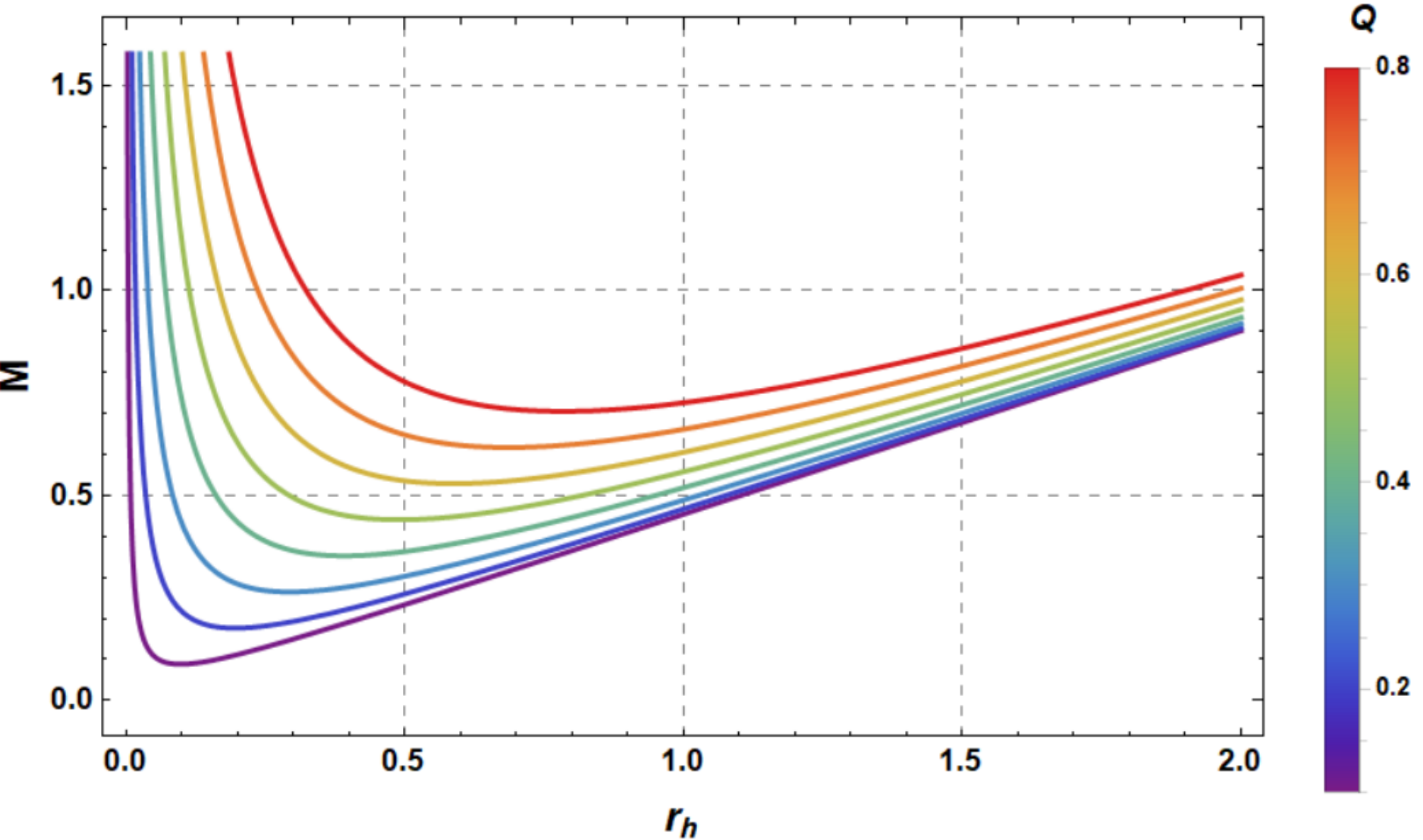}
\includegraphics[height=5cm]{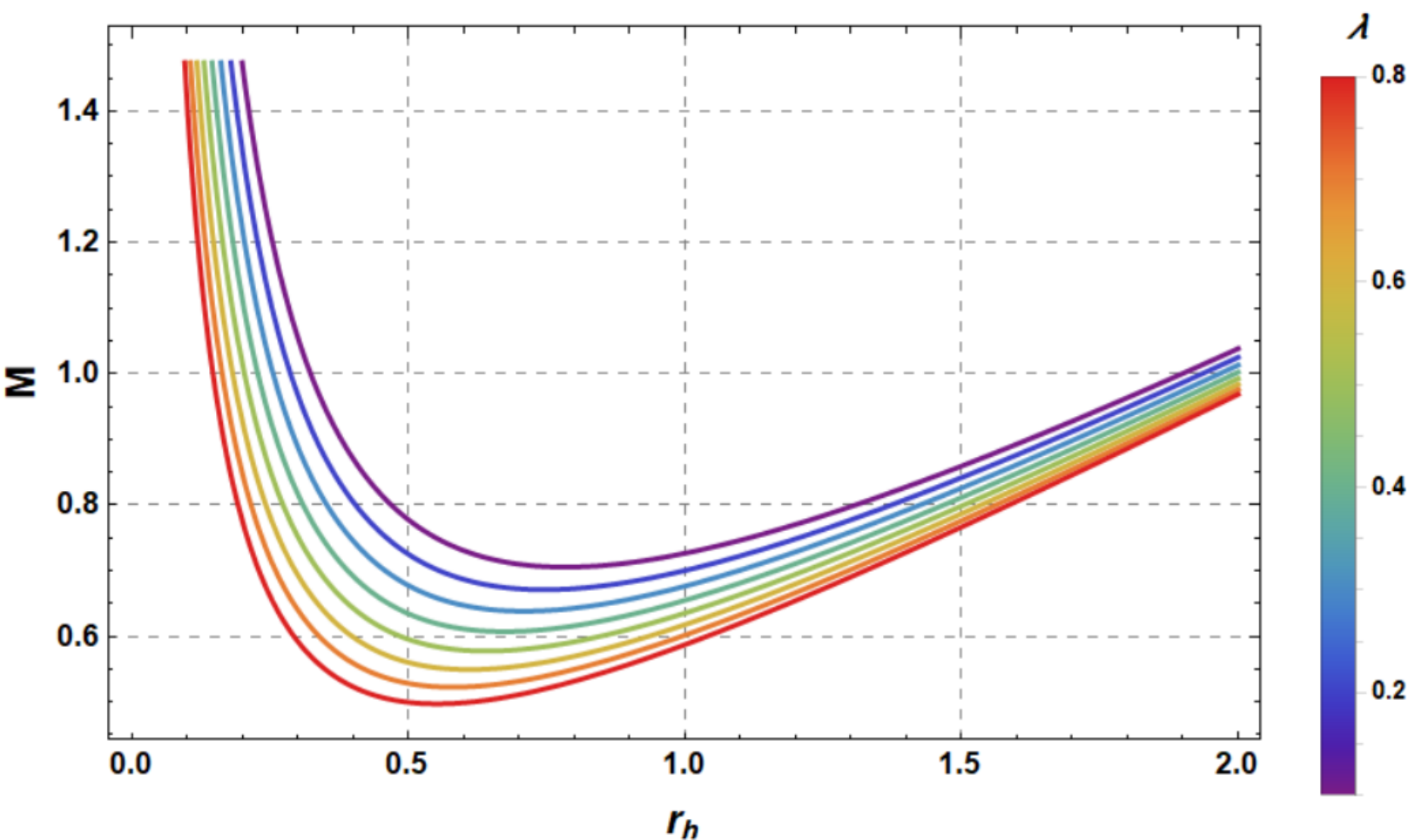}
\\
(c)$\ell=\lambda=0.1$ and $\alpha=0.01$\hspace{5cm}(d)$\ell=Q=0.1$ and $\alpha=0.01$\\ 
\end{tabular}
\end{center}
\vspace{-0.5cm}
\caption{Behavior of $M$ varying the parameters $(\alpha,\ell,Q,\lambda)$.
\label{M}}
\end{figure}

Figure~\ref{M} displays the behavior of the black hole mass $M$ as a function of the event horizon radius $r_h$ for different values of the parameters $(\alpha,\ell,Q,\lambda)$. In all panels, the mass function exhibits a characteristic non-monotonic profile with a well-defined minimum, which corresponds to the extremal configuration where the two horizons coincide. This minimum signals the transition between configurations with two horizons and those without horizons, thus encoding the extremality condition discussed in Eq.~\eqref{eq:extremality_condition}. 
In panel (a), varying the Lorentz-violating parameter $\ell$ shifts the curves upward, indicating that Lorentz violation increases the effective contribution of the charge sector and requires a larger mass to sustain a given horizon radius. In panel (b), increasing the cloud-of-strings parameter $\alpha$ lowers the mass for fixed $r_h$, reflecting the fact that the string cloud weakens the effective gravitational attraction through the factor $(1-\alpha)$, thereby reducing the energy required to form the black hole. Panel (c) shows that the electric charge $Q$ significantly raises the mass and deepens the minimum, reinforcing the role of the electromagnetic sector in stabilizing near-extremal configurations and enlarging the parameter space where two horizons exist. Finally, panel (d) illustrates the effect of the ModMax parameter $\lambda$, where increasing $\lambda$ decreases the mass due to the exponential suppression of the effective charge contribution via $e^{-\lambda}$, making the system approach the uncharged limit.
From a thermodynamic perspective, the position of the minimum of $M(r_h)$ is directly related to the extremal radius and the zero-temperature limit. The dependence of this minimum on $(\alpha,\ell,Q,\lambda)$ reveals how each parameter controls the onset of extremality and the stability of the black hole branches. In particular, parameters that enhance the effective charge contribution, such as $Q$ and $\ell$, tend to deepen the minimum and favor the existence of near-extremal, thermodynamically stable configurations, whereas parameters that suppress it, such as $\lambda$, or weaken gravity, such as $\alpha$, shift the system away from extremality. Therefore, Fig.~\ref{M} provides a clear geometric and thermodynamic visualization of how Lorentz violation, nonlinear electrodynamics, and string cloud effects interplay in determining the mass spectrum and horizon structure of the solution.

For a static metric of the form
\begin{align}
ds^2=-A(r)\,dt^2+\frac{1+\ell}{A(r)}\,dr^2+r^2d\Omega^2,
\end{align}
the surface gravity is
\begin{align}
\kappa=\frac{A'(r_h)}{2\sqrt{1+\ell}},
\end{align}
and therefore the Hawking temperature is
\begin{align}
T_H=\frac{\kappa}{2\pi}
=\frac{A'(r_h)}{4\pi\sqrt{1+\ell}}.
\label{eq:T_from_kappa}
\end{align}
Since
\begin{align}
A'(r)=\frac{2M}{r^2}-\frac{2\beta}{r^3},
\end{align}
using \eqref{eq:mass_rh} we obtain
\begin{align}
A'(r_h)=\frac{1-\alpha}{r_h}-\frac{\beta}{r_h^3},
\end{align}
and hence
\begin{align}
T_H=\frac{1}{4\pi\sqrt{1+\ell}}
\left(
\frac{1-\alpha}{r_h}-\frac{\beta}{r_h^3}
\right).
\label{eq:T_beta}
\end{align}
In terms of the original parameters,
\begin{align}
T_H=
\frac{1}{4\pi\sqrt{1+\ell}}
\left[
\frac{1-\alpha}{r_h}
-\frac{2(1+\ell)\,Q^2 e^{-\lambda}}{(2+\ell)\,r_h^3}
\right].
\label{eq:T_explicit}
\end{align}

The extremal limit is characterized by $T_H=0$, which implies $(1-\alpha)r_h^2=\beta$, in agreement with \eqref{eq:rext}. From \eqref{eq:T_explicit} we see that increasing $\alpha$ lowers the temperature, since it reduces the coefficient of the $1/r_h$ term, while
increasing $Q$ lowers the temperature, as expected from the repulsive contribution of the electric sector. On the other hand, increasing $\lambda$ increases the temperature, because the effective charge contribution is exponentially suppressed by $e^{-\lambda}$, and increasing $\ell$ lowers the temperature both through the overall factor $1/\sqrt{1+\ell}$ and through the charge-dependent term. Therefore, the LV parameter $\ell$ tends to cool the black hole, while the ModMax parameter $\lambda$ weakens the charge sector and tends to heat it up.

\begin{figure}[ht!]
\begin{center}
\begin{tabular}{ccc}
\includegraphics[height=5cm]{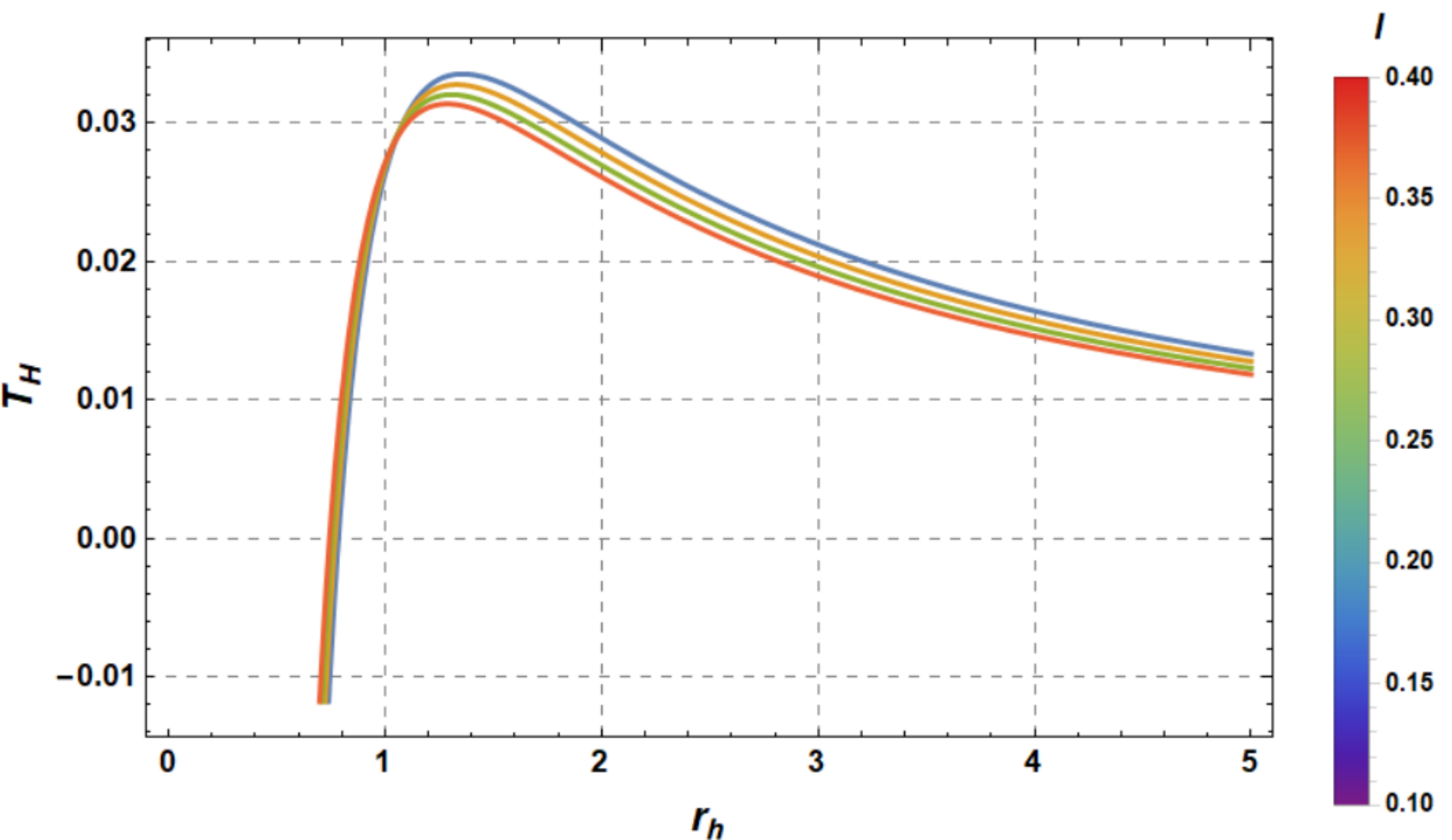} 
\includegraphics[height=5cm]{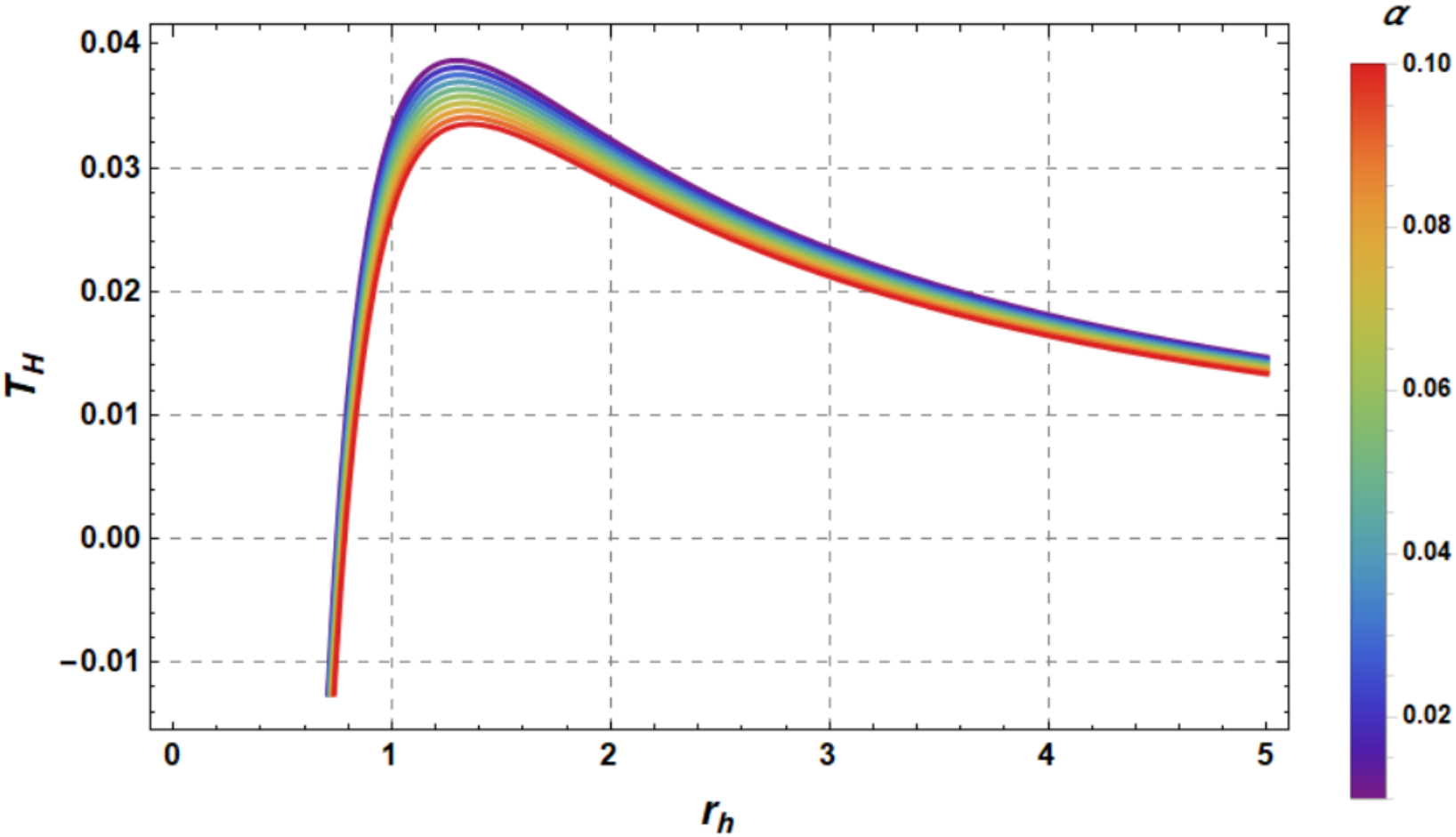}\\
(a) $Q=\lambda=0.1$ and $\alpha=0.01$\hspace{5cm}(b)$\ell=Q=\lambda=0.1$\\
\includegraphics[height=5cm]{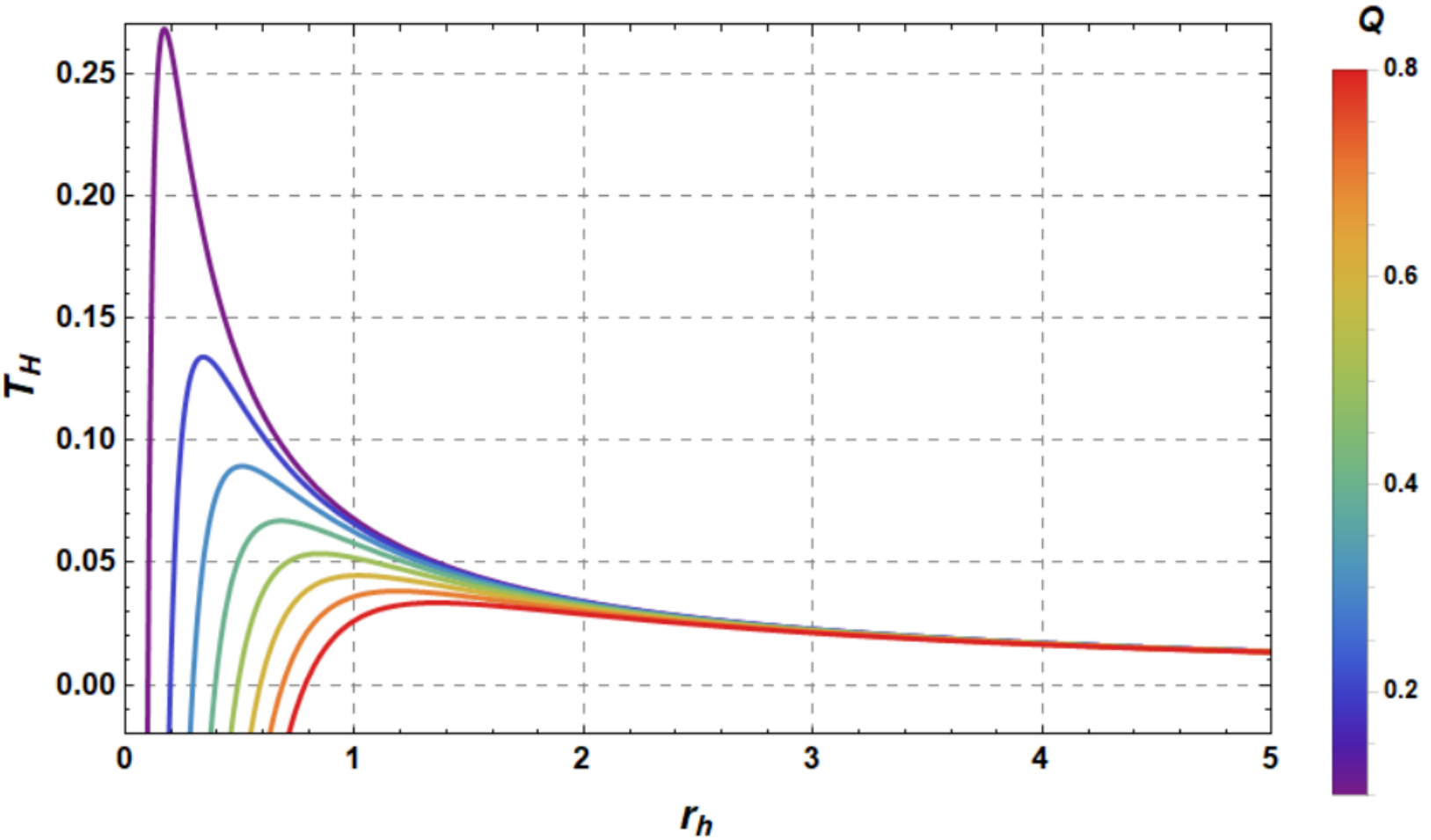}
\includegraphics[height=5cm]{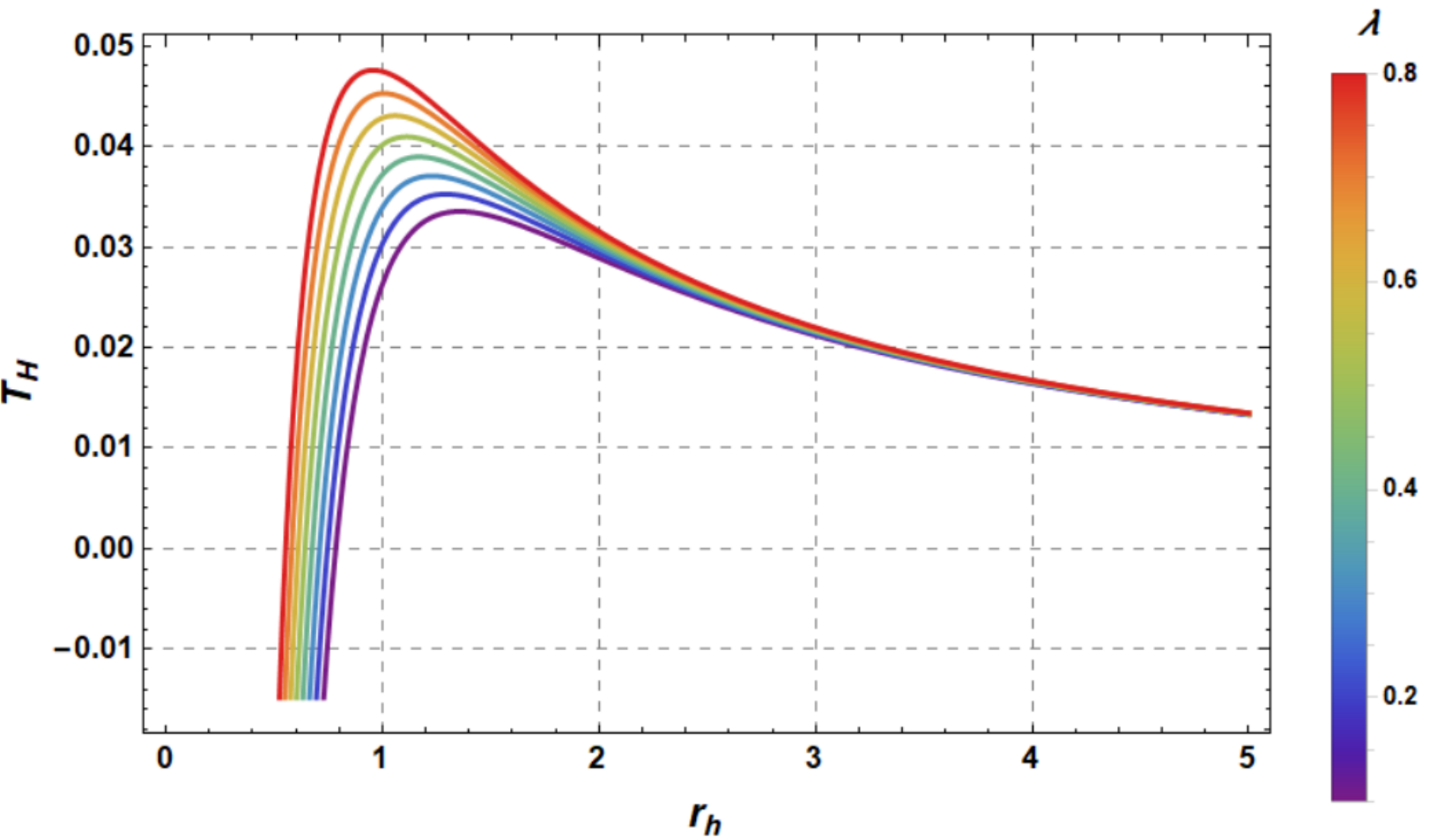}
\\
(c)$\ell=\lambda=0.1$ and $\alpha=0.01$\hspace{5cm}(d)$\ell=Q=0.1$ and $\alpha=0.01$\\  
\end{tabular}
\end{center}
\vspace{-0.5cm}
\caption{Behavior of $T_H$ varying the parameters $(\alpha,\ell,Q,\lambda)$.
\label{TH}}
\end{figure}

Figure~\ref{TH} shows the behavior of the Hawking temperature $T_H$ as a function of the event horizon radius $r_h$ for different values of the parameters $(\alpha,\ell,Q,\lambda)$. In all panels, the temperature exhibits a characteristic non-monotonic profile: starting from zero at the extremal radius, it increases to a maximum and then gradually decreases for large $r_h$. This behavior reflects the existence of a near-extremal cold branch, a thermodynamically active intermediate regime, and a large-black-hole regime where the temperature decays as $T_H \sim 1/r_h$, consistent with asymptotically flat geometries.
In panel (a), increasing the Lorentz-violating parameter $\ell$ lowers the temperature across the entire range of $r_h$. This effect originates from both the overall factor $1/\sqrt{1+\ell}$ in Eq.~\eqref{eq:T_explicit} and the modification of the effective charge sector, indicating that Lorentz violation tends to cool the black hole and suppress its thermal emission. In panel (b), the cloud-of-strings parameter $\alpha$ also reduces the temperature, as it weakens the effective gravitational attraction through the factor $(1-\alpha)$, thereby shifting the curves downward and enlarging the low-temperature regime.
Panel (c) highlights the role of the electric charge $Q$, which strongly suppresses the temperature, especially near the extremal region. As $Q$ increases, the extremal radius shifts to larger values and the maximum temperature decreases, reflecting the well-known effect that charge stabilizes the black hole and drives it toward colder configurations. Finally, panel (d) shows that increasing the ModMax parameter $\lambda$ produces the opposite effect, raising the temperature by exponentially suppressing the effective charge contribution through the factor $e^{-\lambda}$. As a result, the system moves away from extremality and approaches the behavior of a neutral black hole.

To obtain the entropy consistently with the first law, we write
\begin{align}
dM = T_H\,dS + \Phi\,dQ,
\label{eq:firstlaw_basic}
\end{align}
with $\alpha$, $\ell$, and $\lambda$ held fixed. Since
\begin{align}
M(r_h,Q)=\frac{1-\alpha}{2}\,r_h+\frac{\beta}{2r_h},
\end{align}
we have
\begin{align}
\left(\frac{\partial M}{\partial r_h}\right)_Q
=
\frac{1-\alpha}{2}-\frac{\beta}{2r_h^2}.
\label{eq:dMdrh}
\end{align}
Using \eqref{eq:T_beta}, the entropy follows from
\begin{align}
\frac{dS}{dr_h}
=
\frac{1}{T_H}
\left(\frac{\partial M}{\partial r_h}\right)_Q
=
2\pi\sqrt{1+\ell}\,r_h.
\end{align}
Integrating and setting the integration constant to zero, we obtain
\begin{align}
S=\pi\sqrt{1+\ell}\,r_h^2.
\label{eq:entropy}
\end{align}

Thus, the entropy acquires a Lorentz-violating correction through the factor $\sqrt{1+\ell}$. In the limit $\ell\to 0$, one recovers the standard Bekenstein-Hawking form
\begin{align}
S\to \pi r_h^2=\frac{\mathcal{A}_h}{4},
\end{align}
where $\mathcal{A}_h=4\pi r_h^2$ is the horizon area. It is useful to express the horizon radius in terms of the entropy:
\begin{align}
r_h=\sqrt{\frac{S}{\pi\sqrt{1+\ell}}}.
\label{eq:rh_S}
\end{align}

The thermodynamic potential conjugate to $Q$ is
\begin{align}
\Phi=\left(\frac{\partial M}{\partial Q}\right)_S.
\end{align}
At fixed entropy, $r_h$ is fixed, so from \eqref{eq:mass_rh_explicit} one finds
\begin{align}
\Phi=
\frac{2(1+\ell)\,Q\,e^{-\lambda}}{(2+\ell)\,r_h}.
\label{eq:potential}
\end{align}
This shows that the ModMax parameter suppresses the electric potential exponentially, whereas the Lorentz-violating parameter modifies it algebraically. Using \eqref{eq:entropy} and \eqref{eq:potential}, the differential first law reads
\begin{align}
dM=T_H\,dS+\Phi\,dQ,
\qquad
(\alpha,\ell,\lambda=\text{const.})
\label{eq:firstlaw_final}
\end{align}
and one can verify directly that the mass satisfies the Smarr formula
\begin{align}
M=2T_H S+\Phi Q.
\label{eq:smarr}
\end{align}
Indeed,
\begin{align}
2T_HS
&=
2\left[
\frac{1}{4\pi\sqrt{1+\ell}}
\left(
\frac{1-\alpha}{r_h}-\frac{\beta}{r_h^3}
\right)
\right]
\left(
\pi\sqrt{1+\ell}\,r_h^2
\right)
\nonumber\\[1mm]
&=
\frac{(1-\alpha)r_h^2-\beta}{2r_h},
\end{align}
while
\begin{align}
\Phi Q=\frac{2(1+\ell)Q^2e^{-\lambda}}{(2+\ell)r_h}
=\frac{\beta}{r_h}.
\end{align}
Hence,
\begin{align}
2T_HS+\Phi Q
=
\frac{(1-\alpha)r_h^2+\beta}{2r_h}
=M.
\end{align}

The heat capacity at fixed charge is defined by
\begin{align}
C_Q=\left(\frac{\partial M}{\partial T_H}\right)_Q
=
\frac{\left(\frac{\partial M}{\partial r_h}\right)_Q}
{\left(\frac{\partial T_H}{\partial r_h}\right)_Q}.
\end{align}
From \eqref{eq:dMdrh} and \eqref{eq:T_beta}, we get
\begin{align}
\frac{\partial T_H}{\partial r_h}
=
\frac{1}{4\pi\sqrt{1+\ell}}
\left(
-\frac{1-\alpha}{r_h^2}+\frac{3\beta}{r_h^4}
\right),
\end{align}
so that
\begin{align}
C_Q
=
2\pi\sqrt{1+\ell}\,r_h^2\,
\frac{(1-\alpha)r_h^2-\beta}
{3\beta-(1-\alpha)r_h^2}.
\label{eq:CQ}
\end{align}
Equivalently,
\begin{align}
C_Q=
2\pi\sqrt{1+\ell}\,r_h^2\,
\frac{
(1-\alpha)r_h^2-\dfrac{2(1+\ell)Q^2e^{-\lambda}}{(2+\ell)}
}{
\dfrac{6(1+\ell)Q^2e^{-\lambda}}{(2+\ell)}-(1-\alpha)r_h^2
}.
\label{eq:CQ_explicit}
\end{align}

The zeros and divergences of $C_Q$ are physically important. In this case, $C_Q=0$ at the extremal point, namely $(1-\alpha)r_h^2=\beta$, where $T_H=0$ and $C_Q$ diverges at $(1-\alpha)r_h^2=3\beta$, which signals a second-order phase transition of Davies type. In this sense,
the sign of $C_Q$ determines local thermodynamic stability: $\beta<(1-\alpha)r_h^2<3\beta$, with $C_Q>0$, so near-extremal black holes are locally stable, whereas sufficiently large black holes satisfy $(1-\alpha)r_h^2>3\beta$, with $C_Q<0$, and are therefore locally unstable in the canonical ensemble. This behavior is analogous to the Reissner-Nordstr\"om case: the charge stabilizes the black hole near extremality, but the asymptotically flat character of the geometry leads to an unstable large-radius branch.

\begin{figure}[ht!]
\begin{center}
\begin{tabular}{ccc}
\includegraphics[height=5cm]{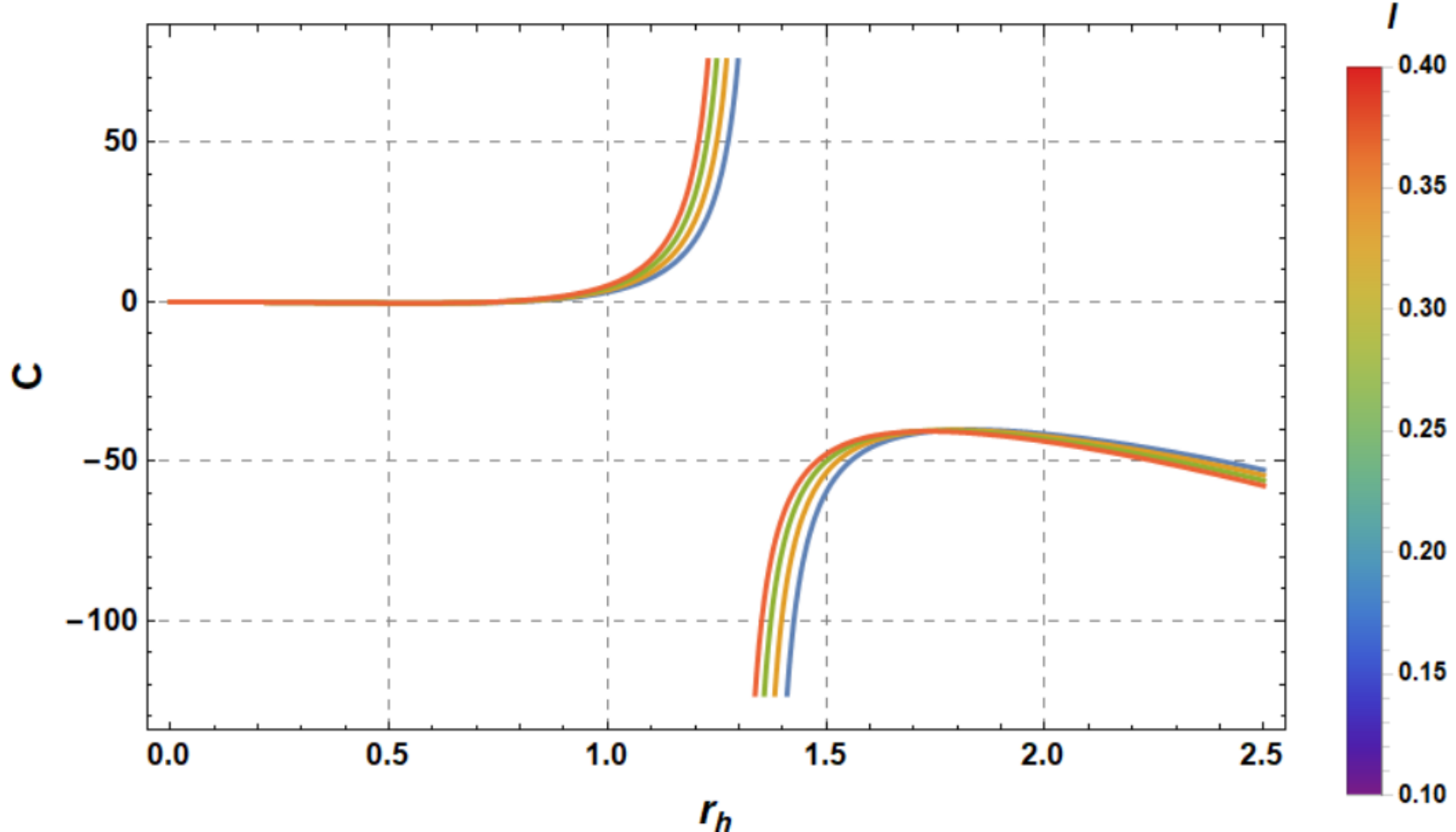} 
\includegraphics[height=5cm]{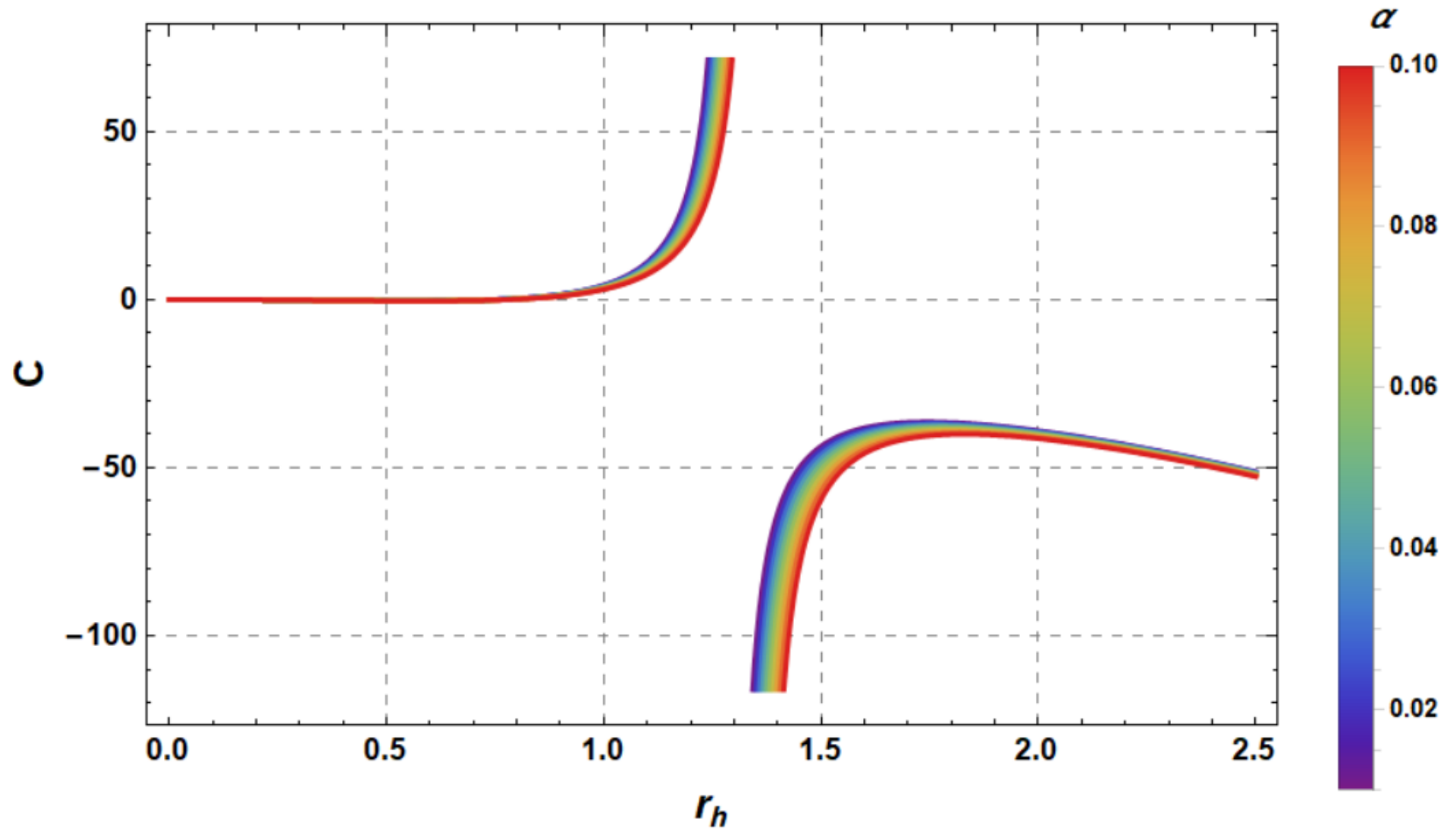}\\
(a) $Q=\lambda=0.1$ and $\alpha=0.01$\hspace{5cm}(b)$\ell=Q=\lambda=0.1$\\
\includegraphics[height=5cm]{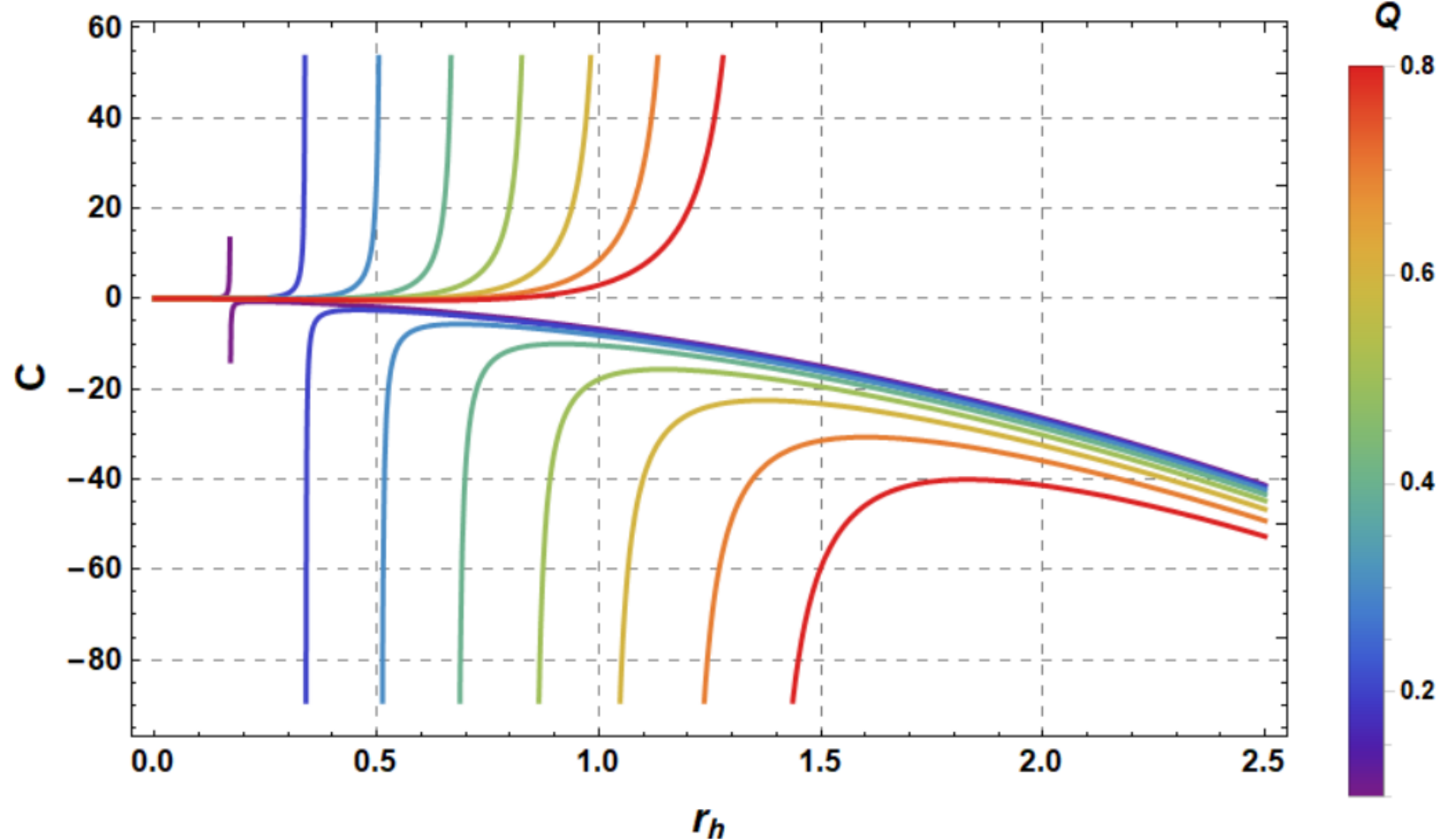}
\includegraphics[height=5cm]{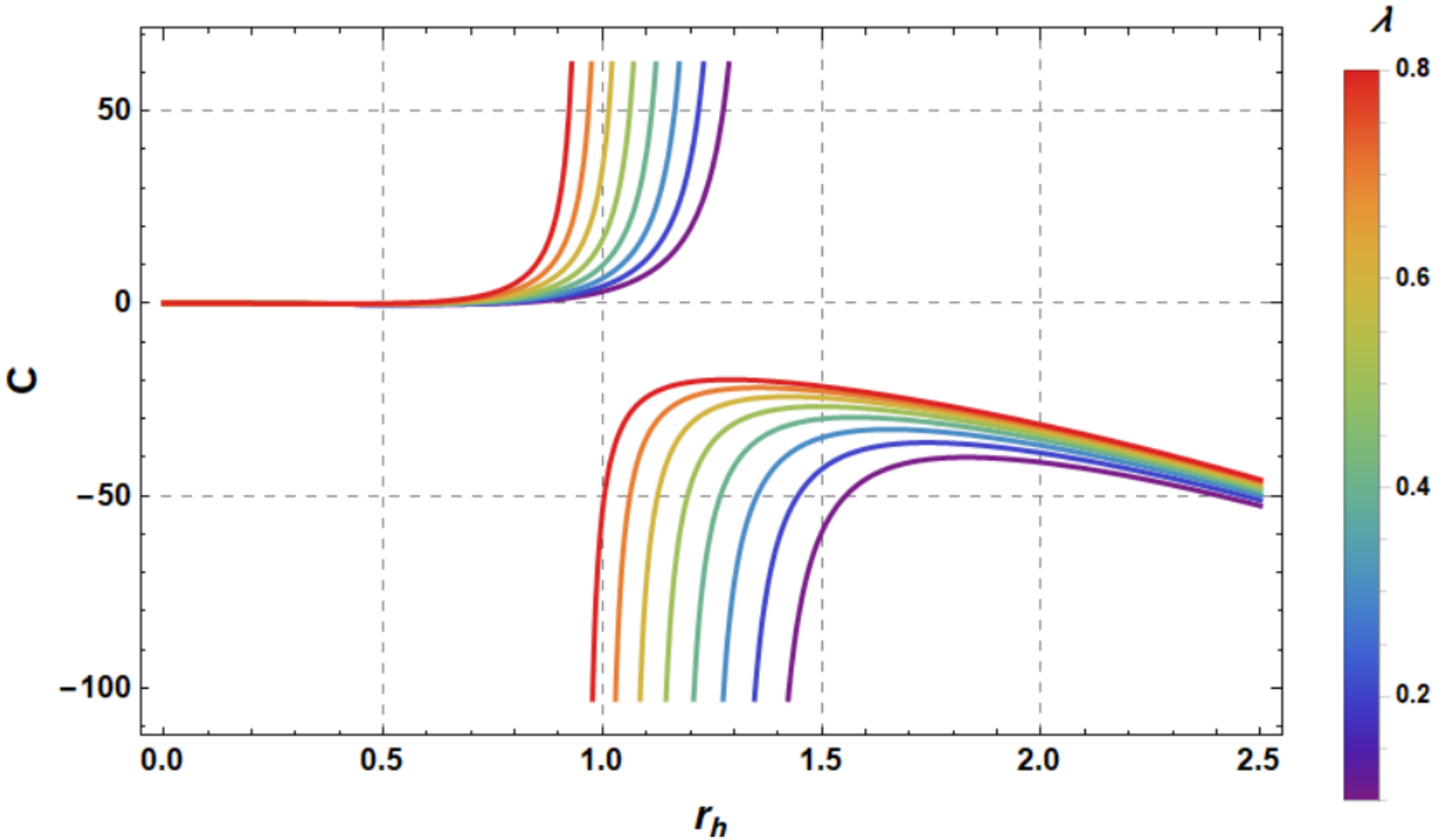}
\\
(c)$\ell=\lambda=0.1$ and $\alpha=0.01$\hspace{5cm}(d)$\ell=Q=0.1$ and $\alpha=0.01$\\  
\end{tabular}
\end{center}
\vspace{-0.5cm}
\caption{Behavior of $C$ varying the parameters $(\alpha,\ell,Q,\lambda)$.
\label{C}}
\end{figure}

Figure~\ref{C} illustrates the behavior of the heat capacity $C_Q$ as a function of the event horizon radius $r_h$ for different values of the parameters $(\alpha,\ell,Q,\lambda)$. In all panels, the heat capacity exhibits a characteristic divergence at a critical radius, separating two distinct thermodynamic branches. This divergence corresponds to the condition $(1-\alpha)r_h^2=3\beta$, signaling a second-order phase transition of Davies type. On one side of this critical point, the heat capacity is positive, indicating a locally stable black hole configuration, whereas on the other side, it becomes negative, corresponding to thermodynamically unstable configurations.
In panel (a), increasing the Lorentz-violating parameter $\ell$ shifts the divergence point and modifies the magnitude of the heat capacity, generally enlarging the region where $C_Q>0$. This indicates that Lorentz violation favors the stabilization of near-extremal configurations by extending the domain of positive heat capacity. In panel (b), the cloud-of-strings parameter $\alpha$ produces a similar qualitative effect, shifting the phase transition point and slightly enlarging the stable branch due to the weakening of the effective gravitational interaction.
Panel (c) highlights the strong influence of the electric charge $Q$, which significantly shifts the divergence toward larger values of $r_h$ and broadens the region of positive heat capacity. This confirms that the electric charge plays a crucial role in stabilizing the black hole, allowing for a wider range of near-extremal configurations with $C_Q>0$. Finally, panel (d) shows that increasing the ModMax parameter $\lambda$ shifts the divergence toward smaller values of $r_h$ and reduces the extent of the stable region. This behavior reflects the exponential suppression of the effective charge contribution, which weakens the stabilizing effect of the electromagnetic sector and drives the system toward the thermodynamic behavior of an uncharged black hole.

In the fixed-charge ensemble, the Helmholtz free energy is
\begin{align}
F=M-T_HS.
\end{align}
Using \eqref{eq:mass_rh} and \eqref{eq:entropy}, one finds
\begin{align}
T_HS=\frac{(1-\alpha)r_h^2-\beta}{4r_h},
\end{align}
and therefore
\begin{align}
F=
\frac{(1-\alpha)r_h^2+3\beta}{4r_h}.
\label{eq:free_energy}
\end{align}
Since $F>0$ for physical black holes with $r_h>0$, there is no Hawking-Page-type transition in this asymptotically non-(A)dS geometry. The relevant thermodynamic transition is instead the Davies transition encoded in the divergence of $C_Q$. 

The thermodynamics of this black hole is controlled by the connection among four parameters: the CoS parameter $\alpha$, the LV parameter $\ell$, the electric charge $Q$, and the ModMax parameter $\lambda$. As we observe, the CoS parameter $\alpha$ appears in the combination $(1-\alpha)$ and therefore weakens the effective gravitational attraction. As $\alpha$ increases, the horizon temperature decreases, the extremal radius increases, and the black hole moves more easily toward the cold regime. In this sense, the CoS parameter acts as a screening contribution in the geometry. The electric charge $Q$ enters quadratically through the effective combination $\beta$. As usual, charge lowers the Hawking temperature and creates an extremal configuration where the two horizons merge. Moreover, the charge is responsible for the existence of a locally stable near-extremal branch with positive heat capacity.

On the other hand, the ModMax parameter $\lambda$ appears only through the factor $e^{-\lambda}$. Therefore, increasing $\lambda$ exponentially suppresses the effective electromagnetic contribution. This has three immediate consequences: the extremal bound becomes weaker, the Hawking temperature increases, and the charge-induced stable branch shrinks. In other words, a large $\lambda$ makes the solution thermodynamically closer to an uncharged string-cloud black hole. The LV parameter $\ell$ has a richer effect. First, it rescales the radial metric component, which produces the overall factor $1/\sqrt{1+\ell}$ in the Hawking temperature. This tends to reduce the temperature. Second, it modifies the effective charge contribution through the factor $(2+\ell)/(1+\ell)$. Third, it changes the entropy itself through the factor $\sqrt{1+\ell}$, showing that Lorentz violation deforms the horizon thermodynamics in a genuinely nontrivial way. Hence, $\ell$ does not simply mimic a redefinition of the charge. This alters the thermal response of the geometry as a whole.

In short, we see that the black hole solution \eqref{A} exhibits the following thermodynamic picture: an outer and an inner horizon exist when the bound \eqref{eq:extremality_condition} is satisfied, the extremal configuration has zero temperature, the entropy is deformed by the LV parameter, and the first law and Smarr relation retain their standard form when $\alpha$, $\ell$, and $\lambda$ are treated as fixed background parameters. In addition, the heat capacity reveals a locally stable near-extremal branch and an unstable large-black-hole branch, separated by a Davies-type transition.

\section{Sparsity of the Hawking radiation}\label{s5}

After analyzing the thermal properties of the black hole solution, the next step is to examine the sparsity of the Hawking radiation.
An important aspect of Hawking emission is that, for many black hole geometries, the radiation is not emitted as a continuous thermal flux in the ordinary thermodynamic sense, but rather as a highly dilute sequence of well-separated quanta. This property is known as the sparsity of Hawking radiation. The basic idea is to compare two characteristic time scales, which are the average time interval between the emission of successive particles and the intrinsic time scale associated with an emitted quantum. If the time gap between quanta is much larger than the oscillation time of a typical Hawking particle, then we say that the radiation is sparse. This means that the black hole emits particles one by one, with long quiet intervals between them, instead of behaving like an ordinary blackbody source.

For a bosonic Hawking flux, a convenient measure of sparsity is obtained by introducing the dimensionless parameter
\begin{align}
\eta \equiv \frac{\tau_{\rm gap}}{\tau_{\rm loc}},
\label{eq:eta_def}
\end{align}
where $\tau_{\rm gap}$ is the average time between two successive emitted quanta and $\tau_{\rm loc}$ is the characteristic localization time of an individual quantum. The emission is said to be sparse when
$\eta \gg 1$. Equivalently, one may write
\begin{align}
\tau_{\rm gap}\sim \frac{1}{\Gamma},
\end{align}
where $\Gamma$ is the total emission rate, while the localization time is estimated from the typical frequency scale of the Hawking spectrum,
$\tau_{\rm loc}\sim \frac{1}{\omega_{\rm peak}}$. Hence, $\eta \sim \frac{\omega_{\rm peak}}{\Gamma}$. To obtain an analytic estimate, one can approximate the Hawking flux by a blackbody spectrum. For massless bosonic quanta in flat-space thermodynamics, the number flux per unit area is
\begin{align}
\Gamma_{\rm bb}=
\frac{g\,\zeta(3)}{4\pi^2}\,T^3,
\label{eq:bb_flux_density}
\end{align}
where $g$ is the spin degeneracy factor and $\zeta(3)$ is the Riemann zeta function. Multiplying by the effective emitting area $A_{\rm eff}$, one obtains
\begin{align}
\Gamma=\frac{g\,\zeta(3)}{4\pi^2}\,A_{\rm eff}\,T_H^3.
\label{eq:Gamma_total}
\end{align}
On the other hand, the peak of the number spectrum satisfies $\omega_{\rm peak}=\xi\, T_H$, where $\xi$ is a numerical constant of order unity determined by the maximum of the distribution. Therefore, Eq.~\eqref{eq:eta_basic} yields
\begin{align}
\eta
\sim
\frac{\xi T_H}{\Gamma}
=
\frac{4\pi^2\,\xi}{g\,\zeta(3)}\,
\frac{1}{A_{\rm eff}\,T_H^2}.
\label{eq:eta_general}
\end{align}
This formula shows that the sparsity is essentially controlled by the dimensionless combination
\begin{align}
A_{\rm eff}\,T_H^2.
\label{eq:sparsity_control}
\end{align}
A hotter black hole or a larger effective emitting area tends to reduce $\eta$, making the radiation less sparse, whereas colder configurations produce larger values of $\eta$, corresponding to a more dilute Hawking flux. 

For the static and spherically symmetric black hole considered in this work, the horizon area is $A_h=4\pi r_h^2$. As a first approximation, one may identify the effective emitting area with the horizon area, e.g. $A_{\rm eff}\approx A_h$. A more refined treatment may replace $A_h$ by an optical or capture cross-sectional area determined by graybody factors, but the horizon-area approximation already captures the main thermodynamic scaling of the sparsity. Using the Hawking temperature derived previously,
\begin{align}
T_H=\frac{1}{4\pi\sqrt{1+\ell}}
\left[\frac{1-\alpha}{r_h}
-\frac{2(1+\ell)Q^2e^{-\lambda}}{(2+\ell)\,r_h^3}
\right],
\label{eq:TH_repeat}
\end{align}
we find
\begin{align}
A_h T_H^2
&=
4\pi r_h^2
\left\{
\frac{1}{4\pi\sqrt{1+\ell}}
\left[
\frac{1-\alpha}{r_h}
-\frac{2(1+\ell)Q^2e^{-\lambda}}{(2+\ell)\,r_h^3}
\right]
\right\}^2
\nonumber\\[1mm]
&=
\frac{1}{4\pi(1+\ell)}
\left[
(1-\alpha)-\frac{2(1+\ell)Q^2e^{-\lambda}}{(2+\ell)\,r_h^2}
\right]^2.
\label{eq:AHT2}
\end{align}
Substituting this into Eq.~\eqref{eq:eta_general}, the sparsity parameter becomes
\begin{align}
\eta
\sim
\frac{16\pi^3\,\xi\,(1+\ell)}
{g\,\zeta(3)}
\left[
(1-\alpha)-\frac{2(1+\ell)Q^2e^{-\lambda}}{(2+\ell)\,r_h^2}
\right]^{-2}.
\label{eq:eta_final}
\end{align}
This is the basic analytical expression governing the sparsity of the Hawking radiation in the present geometry. A particularly important regime is the near-extremal limit. Extremality occurs when we have the following condition
\begin{align}
(1-\alpha)r_h^2=\frac{2(1+\ell)Q^2e^{-\lambda}}{2+\ell},
\label{eq:extremality_repeat}
\end{align}
for which the Hawking temperature vanishes, $T_H\to 0$. From Eqs.~\eqref{eq:Gamma_total} and \eqref{eq:eta_general}, this implies
$\Gamma \propto T_H^3 \to 0$ and $\eta \propto \frac{1}{T_H^2}\to \infty$. Therefore, as the black hole approaches extremality, the Hawking emission becomes extremely sparse. Physically, this means that the black hole emits quanta at an increasingly slow rate, with very long intervals between successive particles. In the exact extremal limit, the semiclassical emission ceases altogether because the temperature vanishes. Equation~\eqref{eq:eta_final} allows one to understand qualitatively how the parameters $\alpha$, $\ell$, $Q$, and $\lambda$ affect the sparsity of the Hawking radiation.

First, the CoS parameter $\alpha$ appears through the factor $(1-\alpha)$. Increasing $\alpha$ reduces the temperature of the black hole and thus decreases the quantity $A_h T_H^2$. As a consequence, the sparsity parameter $\eta$ increases. Hence, a stronger CoS background makes the Hawking radiation more dilute. Second, the electric charge $Q$ contributes through the negative charge-dependent term inside the square brackets in Eq.~\eqref{eq:eta_final}. A larger charge lowers the Hawking temperature and drives the solution toward extremality. Therefore, increasing $Q$ enhances the sparsity of the radiation. This is fully consistent with the usual behavior of charged black holes, where near-extremal configurations are colder and emit particles less frequently.

Third, the ModMax parameter $\lambda$ enters through the exponential factor $e^{-\lambda}$. As $\lambda$ increases, the effective electromagnetic contribution becomes suppressed. This tends to increase the Hawking temperature and move the system away from extremality. Consequently, larger values of $\lambda$ reduce the sparsity, making the Hawking flux less dilute. In this sense, the ModMax parameter weakens the charge-induced suppression of the emission.

Finally, the LV parameter $\ell$ affects the sparsity in two ways. On one hand, the temperature contains the overall factor $1/\sqrt{1+\ell}$, which tends to lower $T_H$ and hence increase $\eta$. On the other hand, $\ell$ also modifies the effective charge term through the factor $(2+\ell)/(1+\ell)$. The net effect is that Lorentz violation generally favors a colder black hole and therefore a more sparse Hawking flux. Thus, increasing $\ell$ tends to enhance the particle-by-particle character of the evaporation process.

The physical meaning of a large sparsity parameter is that the Hawking process is better described as a sequence of independent emission events rather than as a continuous stream of radiation. This has important implications. First, it shows that semiclassical black hole evaporation is intrinsically a low-occupancy process. Second, it emphasizes the relevance of graybody factors, since the actual emission rate can be substantially smaller than the ideal blackbody estimate, which would make the radiation even more sparse. Third, it indicates that any parameter that lowers the temperature or suppresses the transmission probabilities will increase the temporal separation between emitted quanta.

For the black hole studied here, the combined presence of the cloud-of-strings sector, Lorentz violation, electric charge, and ModMax nonlinear electrodynamics leads to a rich behavior of the Hawking sparsity. The parameters $\alpha$ and $Q$ increase the dilution of the Hawking cascade by lowering the temperature, while $\lambda$ acts in the opposite direction by weakening the effective charge contribution. The LV parameter $\ell$ introduces an additional cooling effect and therefore strengthens the sparse character of the radiation.

\section{Greybody factors and Absorption}\label{s6}

This section is dedicated to examining the greybody factor and the absorption cross section for bosonic particles around the black hole geometry presented in the section \ref{s2}. As we discussed in the preceding sections, one of the most prominent outcomes in the physics of black holes is the fact that these objects are capable of emitting what is known as Hawking radiation. In essence, such radiation consists of the emission of particles, such as photons or neutrinos, which originates just outside a black hole's event horizon. This idea was proposed by Stephen Hawking in 1974 and has its origin in quantum effects, where virtual particle-antiparticle pairs are separated, with one falling in and the other escaping. This leads to the progressive evaporation of the black hole over time. However, due to strong curvature around the black hole, a gravitational barrier is formed. As a consequence, some of the Hawking radiation is reflected back into the black hole, but a portion of this radiation manages to escape, reaching an observer at infinity. The quantity of radiation is calculated through the greybody factor. 

In technical terms, we can define a rigorous lower bound on the greybody factor $|T_b|$ as follows
\begin{equation}
|T_b| \geq \mathrm{sech}^{2}\!\left( \int_{-\infty}^{+\infty} G \, dr_{*} \right),
\label{eq:greybody_bound}
\end{equation}
where the function $G$ is defined by
\begin{equation}
G = \frac{\sqrt{\left(\xi'\right)^{2} + \left(\omega^{2} - V - \xi^{2}\right)^{2}}}{2\,\xi}.
\label{eq:G_def}
\end{equation}
In the expression above, $\omega$ denotes the mode frequency, $V$ is the effective potential, $r_{*}$ is the tortoise coordinate, and $\xi(r_{*})$ is a strictly positive auxiliary function satisfying the asymptotic conditions $\xi(-\infty) = \xi(+\infty) = \omega$. Note that
when $\xi$ is chosen to be constant and equal to $\omega$, the bound in Eq.~\eqref{eq:greybody_bound} simplifies considerably, yielding
\begin{equation}
|T_b| \geq \mathrm{sech}^{2}\!\left( \int_{-\infty}^{+\infty} \frac{V}{2\omega} \, dr_{*} \right)
\geq \mathrm{sech}^{2}\!\left( \int_{r_h}^{+\infty} \frac{V}{2\omega\, f(r)} \, dr \right),
\label{eq:greybody_simplified}
\end{equation}
where $r_h$ denotes the event horizon radius and $f(r)$ is the metric function relating the radial coordinate $r$ to the tortoise coordinate via $dr_{*} = dr/f(r)$.

In this context, another important quantity is the absorption cross-section. The absorption of bosonic particles by a black hole is an important physical process because it provides direct information about how the geometry interacts with external perturbations and fields. In particular, the absorption probability and the absorption cross section encode how scalar, electromagnetic, or gravitational waves propagate through the effective potential generated by the spacetime and how much of the incoming radiation is captured by the event horizon. These quantities are especially relevant in black hole physics since they are able to connect classical scattering theory, horizon dynamics, and quantum emission processes through the relation between absorption and Hawking radiation. Moreover, this study offers a powerful way to probe the effects of additional parameters present in generalized black hole solutions, such as electric charge, LV corrections, nonlinear electrodynamics, or matter sources like a cloud of strings. In this sense, changes in these parameters modify the effective potential barrier surrounding the black hole and consequently alter the transmission rates and spectral profile of the absorption. Therefore,
we seek to analyze the absorption of bosonic particles in order to understand the response of the black hole to external fields, the stability of wave propagation in the background geometry, and the possible observational signatures associated with deviations from the standard black hole scenarios.

Mathematically, we can calculate the absorption cross-section using the greybody factor through
\begin{align}
    \sigma(\omega)=\frac{\pi\,(2m+1)}{\omega^2}\,T_b(\omega)
\end{align}
In the next subsections, we will utilize the formalism presented here to analyze the greybody factors and absorption cross section for particles with spins $s=0$, $1$, and $2$. We intend to show how the structure of the effective potential modifies the transmission probability and absorption cross section in each case.

From a physical point of view, the greybody factor acts as the bridge between classical scattering and quantum evaporation. A larger greybody factor means that the black hole is more transparent to a given bosonic mode, which increases both the absorption probability in the scattering problem and the emission probability in the Hawking process. As a consequence, the absorption cross section and the Hawking number flux grow together, while the sparsity decreases because the emitted quanta become less separated in time. Conversely, when the potential barrier is high and the greybody factor is small, the black hole absorbs less incoming radiation and also emits fewer Hawking quanta to infinity, leading to a sparser radiation process. Therefore, the absorption cross section and the sparsity are not independent observables; both are governed by the same transmission properties of the black-hole potential.

\subsection{Spin 0}\label{s6-1}

Initially, in the case of spin 0 wave propagation, we start from the Klein-Gordon equation in curved spacetime, namely
\begin{equation}
\frac{1}{\sqrt{-g}} \partial_\mu \Big( \sqrt{-g} \, g^{\mu\nu} \partial_\nu \Phi \Big) = 0. \label{eq:KG}
\end{equation}
By means of the spherical harmonics, we decompose the scalar field as follows
\begin{equation}
\Phi(t, r, \theta, \phi) = \frac{1}{r} \sum_{l,m} \psi_l(t,r) Y_{lm}(\theta, \phi), \label{eq:scalar_decomp}
\end{equation}
where \(\psi_l(t,r)\) is the radial time-dependent wave function, and \(l\) and \(m\) are the indices of the spherical harmonics. Substituting this decomposition into equation \eqref{eq:KG}, we obtain
\begin{equation}
\frac{d^2 \psi_l(r_*)}{d r_*^2} + \omega^2 \psi_l(r_*) = V_s(r) \psi_l(r_*), \label{eq:radial_eq}
\end{equation}
where \(r_*\) is the tortoise coordinate defined by
\begin{equation}
\frac{d r_*}{d r} = \sqrt{ \frac{g_{rr}}{|g_{tt}|} }, \label{eq:tortoise}
\end{equation}
and \(V_s(r)\) is the effective potential of the field, which can be expressed as
\begin{equation}
V_{spin\,0}(r) = |g_{tt}| \left[ \frac{l(l+1)}{r^2} + \frac{1}{r \sqrt{|g_{tt}| g_{rr}}} \frac{d}{dr} \sqrt{ \frac{|g_{tt}|}{g_{rr}} } \right]. \label{eq:scalar_potential}
\end{equation}
Here, \(l\) is known as the multipole moment of the quasinormal modes of the black hole. Using $|g_{tt}|=A(r)$ and $g_{rr}=\frac{1+l}{A(r)}$, the effective potential \eqref{eq:scalar_potential} becomes
\begin{equation}
V_{spin\,0}(r) = A(r) \left[ \frac{l(l+1)}{r^2} + \frac{1}{r (1+\ell)} \frac{dA}{dr} \right]. 
\label{eq:scalar_potentialB}
\end{equation}
Thus, for black hole solution \eqref{eq:ModMax-A-bumblebee}, the effective potential takes the following form
\begin{align}
 V_{spin\,0}(r)&=\frac{12 e^{-\lambda } M Q^2}{(l+2) r^5}-\frac{8 e^{-2 \lambda } (l+1) Q^4}{(l+2)^2 r^6}\nonumber\\&-\frac{e^{-2 \lambda } \left(-2 e^{\lambda } (l+1)^2 (l+2) m (m+1) Q^2+4 e^{2 \lambda } (l+2)^2 M^2-4 (\alpha -1) e^{\lambda } (l+1) (l+2) Q^2\right)}{(l+1) (l+2)^2 r^4}\nonumber\\&-\frac{(\alpha -1) m (m+1)}{r^2}-\frac{e^{-2 \lambda } \left(2 e^{2 \lambda } (l+1) (l+2)^2 m (m+1) M+2 (\alpha -1) e^{2 \lambda } (l+2)^2 M\right)}{(l+1) (l+2)^2 r^3}   
\end{align}
Now, we can use the expression for the effective potential to integrate, thereby obtaining the greybody factor of scalar perturbation as follows
\begin{align}
  T_{spin\,0}(\omega)=  \mathrm{sech}^{2}\bigg[\frac{\sqrt{1+\ell}}{\omega}\Sigma_{spin\,0}\bigg],
\end{align}
where
\begin{align}
    \Sigma_{spin\,0}=\frac{1-\alpha}{3 \Delta ^3} \left(-\frac{3 (\alpha -1)M \Delta  }{l+1}-\frac{4 (\alpha -1)^2 e^{-\lambda } Q^2}{l+2}+3 \Delta ^2 m (m+1)\right),
\end{align}
with
\begin{align}
    \Delta=M+\sqrt{M^2-\frac{2(1-\alpha) (1+\ell) Q^2\,e^{-\lambda}}{ (2+\ell)}}.
\end{align}

\begin{figure}[ht!]
\begin{center}
\begin{tabular}{ccc}
\includegraphics[height=5cm]{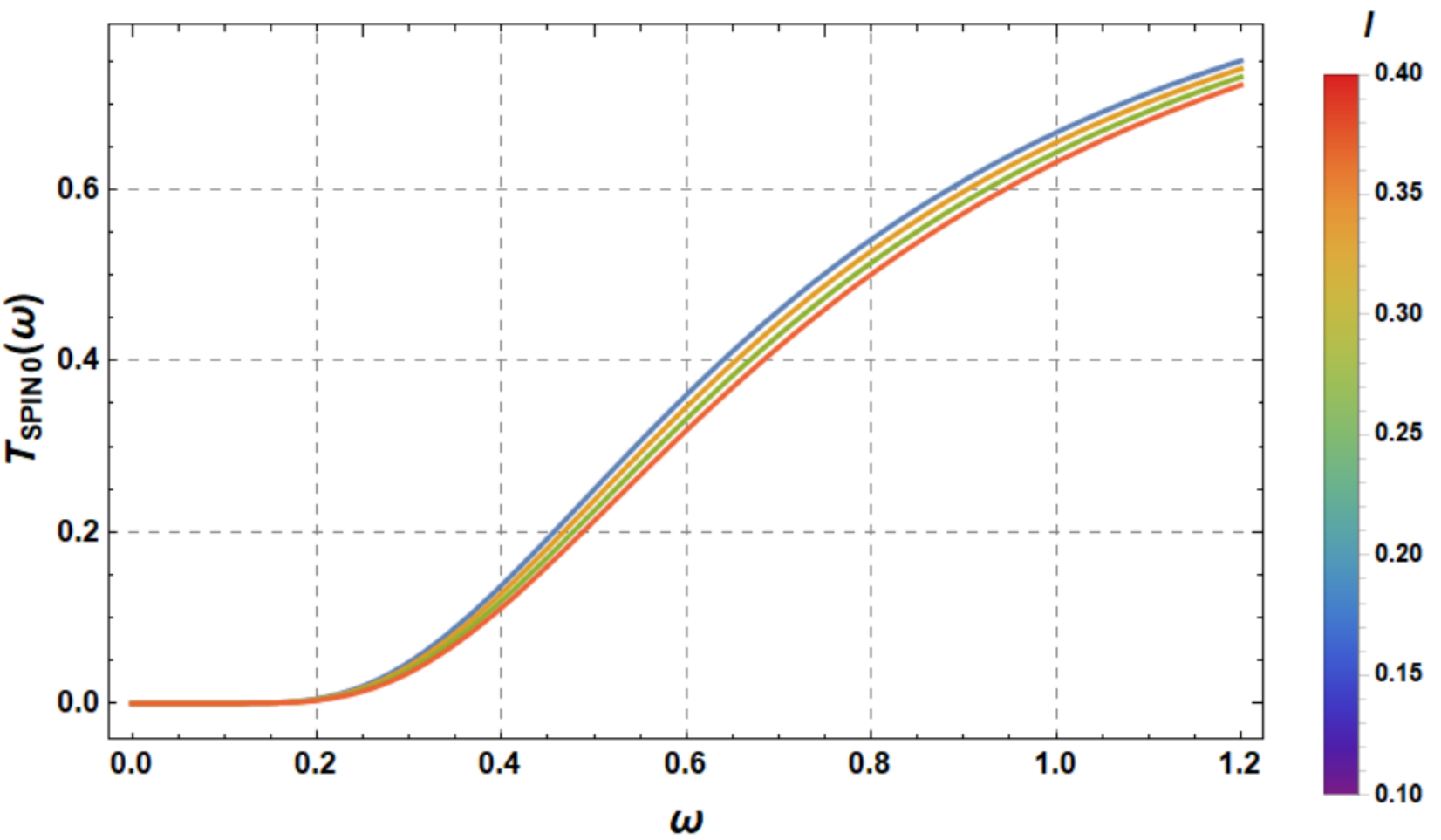} 
\includegraphics[height=5cm]{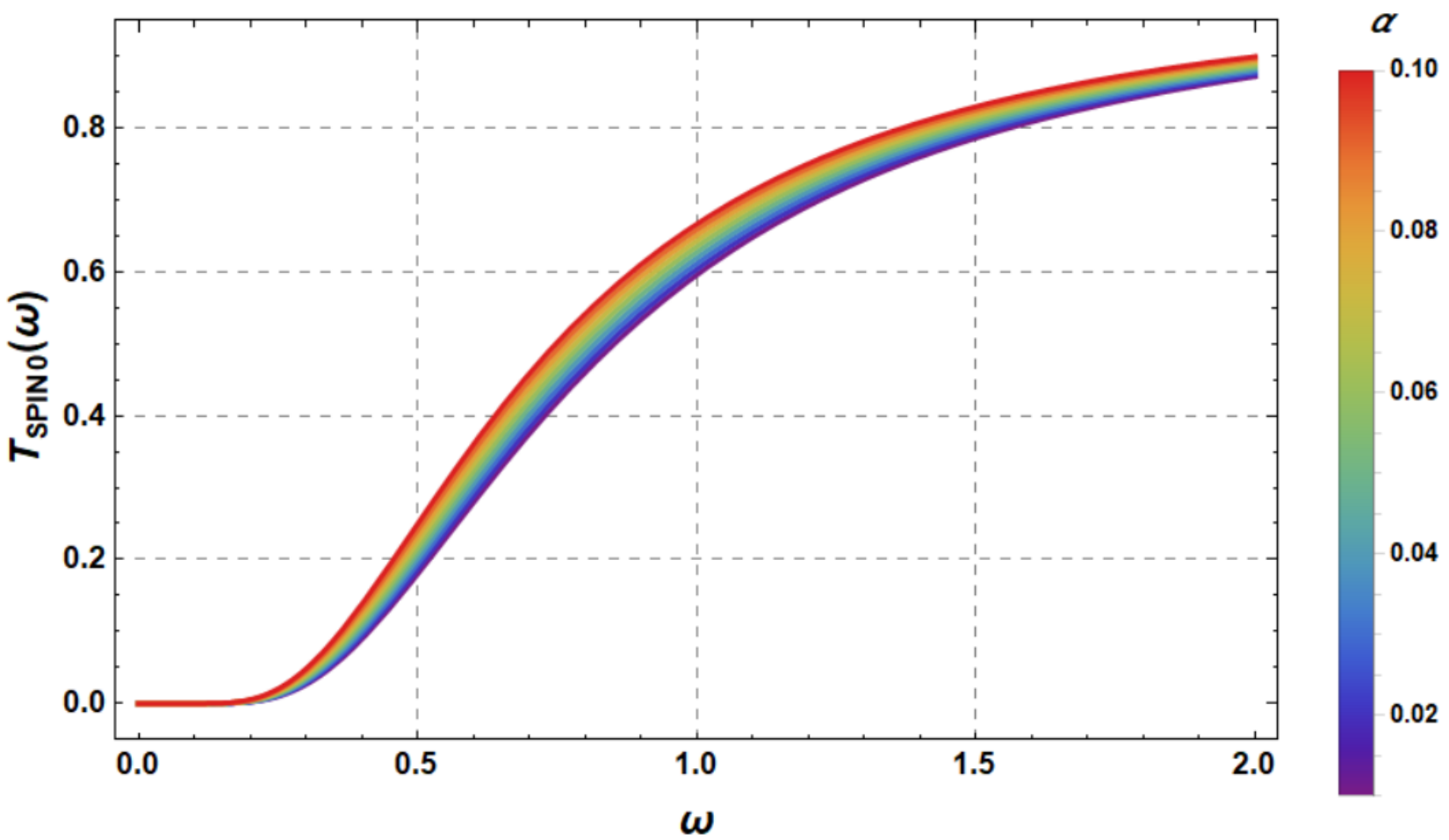}\\
(a) $Q=\lambda=0.1$ and $\alpha=0.01$\hspace{5cm}(b)$\ell=Q=\lambda=0.1$\\
\includegraphics[height=5cm]{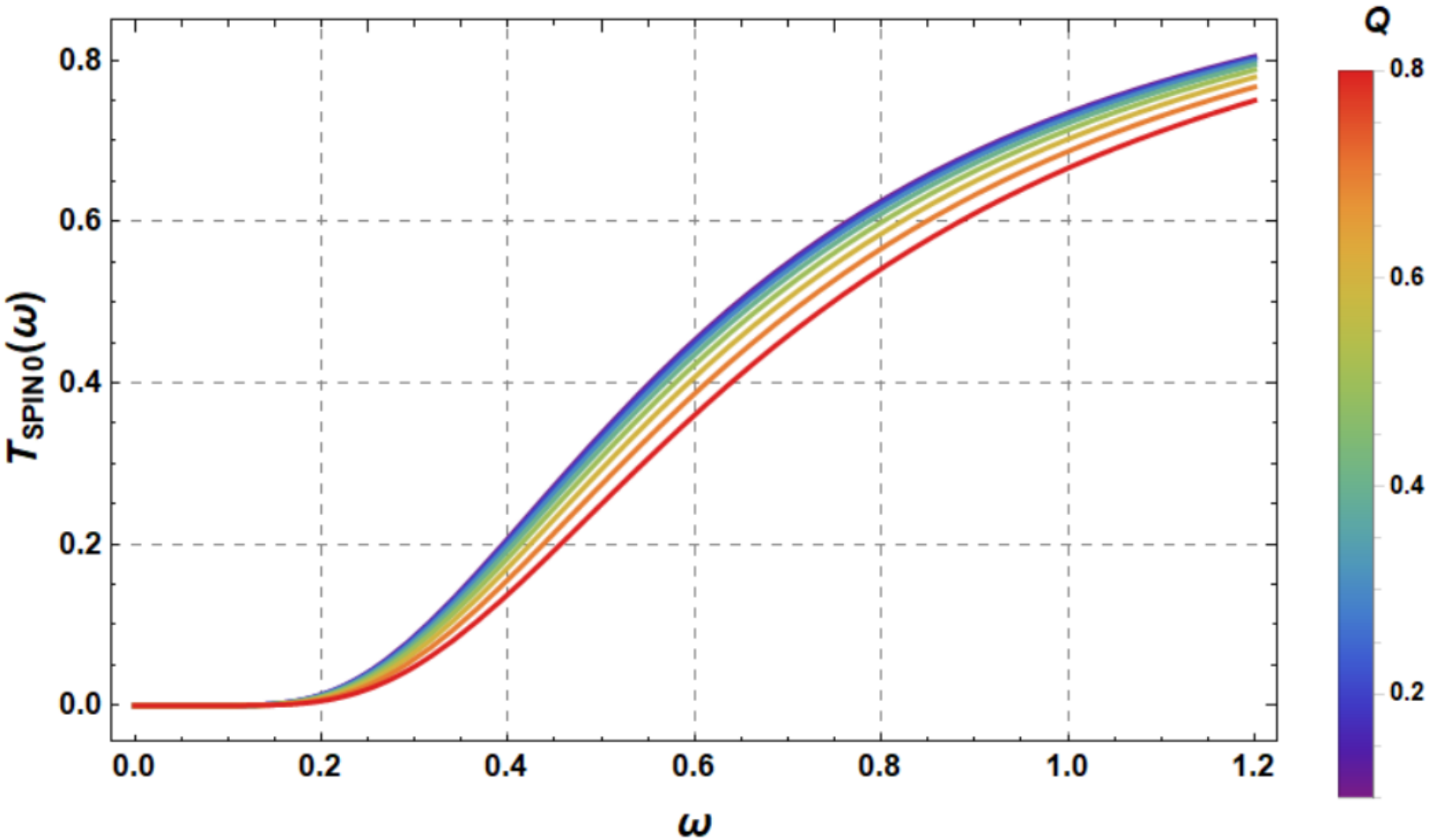}
\includegraphics[height=5cm]{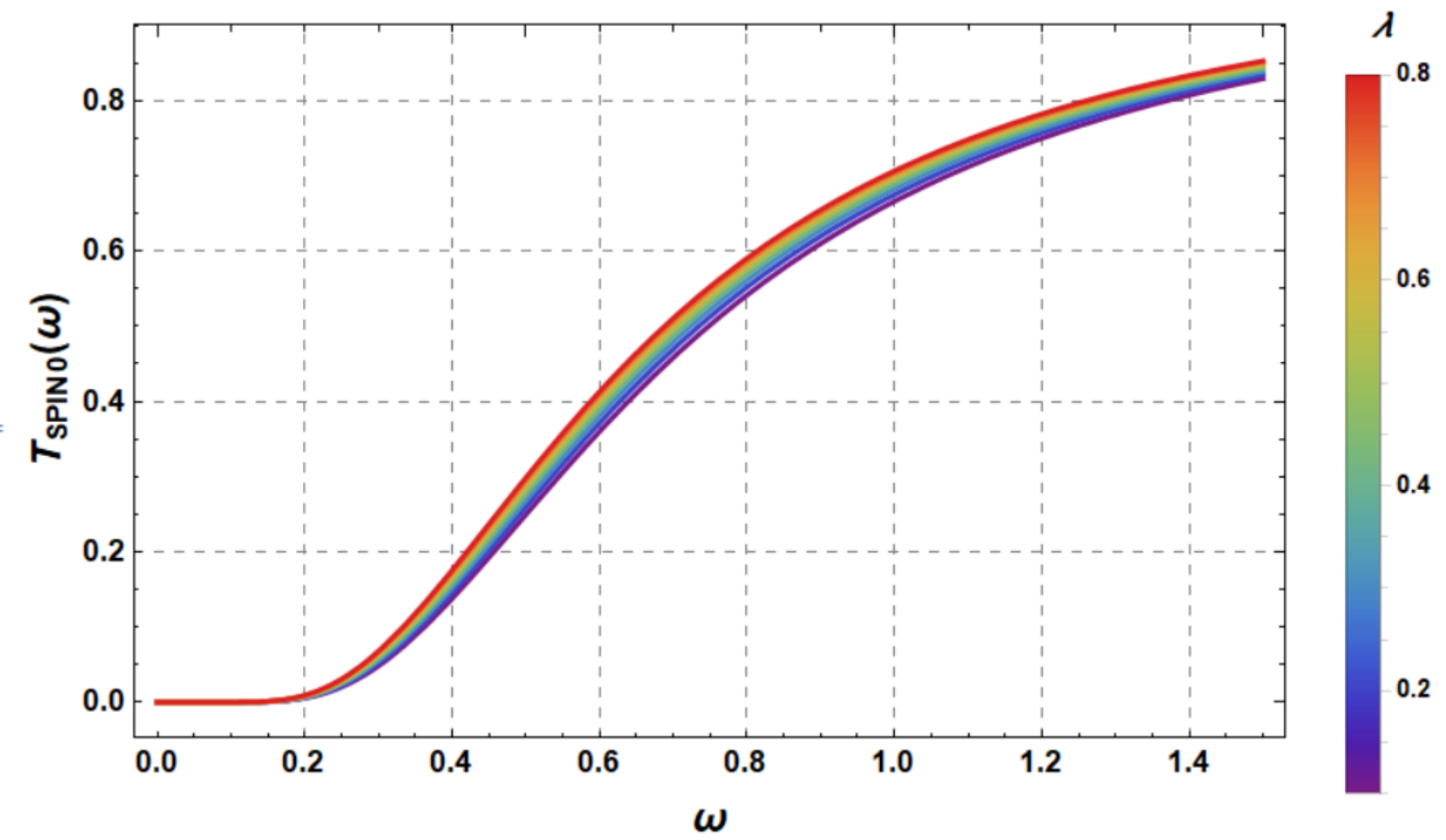}
\\
(c)$\ell=\lambda=0.1$ and $\alpha=0.01$\hspace{5cm}(d)$\ell=Q=0.1$ and $\alpha=0.01$\\ 
\end{tabular}
\end{center}
\vspace{-0.5cm}
\caption{Behavior of the greybody factor for spin 0.
\label{gf0}}
\end{figure}

Once we have obtained the analytical expressions for the effective potential and the greybody factor for spin 0 particles, let us discuss the physical meanings of these expressions. For scalar perturbations, the greybody factor shown in Fig.~\ref{gf0} increases monotonically with frequency in every parameter scan. The LV parameter $l$ lowers the entire family of curves, indicating that LV effects tend to increase the effective barrier and reduce transmission. The electric charge $Q$ acts in the same direction and produces an even clearer suppression over the plotted range. In contrast, larger values of $\alpha$ and $\lambda$ shift the curves upward, so the scalar mode is transmitted more efficiently when the CoS contribution or the ModMax nonlinearity is strengthened.

\begin{figure}[ht!]
\begin{center}
\begin{tabular}{ccc}
\includegraphics[height=5cm]{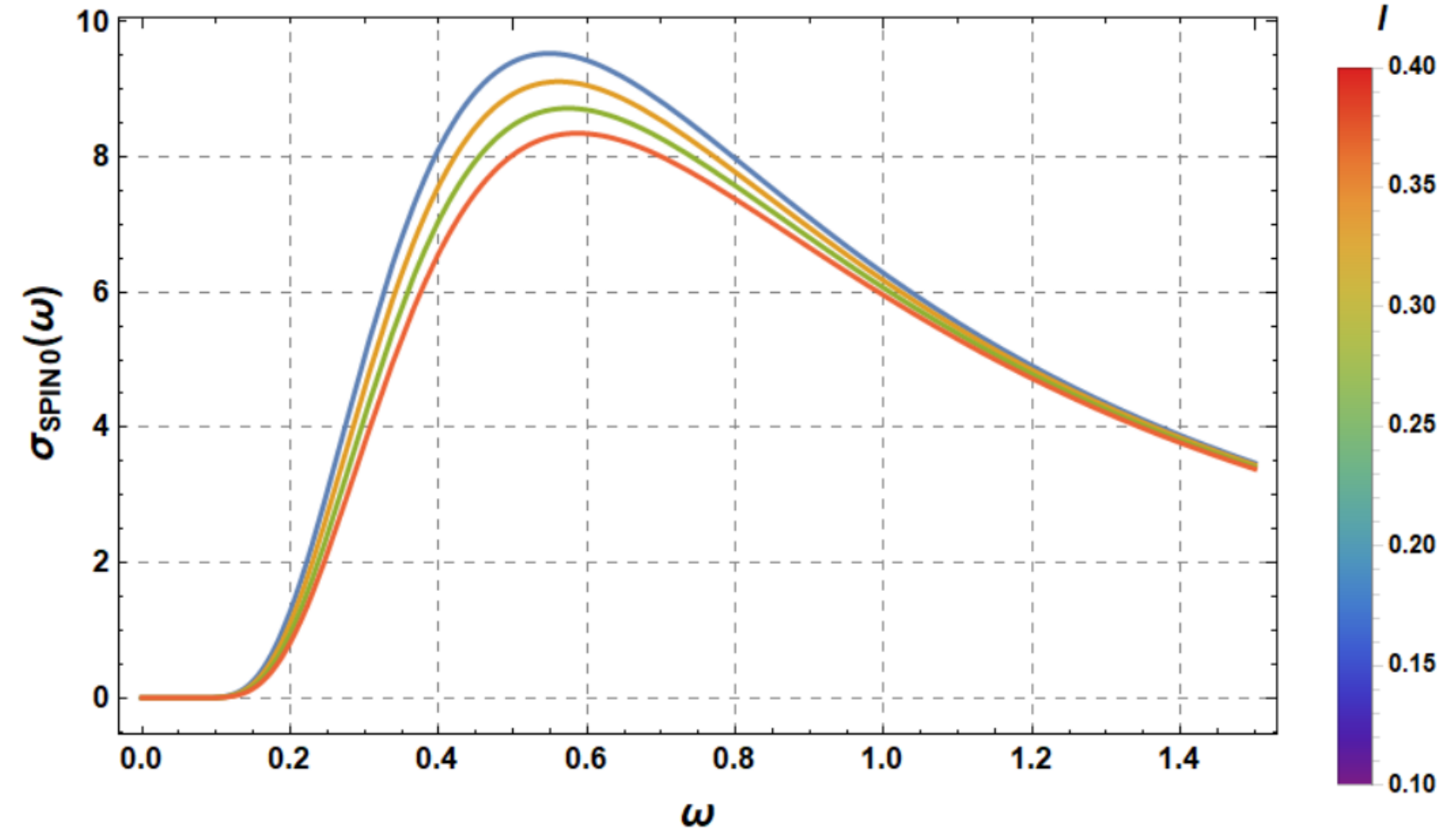} 
\includegraphics[height=5cm]{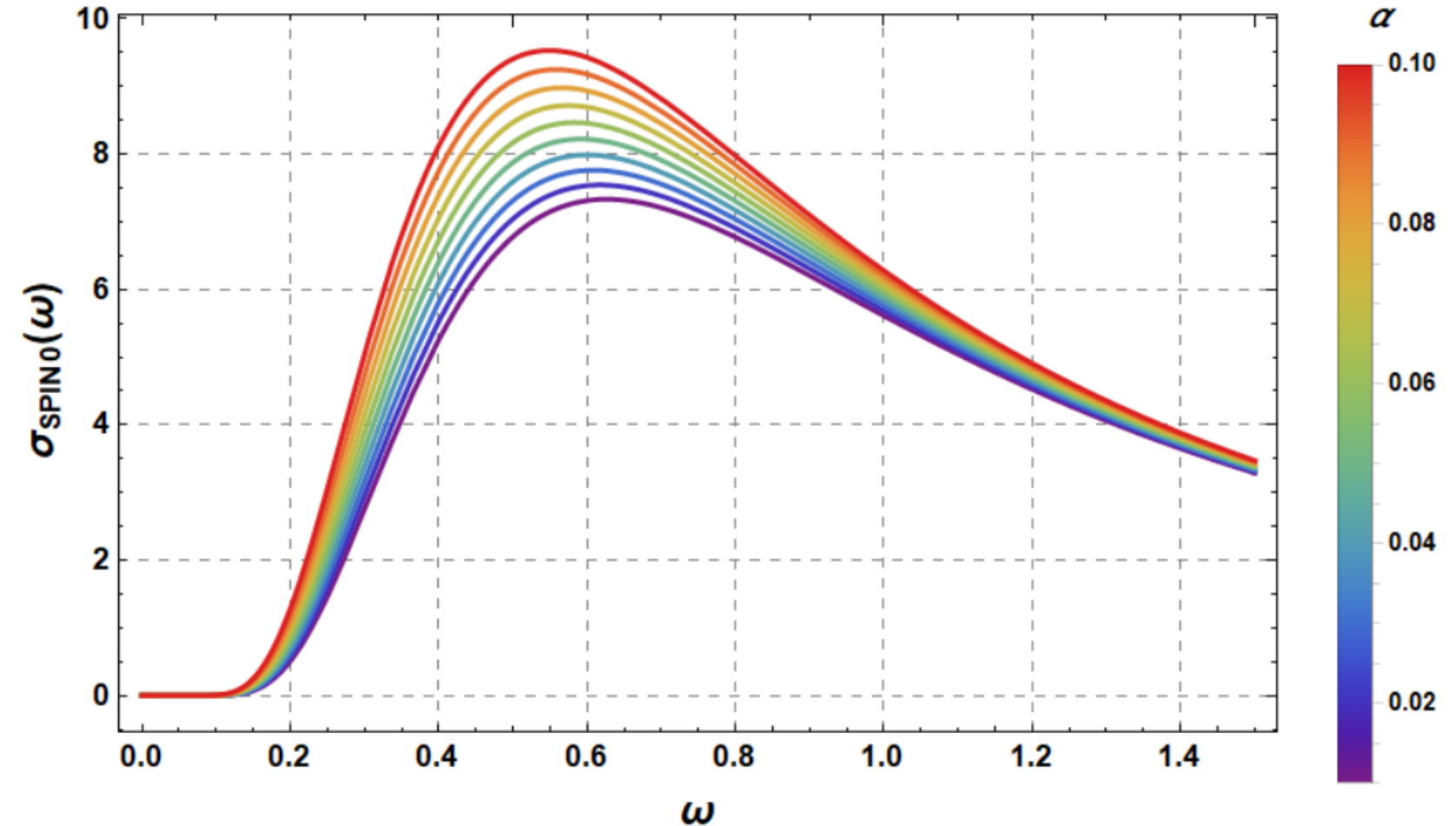}\\
(a) $Q=\lambda=0.1$ and $\alpha=0.01$\hspace{5cm}(b)$\ell=Q=\lambda=0.1$\\
\includegraphics[height=5cm]{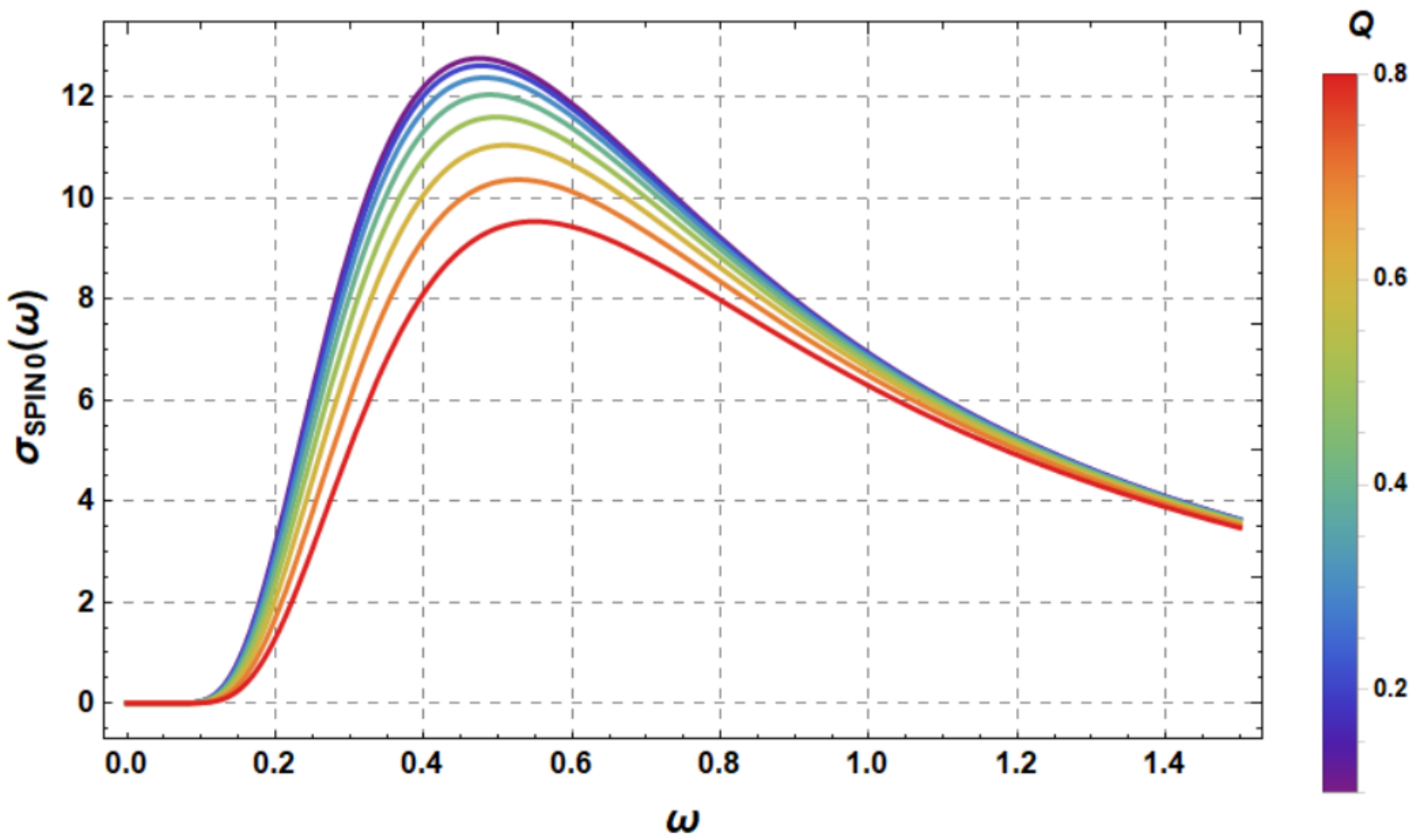}
\includegraphics[height=5cm]{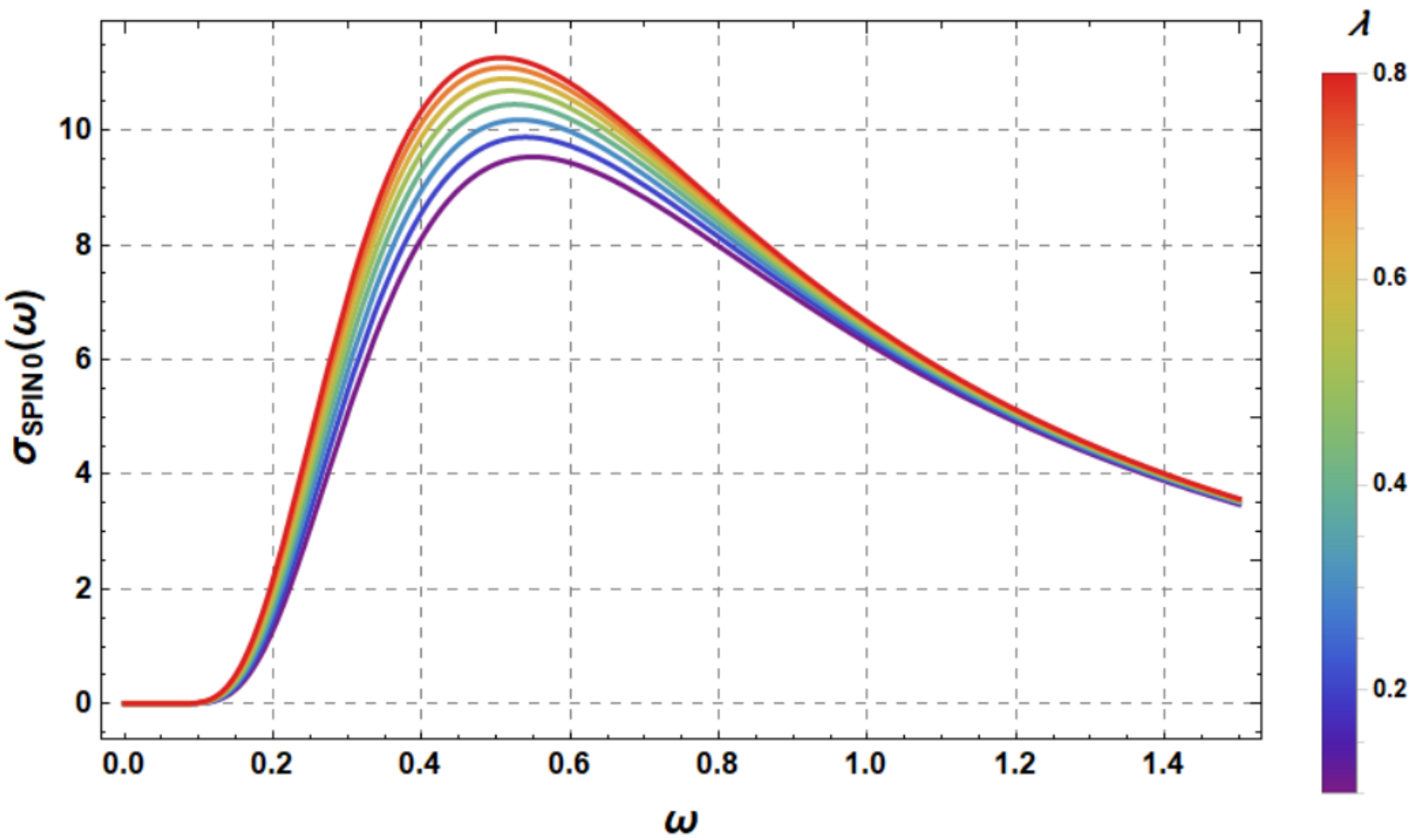}
\\
(c)$\ell=\lambda=0.1$ and $\alpha=0.01$\hspace{5cm}(d)$\ell=Q=0.1$ and $\alpha=0.01$\\
\end{tabular}
\end{center}
\vspace{-0.5cm}
\caption{Behavior of absorption cross section for spin 0.
\label{abs0}}
\end{figure}

The corresponding scalar absorption cross sections in Fig.~\ref{abs0} have a pronounced peak around intermediate frequency, with the maximum occurring near $\omega\approx0.55$ for the representative curves extracted from the notebook. The same parameter hierarchy observed in $T_0(\omega)$ is preserved in $\sigma_0(\omega)$: increasing $l$ or $Q$ lowers the peak and the high-frequency tail, whereas increasing $\alpha$ or $\lambda$ raises them. This is physically reasonable because any parameter that reduces the barrier transmission also reduces the portion of the incoming flux that is captured by the black hole.

\subsection{Spin 1}\label{s6-2}

In the case of a spin-1 particle, we need to use the tetrad formalism, in which a basis \(e_\mu^{(a)}\) is defined associated with the metric \(g_{\mu\nu}\). The basis should satisfy the following relations
\begin{align}
e_{(a)}^\mu e_\mu^{(b)} &= \delta_{(a)}^{(b)}, \\
e_{(a)}^\mu e_\nu^{(a)} &= \delta_\nu^\mu, \\
e_{(a)}^\mu &= g^{\mu\nu} \eta_{(a)(b)} e_\nu^{(b)}, \\
g_{\mu\nu} &= \eta_{(a)(b)} e_\mu^{(a)} e_\nu^{(b)} = e_{(a)\mu} e^{(a)}_\nu. \label{eq:tetrad_metric}
\end{align}
In terms of this basis, the tensor fields can be expressed as follows
\begin{align}
S_\mu &= e_\mu^{(a)} S_{(a)}, & S_{(a)} &= e_{(a)}^\mu S_\mu, \\
P_{\mu\nu} &= e_\mu^{(a)} e_\nu^{(b)} P_{(a)(b)}, & P_{(a)(b)} &= e_{(a)}^\mu e_{(b)}^\nu P_{\mu\nu}.
\end{align}
One should note that in the tetrad formalism, the covariant derivative in the actual coordinate system is replaced with the intrinsic derivative in the tetrad frame, as shown below
\begin{equation}
K_{(a)(b)|(c)} \equiv e^\lambda_{(c)} K_{\mu\nu;\lambda} e^\mu_{(a)} e^\nu_{(b)}
= K_{(a)(b),(c)} - \eta^{(m)(n)} \left( \gamma_{(n)(a)(c)} K_{(m)(b)} + \gamma_{(n)(b)(c)} K_{(a)(m)} \right), \label{eq:ricci_rot}
\end{equation}
where the Ricci rotation coefficients are given by
\begin{equation}
\gamma_{(c)(a)(b)} \equiv e^\mu_{(b)} e^\nu_{(a);\mu} e_\nu^{(c)}.
\end{equation}
The vertical bar and the comma denote the intrinsic and directional derivative, respectively, in the tetrad basis. Now, for the electromagnetic perturbation in the tetrad formalism, the Bianchi identity of the field strength \(F_{[(a)(b)(c)]} = 0\) gives
\begin{align}
\left[ r \sqrt{|g_{tt}|} F^{(t)(\phi)} \right]_{,r} + r \sqrt{g_{rr}} F^{(\phi)(r)}_{,t} &= 0, \label{eq:bianchi1} \\
\left[ \frac{r \sqrt{|g_{tt}|}}{\sin\theta} F^{(t)(\phi)} \right]_{,\theta} + \frac{r^2}{\sin\theta} F^{(\phi)(r)}_{,t} &= 0. \label{eq:bianchi2}
\end{align}
The conservation equation is
\begin{equation}
\eta^{(b)(c)} F_{(a)(b)|(c)} = 0. \label{eq:conservation}
\end{equation}
The above equation can be further written as
\begin{equation}
\left[ r \sqrt{|g_{tt}|} F^{(\phi)(r)} \right]_{,r} + \sqrt{|g_{tt}| g_{rr}} F^{(\phi)(\theta)}_{,\theta} + r \sqrt{g_{rr}} F^{(t)(\phi)}_{,t} = 0. \label{eq:conservation2}
\end{equation}
Differentiating equation \eqref{eq:conservation2} with respect to \(t\) and using equations \eqref{eq:bianchi1} and \eqref{eq:bianchi2}, we get
\begin{equation}
\partial_r \left[ \frac{|g_{tt}|}{g_{rr}} \partial_r \left( r \sqrt{|g_{tt}|} F \right) \right] 
+ \frac{|g_{tt}| \sqrt{g_{rr}}}{r} \left[ \frac{1}{\sin\theta} \partial_\theta \left( \sin\theta \, F_{,\theta} \right) \right]
- r \sqrt{g_{rr}} F_{,tt} = 0, \label{eq:wave_eq}
\end{equation}
where we have considered \(F = F^{(t)(\phi)} \sin\theta\). Using the Fourier decomposition \((\partial_t \to -i\omega)\) and field decomposition \(F(r,\theta) = F(r) \frac{Y_{,\theta}}{\sin\theta}\), where \(Y(\theta)\) is the Gegenbauer function satisfying
\begin{equation}
\frac{1}{\sin\theta} \frac{d}{d\theta} \left( \frac{1}{\sin\theta} \frac{d}{d\theta} Y_{,\theta} \right) = -l(l+1) \frac{Y_{,\theta}}{\sin\theta}, \label{eq:gegenbauer}
\end{equation}
we can write equation \eqref{eq:wave_eq} in the form
\begin{equation}
\partial_r \left[ \frac{|g_{tt}|}{g_{rr}} \partial_r \left( r \sqrt{|g_{tt}|} F \right) \right] + \omega^2 r \sqrt{g_{rr}} F - \frac{|g_{tt}| \sqrt{g_{rr}}}{r} l(l+1) F = 0. \label{eq:wave_eq2}
\end{equation}
Finally, using the tortoise coordinate \(r_*\) and redefining \(\psi_e \equiv r \sqrt{|g_{tt}|} F\), equation \eqref{eq:wave_eq2} can be written in the Schrödinger-like form
\begin{equation}
\frac{d^2 \psi_e}{dr_*^2} + \omega^2 \psi_e = V_e(r) \psi_e, \label{eq:schrodinger_like}
\end{equation}
where the effective potential is given by
\begin{equation}
V_{spin\, 1}(r) = |g_{tt}| \frac{l(l+1)}{r^2}.
\end{equation}
Considering $|g_{tt}| = A(r)$, one has
\begin{equation}
V_{spin\, 1}(r) = A(r) \frac{l(l+1)}{r^2}.
\end{equation}
We write $V_{spin\, 1}(r)$ for black hole solution \eqref{eq:ModMax-A-bumblebee} as follows
\begin{align}
V_{spin\, 1}(r) = -\frac{2 m (m+1) M}{r^3}+\frac{e^{-\lambda } m (m+1) Q^2}{r^4}+\frac{m (m+1)}{r^2} 
\end{align}
With this potential, after integrating, we find the following expression for the greybody factor of spin 1 particles
\begin{align}
  T_{spin\,1}(\omega)=  \mathrm{sech}^{2}\bigg[\frac{\sqrt{1+\ell}}{\omega}\Sigma_{spin\,1}\bigg],
\end{align}
where
\begin{align}
    \Sigma_{spin\,1}=\frac{(1-\alpha ) m (m+1)}{\Delta}.
\end{align}

\begin{figure}[ht!]
\begin{center}
\begin{tabular}{ccc}
\includegraphics[height=5cm]{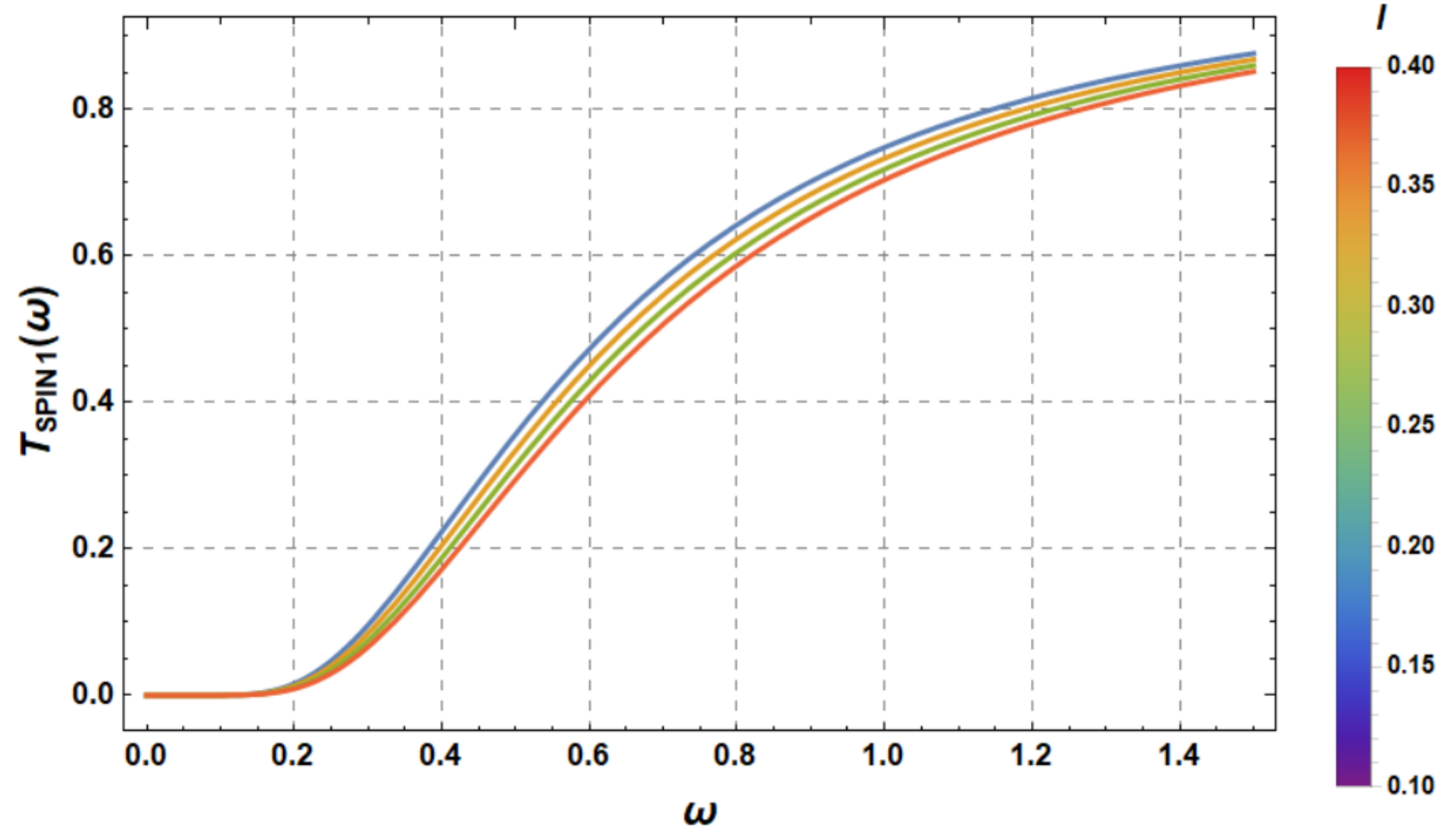} 
\includegraphics[height=5cm]{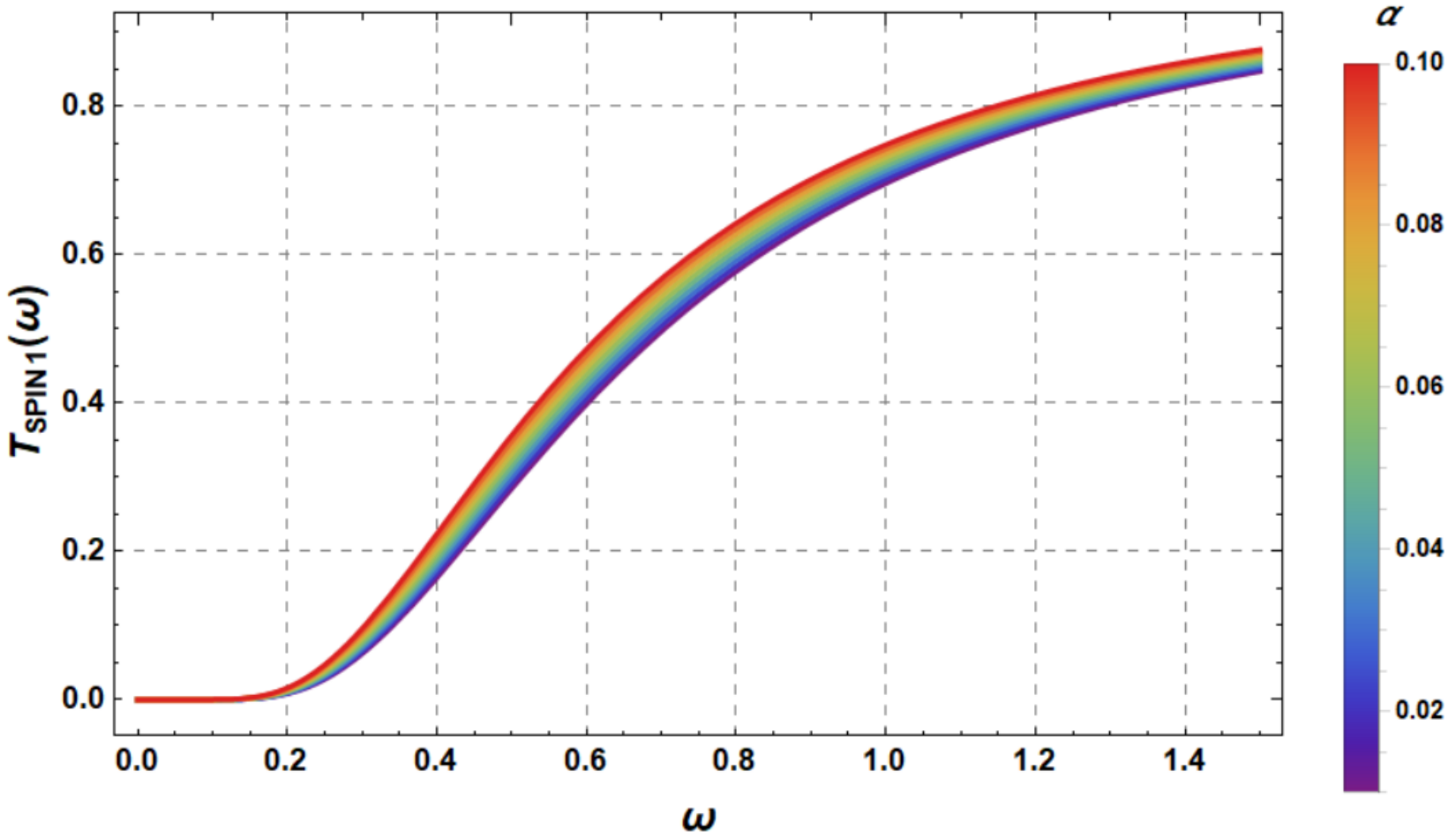}\\
(a) $Q=\lambda=0.1$ and $\alpha=0.01$\hspace{5cm}(b)$\ell=Q=\lambda=0.1$\\
\includegraphics[height=5cm]{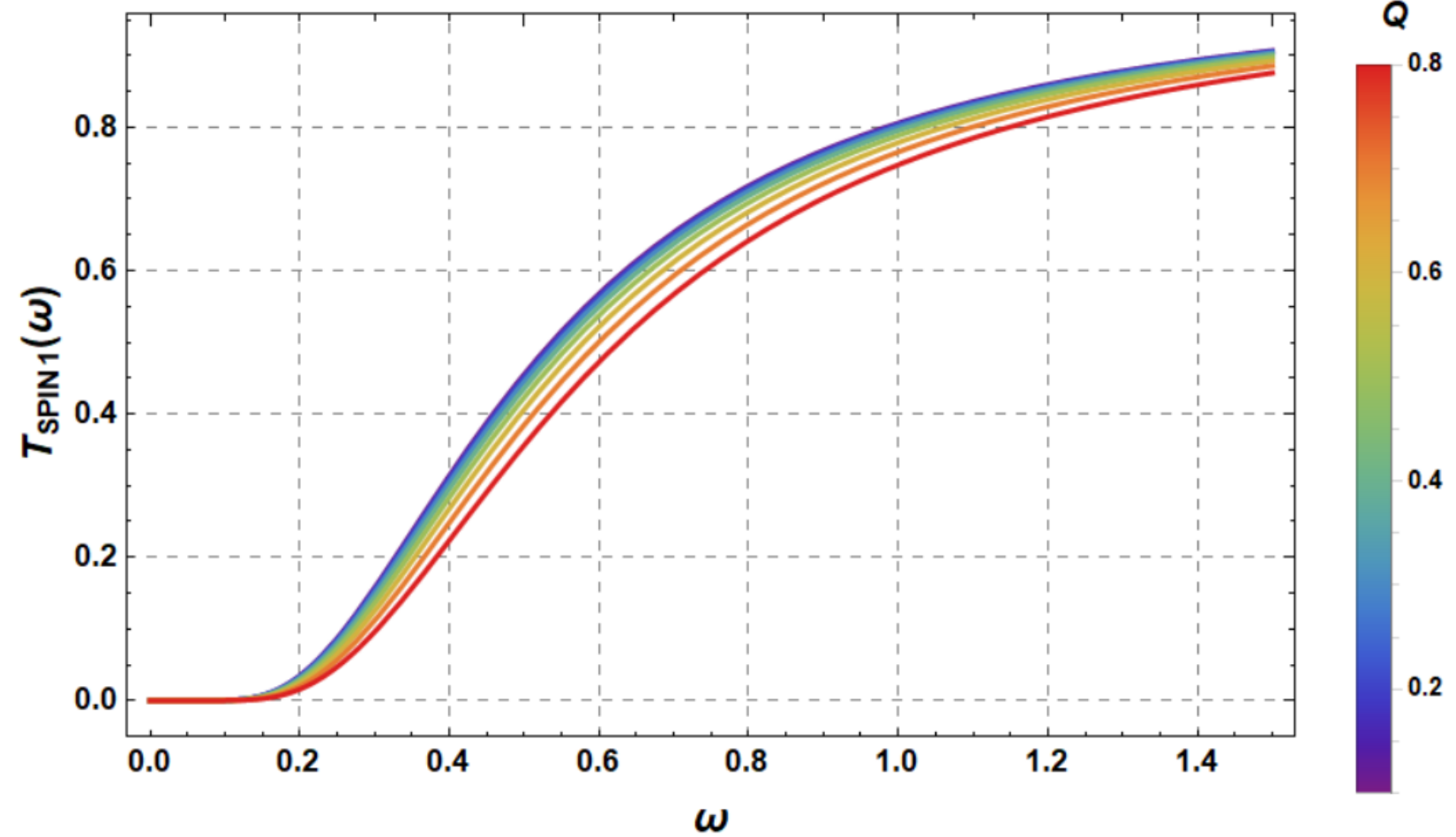}
\includegraphics[height=5cm]{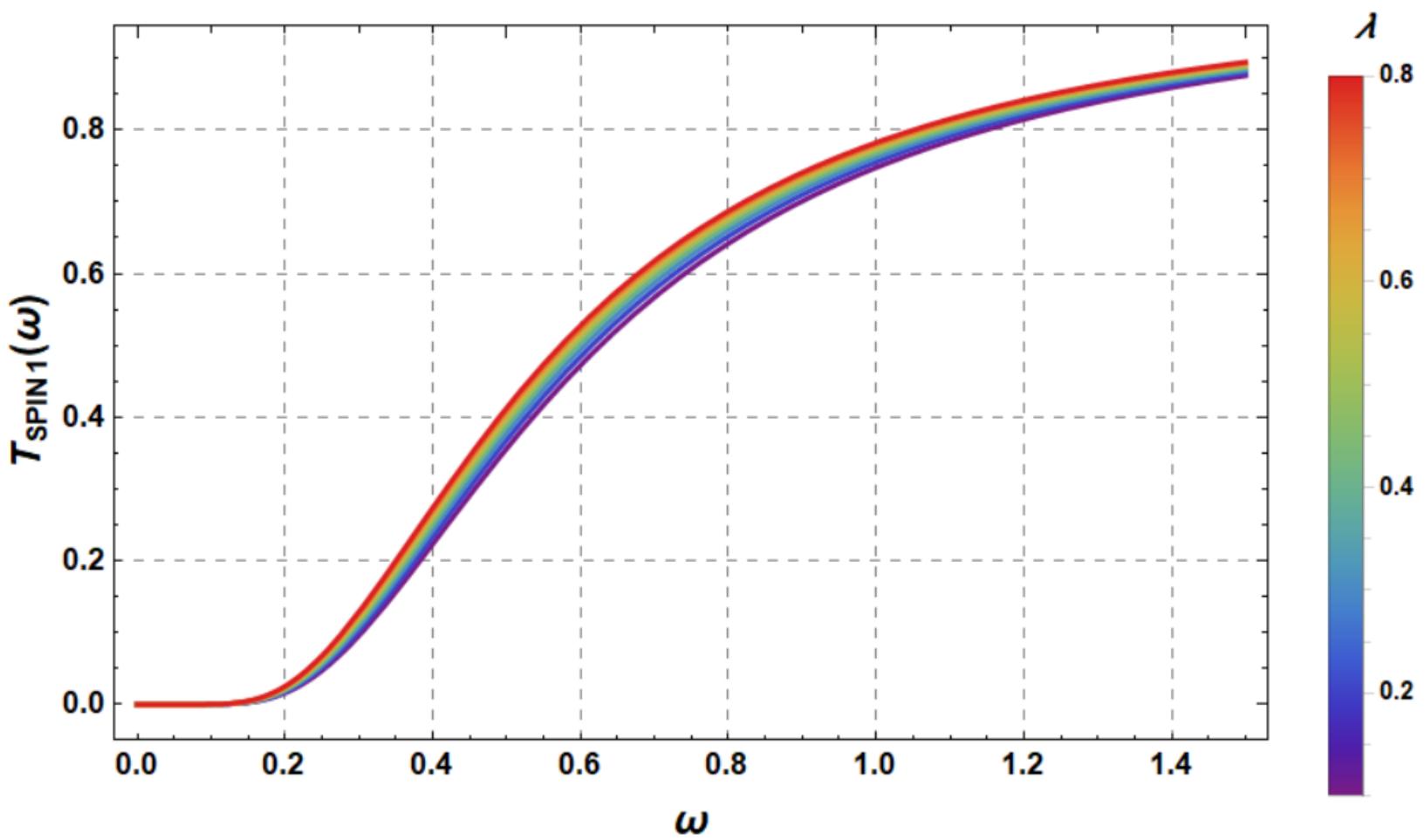}
\\
(c)$\ell=\lambda=0.1$ and $\alpha=0.01$\hspace{5cm}(d)$\ell=Q=0.1$ and $\alpha=0.01$\\
\end{tabular}
\end{center}
\vspace{-0.5cm}
\caption{Behavior of the greybody factor for spin 1.
\label{gf1}}
\end{figure}

We depict the behavior of $T_{spin\, 1}(r)$ in Fig.\ref{gf1}. Note that the graphical behaviors of the greybody factor in relation to the trends of the parameters $(\ell,\alpha,Q,\lambda)$ are quite similar to the spin 0 case. Then, in Fig.~\ref{gf1}, see that larger $l$ and $Q$ reduce $T_1(\omega)$, whereas larger $\alpha$ and $\lambda$ increase it. Quantitatively, the spin-1 greybody factor rises rather quickly with frequency and approaches its asymptotic regime in the same window $\omega\lesssim1.5$ used for the scalar case. This indicates that the spin-1 barrier is not drastically broader than the spin-0 barrier, although the detailed saturation level and curvature of the profiles are spin dependent.

\begin{figure}[ht!]
\begin{center}
\begin{tabular}{ccc}
\includegraphics[height=5cm]{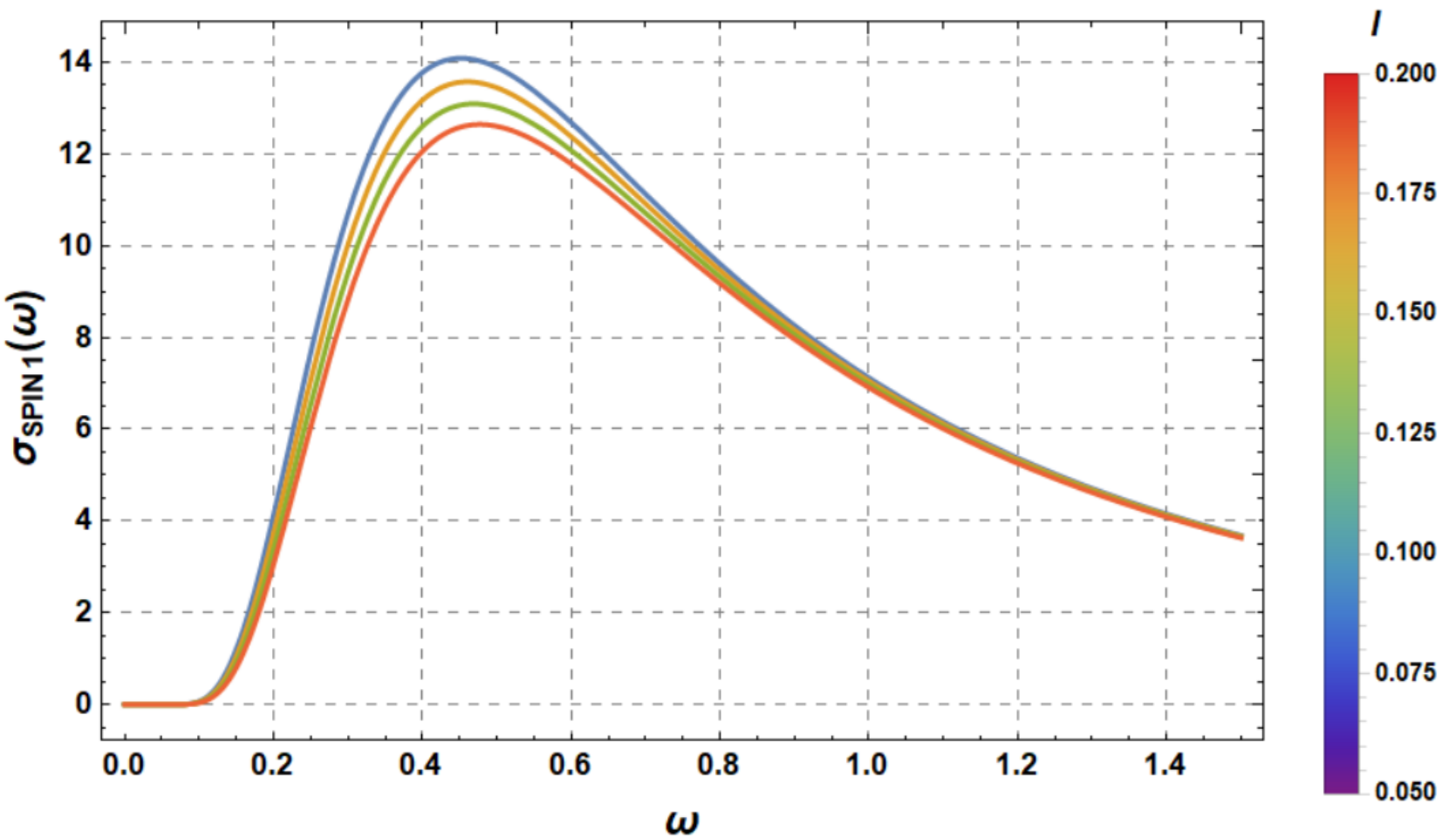} 
\includegraphics[height=5cm]{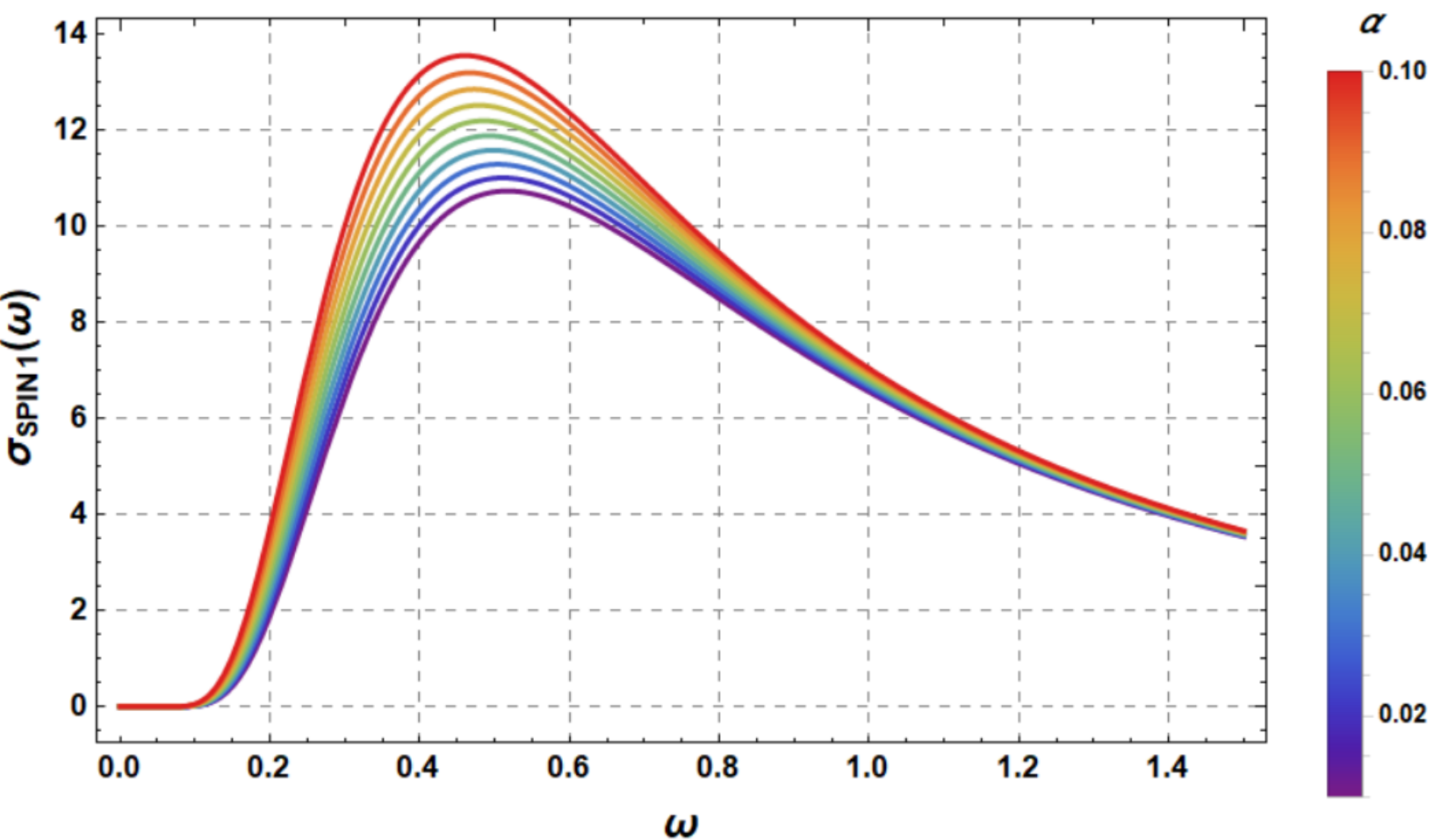}\\
(a) $Q=\lambda=0.1$ and $\alpha=0.01$\hspace{5cm}(b)$\ell=Q=\lambda=0.1$\\
\includegraphics[height=5cm]{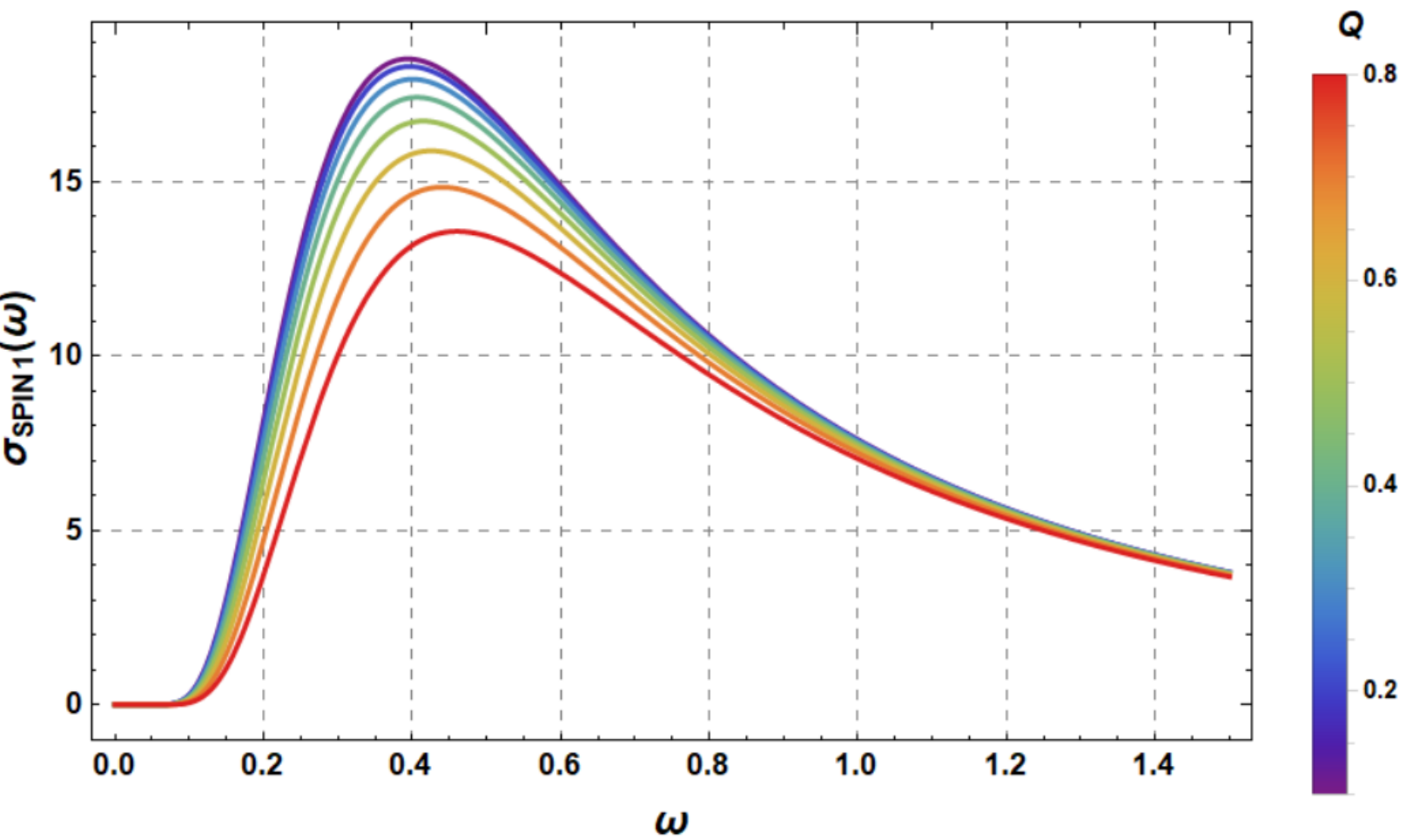}
\includegraphics[height=5cm]{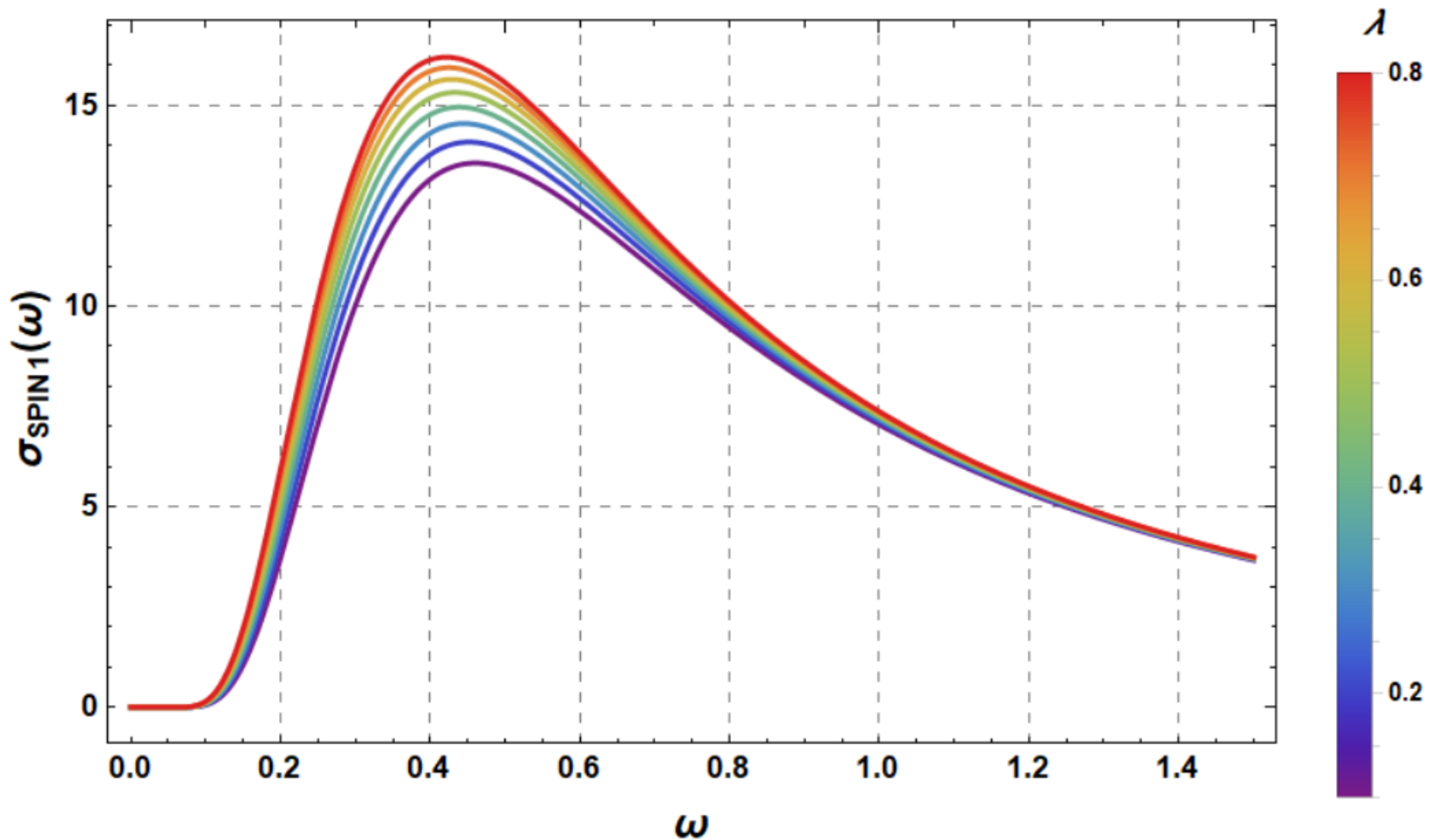}
\\
(c)$\ell=\lambda=0.1$ and $\alpha=0.01$\hspace{5cm}(d)$\ell=Q=0.1$ and $\alpha=0.01$\\
\end{tabular}
\end{center}
\vspace{-0.5cm}
\caption{Behavior of absorption cross section for spin 1.
\label{abs1}}
\end{figure}

The absorption curves in Fig.~\ref{abs1} are especially pronounced and reach the largest peak values among the three spin sectors shown in the notebook. For the representative extracted curve, the peak occurs near $\omega\approx0.45$. Thus, the spin-1 mode is absorbed most efficiently in a relatively low-to-intermediate frequency band. Again, $l$ and $Q$ lower the peak and the tail, while $\alpha$ and $\lambda$ increase them. The fact that the same ordering survives in both $T_1$ and $\sigma_1$ strongly supports the interpretation that all four parameters primarily act through deformations of the effective potential barrier rather than through unrelated kinematic effects.

\subsection{Spin 2}\label{s6-3}

The last case is the axial gravitational perturbation. Then, like the preceding cases, We seek to find a Schrödinger-like equation with an effective potential \(V_g\). This expression will help us calculate the greybody factor ahead. It is shown in Ref. that in the case of axial perturbation, the axial components of the perturbed energy-momentum tensor for an anisotropic fluid are zero, which allows us to write in the tetrad formalism
\begin{equation}
R_{(a)(b)} = 0. \label{eq:axial_r_ab}
\end{equation}
The \(\theta\phi\) and \(r\phi\) components of this equation give
\begin{align}
\left[ r^2 \sqrt{|g_{tt}|} g^{-1}_{rr} \left( a_{2,\theta} - a_{3,r} \right) \right]_{,r} 
&= \left[ r^2 \sqrt{|g_{tt}|}^{-1} g_{rr} (a_{1,\theta} - a_{3,t}) \right]_{,t}, \label{eq:axial_theta_phi} \\
\left[ r^2 \sqrt{|g_{tt}|} g^{-1}_{rr} (a_{3,r} - a_{2,\theta}) \sin^3 \theta \right]_{,\theta} 
&= \left[ \frac{r^4}{\sin^3 \theta} \sqrt{|g_{tt}|} g_{rr} (a_{1,r} - a_{2,t}) \right]_{,t}. \label{eq:axial_r_phi}
\end{align}
Now, using 
\begin{equation}
F_g(r, \theta) = F_g(r) Y(\theta),
\end{equation}
where \(Y(\theta)\) satisfies 
\begin{equation}
\frac{d}{d\theta} \left( \sin^{-3} \theta \frac{dY}{d\theta} \right) = - \left\{ l(l+1) - 2 \right\} Y \sin^{-3} \theta,
\end{equation}
we can simplify equations \eqref{eq:axial_theta_phi} and \eqref{eq:axial_r_phi} to obtain the Schrödinger-like equation:
\begin{equation}
\frac{d^2 \psi_g}{dr_*^2} + \omega^2 \psi_g = V_g(r) \psi_g, \label{eq:schrodinger_axial}
\end{equation}
where we used \(\psi_g = r F_g\) and \(r_*\) is the tortoise coordinate defined in equation (4.6). The effective potential in this expression is given by
\begin{equation}
V_{spin\, 2}(r) = |g_{tt}| \left[
\frac{2}{r^2} \left( \frac{1}{g_{rr}} - 1 \right) + \frac{l(l+1)}{r^2} - \frac{1}{r \sqrt{|g_{tt}| g_{rr}}} \frac{d}{dr} \sqrt{|g_{tt}| g^{-1}_{rr}}
\right].
\end{equation}
Thus, with $|g_{tt}|=A(r)$ and $g_{rr}=\frac{1+l}{A(r)}$, one arrives at
\begin{equation}
V_{spin\, 2}(r) = A(r) \left[ \frac{2}{r^2} \bigg(\frac{A(r)}{1+\ell} - 1\bigg) + \frac{l(l+1)}{r^2} - \frac{A'(r)}{r (1+\ell)} \right]
\end{equation}
Considering \eqref{eq:ModMax-A-bumblebee}, the effective potential reads
\begin{align}
V_{spin\, 2}(r)&=\frac{16 e^{-2 \lambda } (l+1) Q^4}{(l+2)^2 r^6}-\frac{28 e^{-\lambda } M Q^2}{(l+2) r^5}\nonumber\\&+\frac{2 e^{-\lambda } (l+1) (l+2) Q^2 \left(-2 \alpha +l \left(m^2+m-2\right)+m^2+m\right)}{(l+1) (l+2)^2 r^4}\nonumber\\&+\frac{12 (l+2)^2 M^2-8 (\alpha -1) e^{\lambda } (l+1) (l+2) Q^2}{(l+1) (l+2)^2 r^4}\nonumber\\&+\frac{e^{-2 \lambda } \left(6 (\alpha -1) e^{2 \lambda } (l+2)^2 M-2 e^{2 \lambda } (l+2)^2 M \left(-2 \alpha +l \left(m^2+m-2\right)+m^2+m\right)\right)}{(l+1) (l+2)^2 r^3}\nonumber\\& -\frac{(\alpha -1) \left(-2 \alpha +l \left(m^2+m-2\right)+m^2+m\right)}{(l+1) r^2}   
\end{align}
Using this potential, we obtain the following greybody factor for spin 2 perturbations
\begin{align}
  T_{spin\,2}(\omega)=  \mathrm{sech}^{2}\bigg[\frac{\sqrt{1+\ell}}{\omega}\Sigma_{spin\,2}\bigg],
\end{align}
where
\begin{align}
    \Sigma_{spin\,2}&=\frac{1-\alpha}{3\Delta^3}\bigg(\frac{3 \Delta  M \left(-\alpha +2 l \left(m^2+m-2\right)+2 m (m+1)-3\right)}{l+1}\nonumber\\&+\frac{2 (\alpha -1) e^{-\lambda } Q^2 \left(-2 (\alpha +2)+3 l \left(m^2+m-2\right)+3 m (m+1)\right)}{l+2}\bigg)
\end{align}

\begin{figure}[ht!]
\begin{center}
\begin{tabular}{ccc}
\includegraphics[height=5cm]{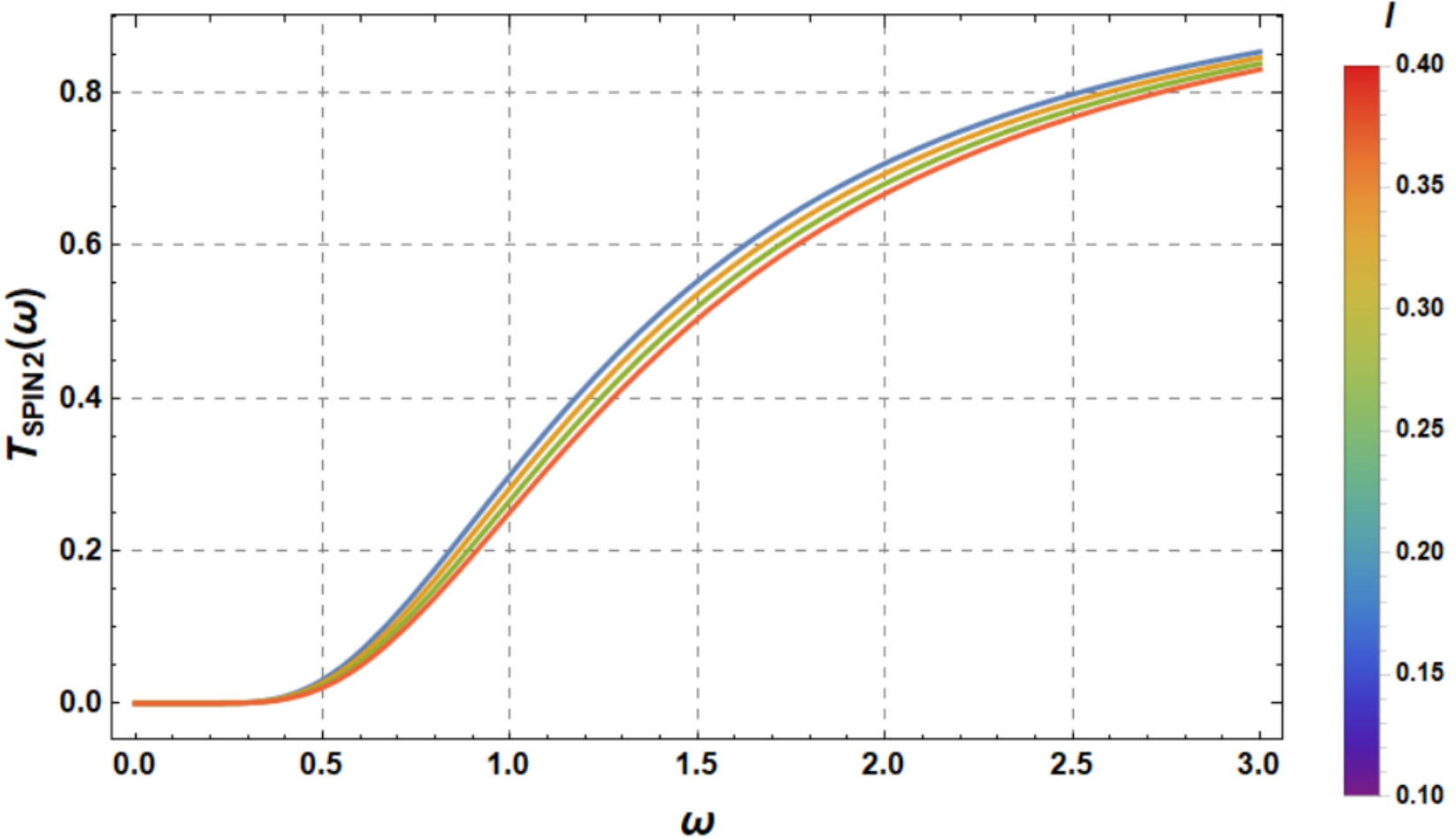} 
\includegraphics[height=5cm]{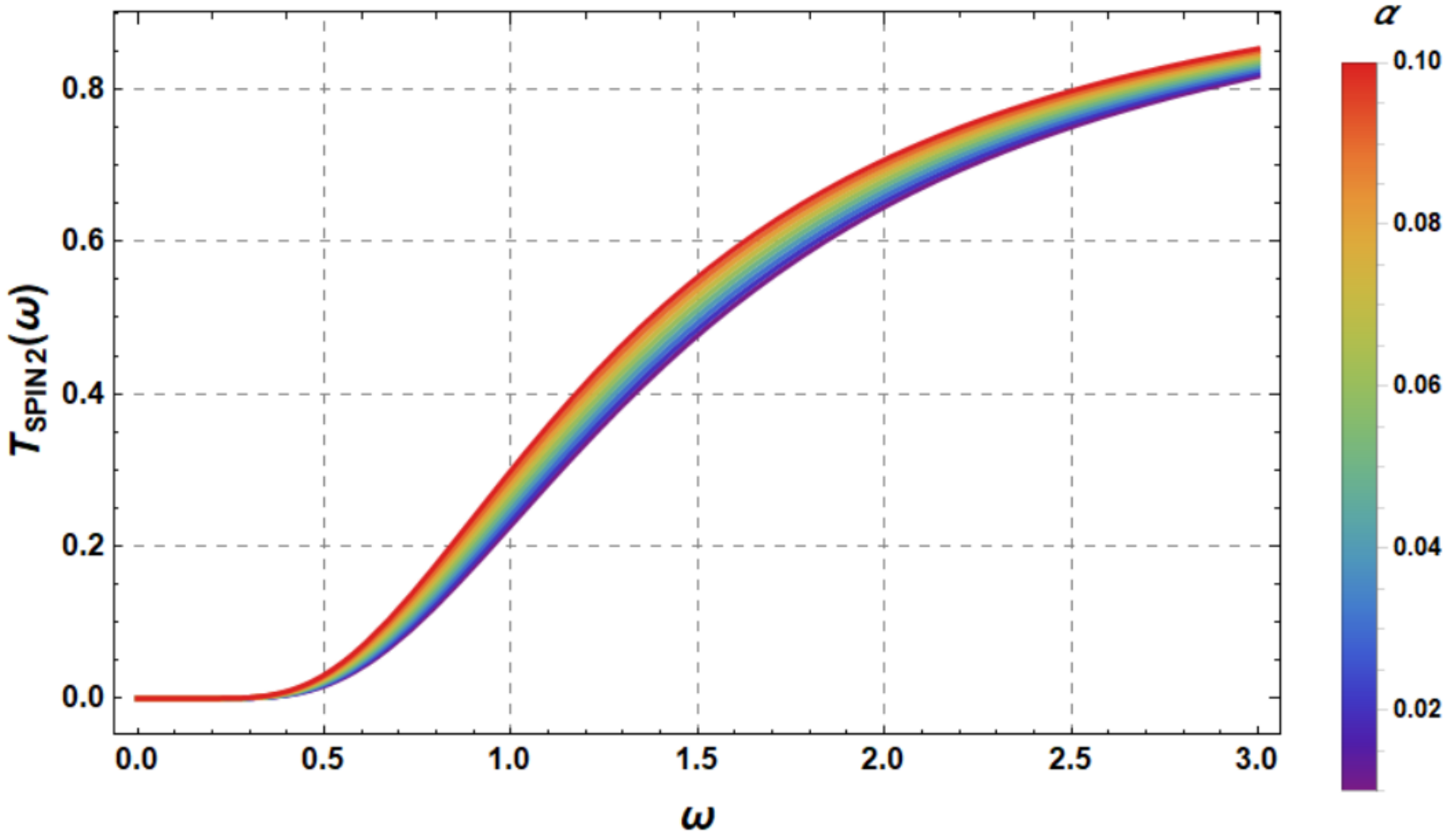}\\
(a) $Q=\lambda=0.1$ and $\alpha=0.01$\hspace{5cm}(b)$\ell=Q=\lambda=0.1$\\
\includegraphics[height=5cm]{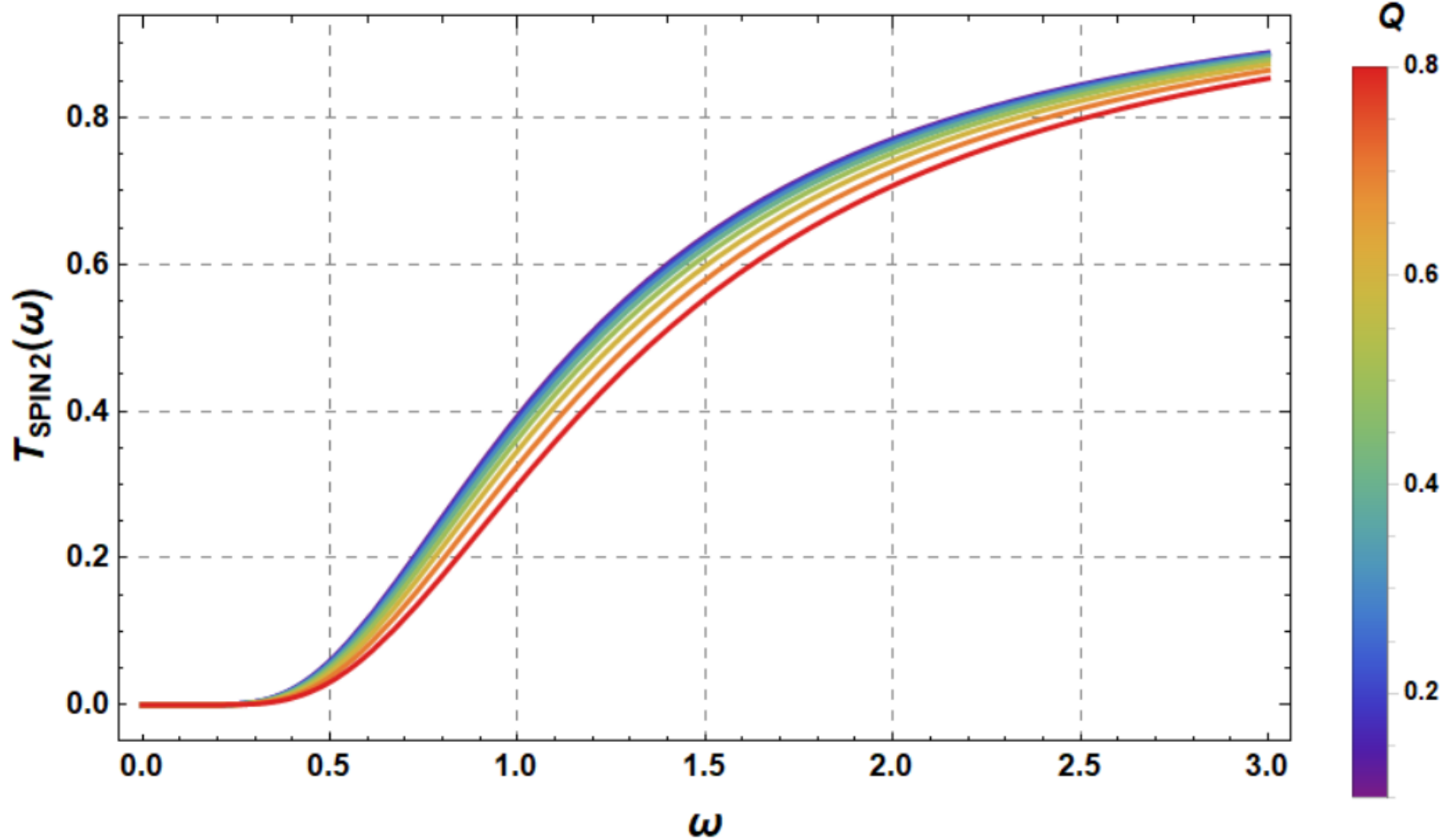}
\includegraphics[height=5cm]{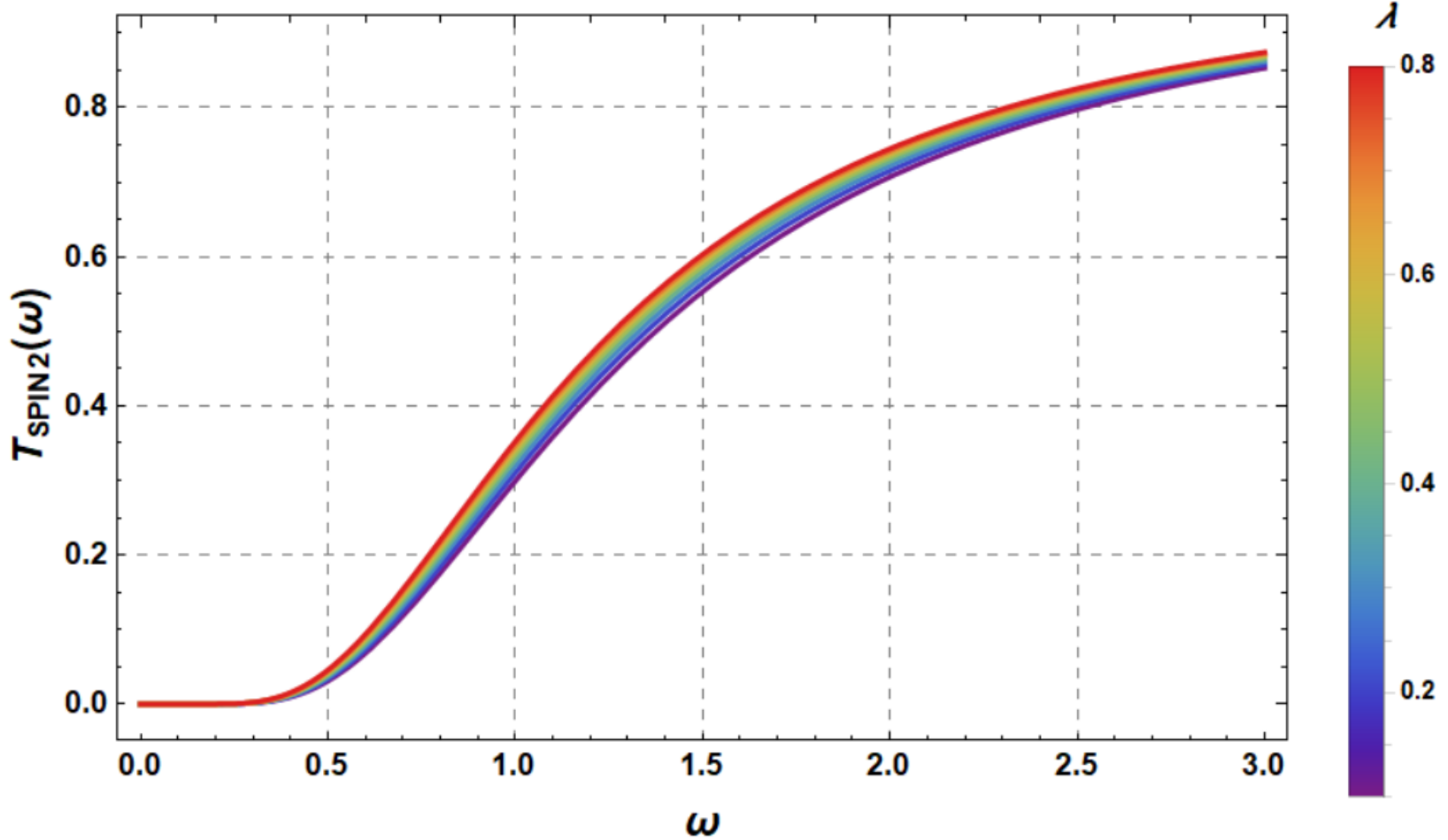}
\\
(c)$\ell=\lambda=0.1$ and $\alpha=0.01$\hspace{5cm}(d)$\ell=Q=0.1$ and $\alpha=0.01$\\
\end{tabular}
\end{center}
\vspace{-0.5cm}
\caption{Behavior of the greybody factor for spin 2.
\label{gf2}}
\end{figure}

\begin{figure}[ht!]
\begin{center}
\begin{tabular}{ccc}
\includegraphics[height=5cm]{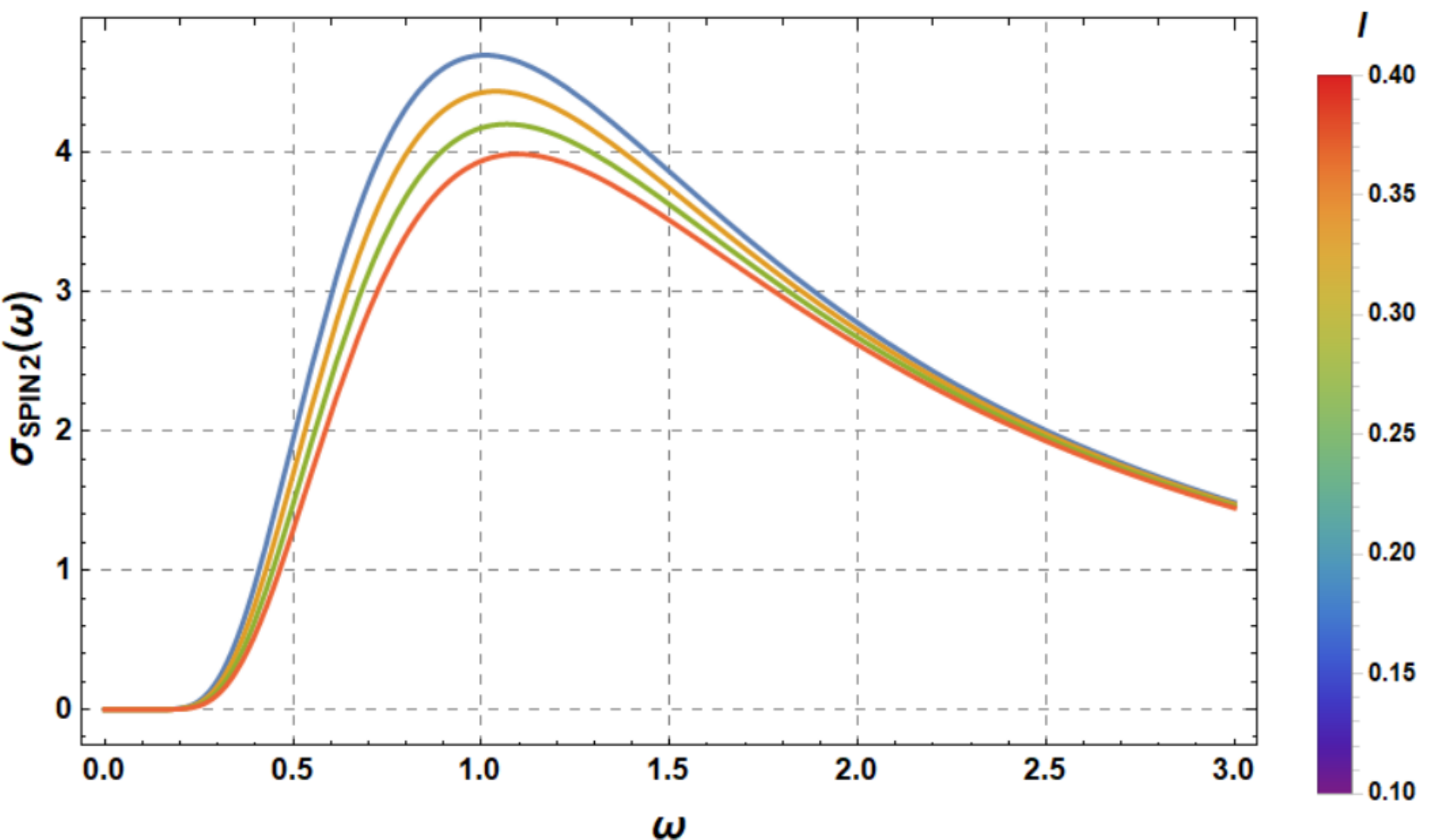} 
\includegraphics[height=5cm]{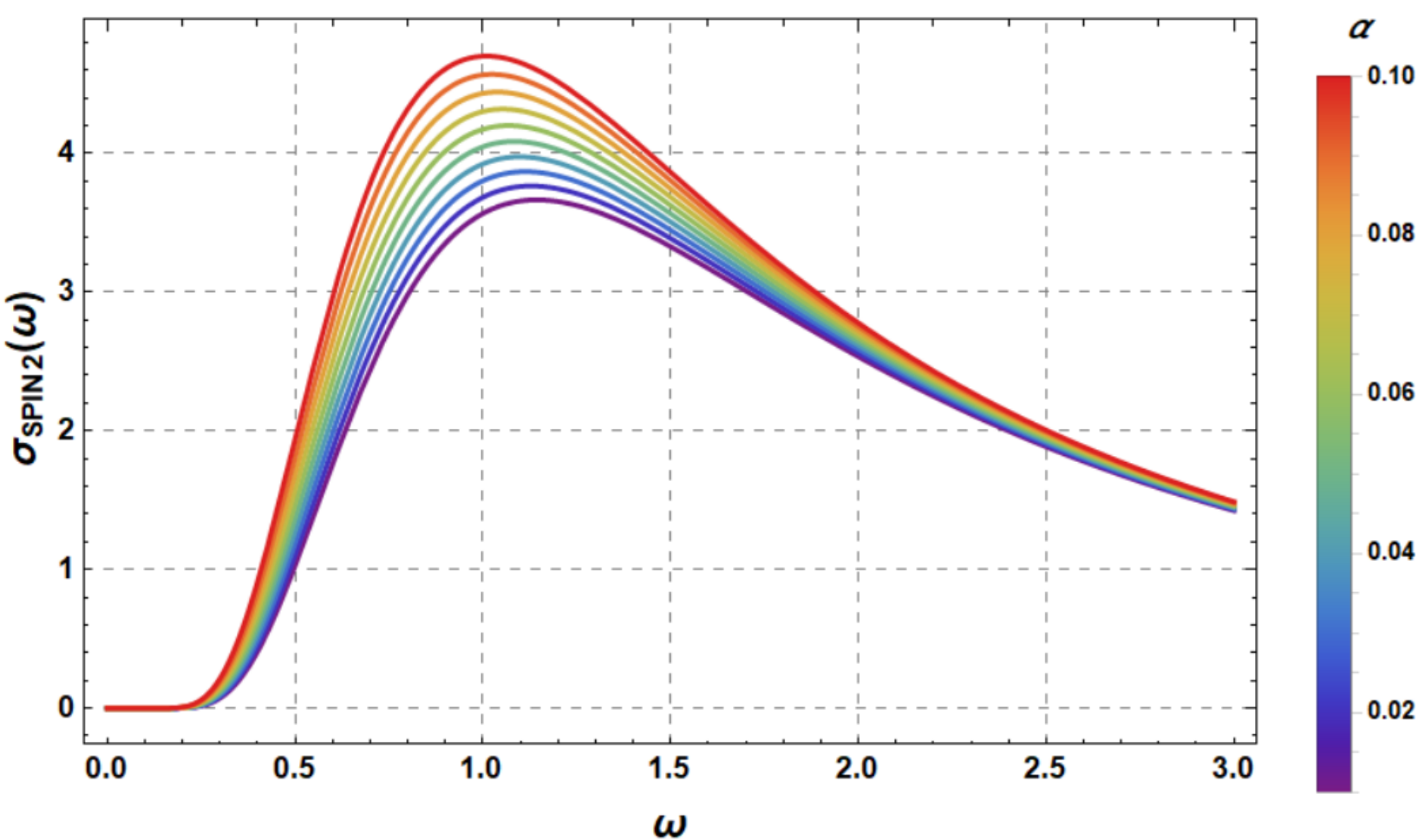}\\
(a) $Q=\lambda=0.1$ and $\alpha=0.01$\hspace{5cm}(b)$\ell=Q=\lambda=0.1$\\
\includegraphics[height=5cm]{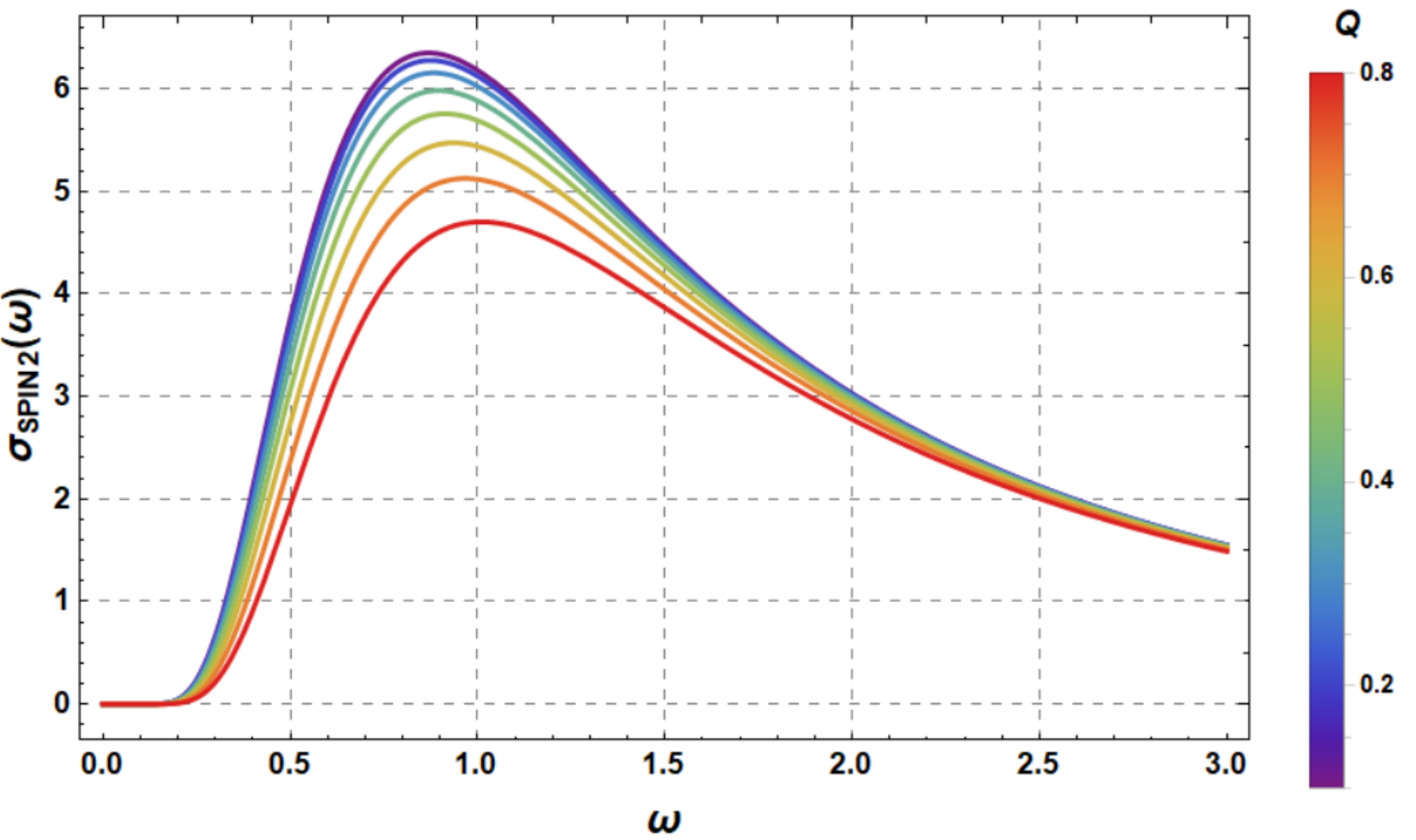}
\includegraphics[height=5cm]{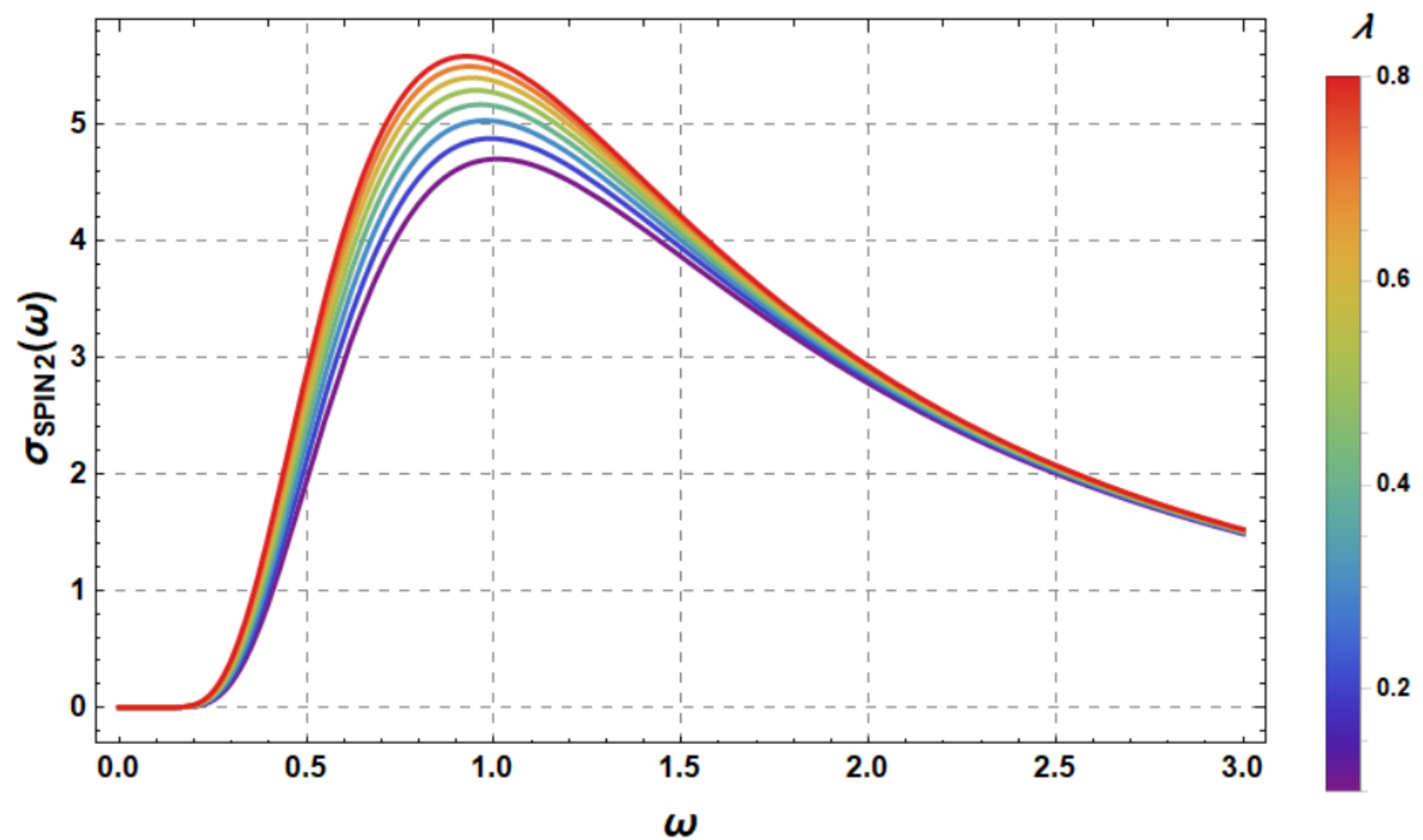}
\\
(c)$\ell=\lambda=0.1$ and $\alpha=0.01$\hspace{5cm}(d)$\ell=Q=0.1$ and $\alpha=0.01$\\
\end{tabular}
\end{center}
\vspace{-0.5cm}
\caption{Behavior of absorption cross section for spin 2.
\label{abs2}}
\end{figure}

In Figures \ref{gf2} and \ref{abs2}, the greybody factor and the absorption cross are depicted, showing a similar trend in the parameters $(\ell,Q,\lambda)$ to the preceding cases. The graviton-like sector retains the same parameter hierarchy, but it is distinguished by a broader frequency scale. In Fig.~\ref{gf2}, the plotted domain extends to $\omega\approx3$, and the transmission coefficient grows more slowly before reaching its large-$\omega$ regime. This suggests that the spin-2 effective barrier remains relevant over a wider energy interval. Even so, the response to the four background parameters is unchanged: raising $l$ or $Q$ suppresses $T_2(\omega)$, while raising $\alpha$ or $\lambda$ enhances it.

The spin-2 absorption cross section in Fig.~\ref{abs2} also differs from the previous cases by being broader and peaking at a larger frequency, around $\omega\approx1.0$ for the representative extracted curve. Its maximum is lower than the spin-1 peak in the plotted window, but the tail extends over a wider range. This is consistent with a channel that opens more gradually and distributes the absorption over a broader band of frequencies. Once more, the LV parameter and the electric charge diminish the absorption efficiency, whereas the CoS and ModMax parameters reinforce it.

Therefore, after the analysis performed in this section, we arrive at a physically transparent conclusion. The LV parameter $l$ and the electric charge $Q$ act as suppressing agents for both transmission and absorption. In this context, the effective potentials for each case play an essential role since they make the barrier harder to penetrate and therefore reduce the fraction of the wave that reaches the horizon. By contrast, the CoS parameter $\alpha$ and the ModMax parameter $\lambda$ act in the opposite direction, where they lower the net obstruction to propagation and enhance both observables.

\begin{figure}[ht!]
\begin{center}
\begin{tabular}{ccc}
\includegraphics[height=5.0cm]{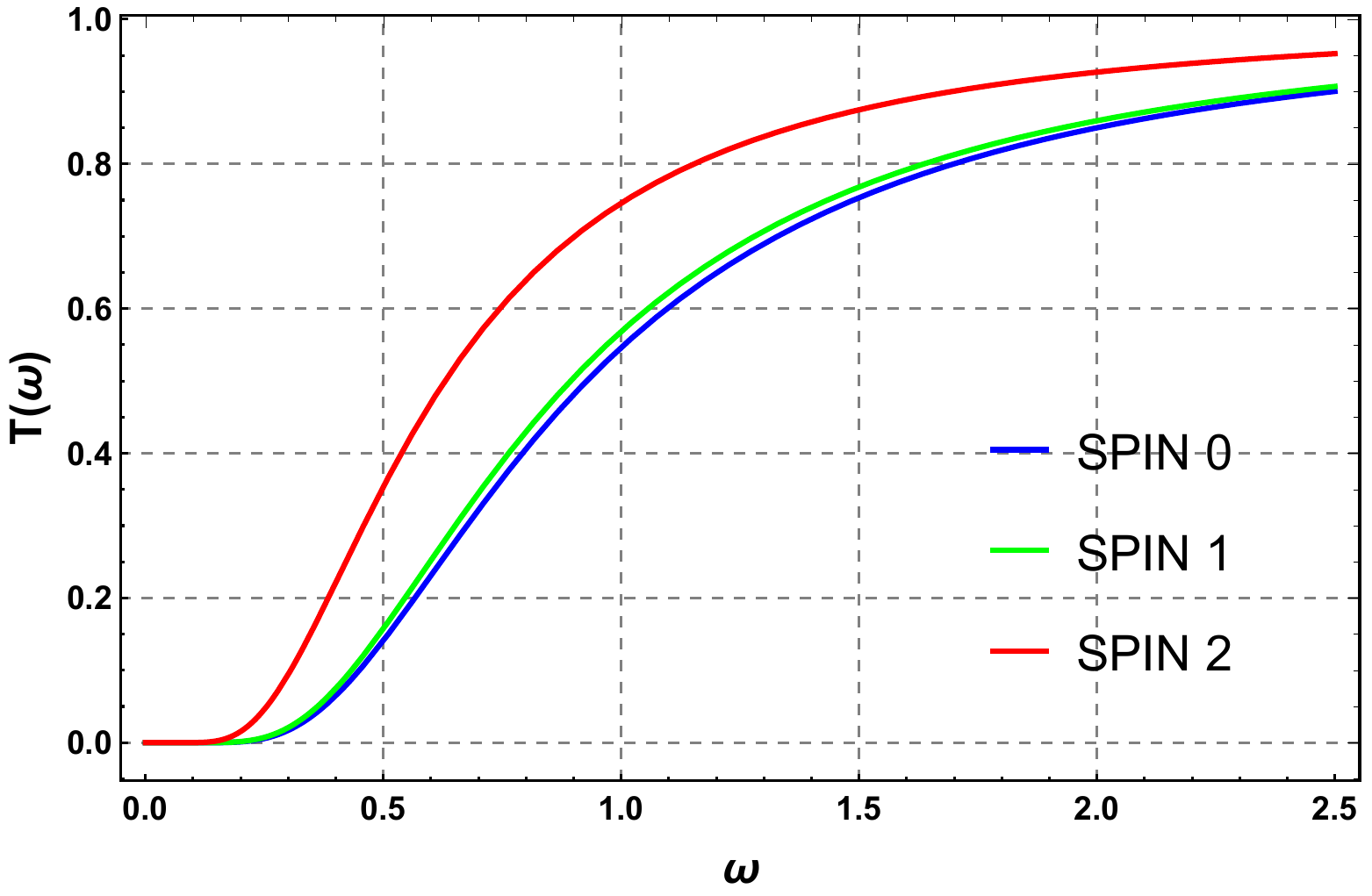} \qquad
\includegraphics[height=5.0cm]{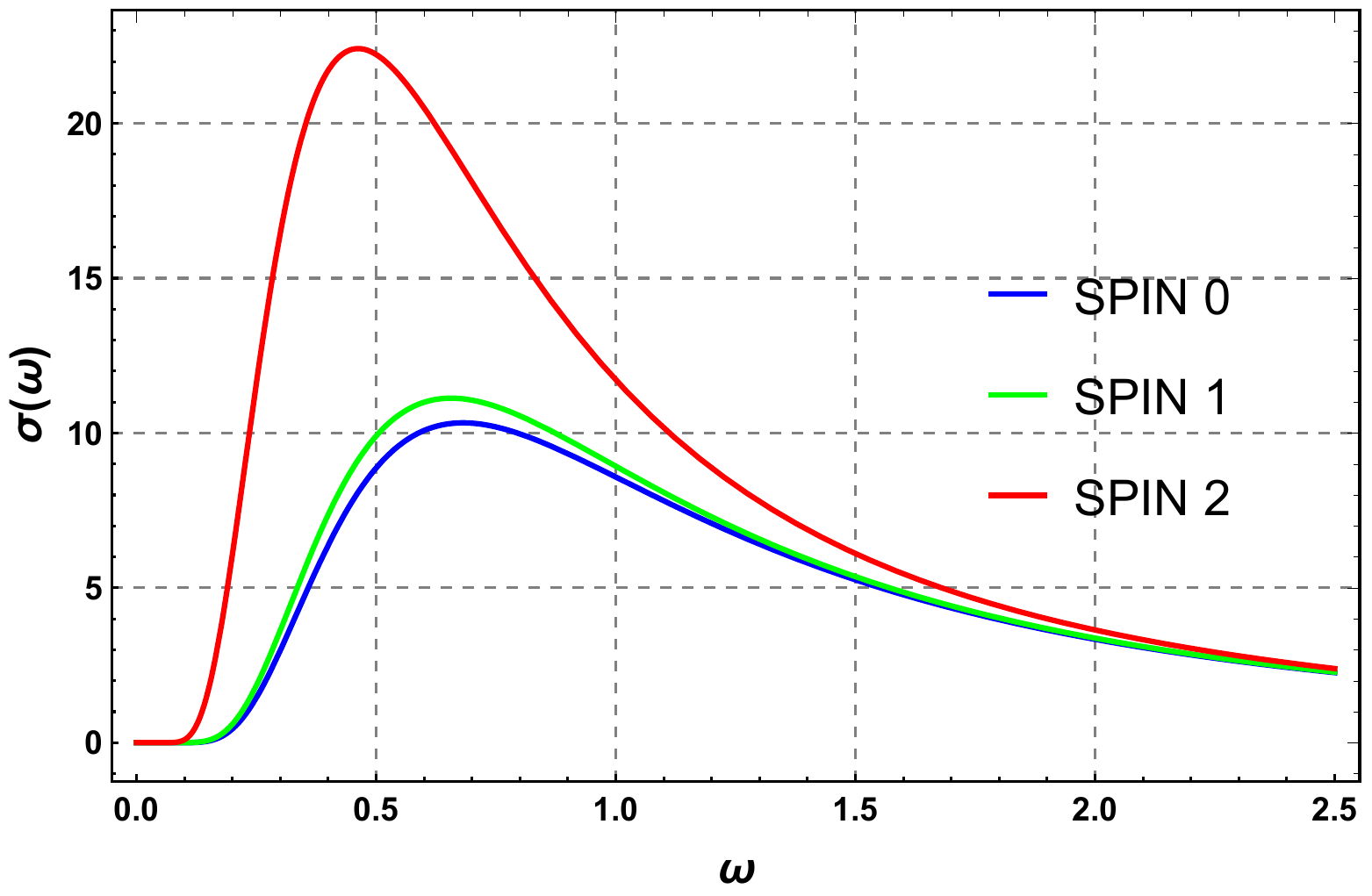}\\
(a)\hspace{6.7cm}(b)\\
\end{tabular}
\end{center}
\vspace{-0.5cm}
\caption{Comparison between (a) greybody factors and (b) absorption cross section for different spins.
\label{com}}
\end{figure}

Another relevant point is related to the spin dependence, which mainly affects the frequency scale and the shape of both the greybody factor and absorption spectrum as seen in Fig.\ref{com}. We realize that the greater the spin, the greater the transmission amplitude and absorption cross section.

Furthermore, it is pivotal to point out that the greybody factor and the absorption cross section discussed in this section, along with the sparsity of Hawking radiation analyzed previously, represent connected aspects of the interaction between a black hole and external bosonic fields. In the black hole background considered in this work, the Hawking spectrum measured at infinity is not exactly Planckian, because the emitted quanta must cross the effective curvature potential surrounding the horizon. This propagation effect is encoded in the greybody factor $\gamma_\ell(\omega)$, which gives the transmission probability for each partial mode of frequency $\omega$. The absorption cross section $\sigma_{\rm abs}(\omega)$ is directly built from these transmission coefficients, and therefore quantifies how efficiently the black hole captures or emits bosonic waves. In this sense, the greybody factor controls the spectral filtering of the Hawking radiation, while the absorption cross section determines the effective emitting area of the black hole.

\section{Final remarks}\label{S7}

In this work, we have studied a charged black hole solution with ModMax electrodynamics and a cloud of strings in Bumblebee gravity. We have shown that the combined presence of the cloud of strings sector, Lorentz violation, and ModMax nonlinearity produces a thermodynamic structure that generalizes the Reissner-Nordstr\"om black hole in a physically clear way, where $\alpha$ weakens the gravitational sector, $\lambda$ suppresses the electromagnetic contribution, and $\ell$ deforms both the temperature and the entropy, leaving a distinctive imprint on the black-hole thermodynamics.

The analysis of sparsity provides a complementary perspective on Hawking radiation beyond the usual thermodynamic quantities such as temperature and entropy. While the temperature measures the characteristic energy scale of the emitted quanta, the sparsity parameter measures how frequently those quanta are emitted. In the present black-hole background, the Hawking cascade is expected to be highly sparse over a wide region of parameter space, and it becomes infinitely sparse in the extremal limit. Therefore, the study of sparsity reinforces the interpretation that the evaporation of this black hole is not a continuous thermal flow, but rather a dilute sequence of individual quantum emission events, strongly modulated by the CoS parameter, the electric charge, the LV parameter, and the ModMax nonlinearity.

We depicted both the greybody factor and the absorption cross section as a function of frequency of mode. With these graphs, we observe that as we increase the LV parameter $\ell$, the greybody factor. Likewise, the electric charge $Q$ causes the height of the gravitational barrier and the greybody factor. Besides, the  ModMax parameter $\lambda$ has a opposite behavior. In this sense, for the LV ModMax black hole with a cloud of strings, the greybody factor provides the fundamental quantity that links wave absorption to Hawking evaporation. The absorption cross section measures the effective capture of bosonic fields, while the Hawking sparsity measures how diluted the emitted flux is in time. Since both depend on the same scattering barrier determined by the metric parameters, any change in $\alpha$, $\ell$, $Q$, or $\lambda$ simultaneously affects the absorption behavior, the greybody spectrum, and the sparsity of the Hawking radiation. This establishes a unified picture in which scattering, absorption, and evaporation are different manifestations of the same underlying geometry.

This connection becomes especially important in the discussion of Hawking sparsity. In the ideal blackbody approximation, the sparsity parameter is determined by the ratio between the time gap separating successive emitted quanta and the characteristic time scale associated with each quantum. However, once greybody effects are taken into account, the total emission rate is reduced with respect to the pure blackbody case, since only a fraction of the near-horizon thermal quanta can tunnel through the potential barrier and reach infinity. Therefore, the presence of nontrivial greybody factors makes the Hawking flux even more dilute, increasing the sparsity of the radiation. In other words, the stronger the suppression of the transmission coefficients, the smaller the number flux at infinity, and the larger the average temporal separation between emitted particles.

\section*{ Acknowledgments}

\hspace{0.5cm} The author Fernando Belchior would like to to express gratitude to the Conselho Nacional de Desenvolvimento Cient\'{i}fico e Tecnol\'{o}gico CNPq for grant No. 151845/2025-5. The author Faizuddin Ahmed acknowledges the Inter University Centre for Astronomy and Astrophysics (IUCAA), Pune, India for granting visiting associateship.

\section*{Author Contribution Statement }
{\bf F. M. Belchior} and  {\bf  A. R. P. Moreira}: Conceptualization; Formal analysis; Methodology; and Writing - Original Draft. {\bf A. Bouzenada} and  {\bf F. Ahmed}: Investigation; Validation, and Writing - Review \& Editing.

\section*{Data Availability}

No new data were generated or analyzed in this study.

\section*{Conflict of Interests}

Authors declares there is no conflict of interests.

\section*{Code/Software}

No code/software were developed in this article [Authors comment: No code/software were developed in this article].

\section*{Funding Statement}

No fund has received for this study.


\begin{thebibliography}{99}


\bibitem{BHM} C. L. Fryer, Astrophys. J., 522(1), 413 (1999).

\bibitem{AbbottPRL2016} B. P. Abbott, et al. [LIGO Scientific and Virgo Collaborations], Phys. Rev. Lett. \textbf{116}, 061102 (2016).

\bibitem{AbbottPRD2020} R. Abbott, et al. [LIGO Scientific and Virgo Collaborations], Phys. Rev. \textbf{D 102}, 043015 (2020).

\bibitem{AbbottPRL2020} R. Abbott, et al. [LIGO Scientific and Virgo Collaborations], Phys. Rev. Lett. \textbf{125}, 101102 (2020).

\bibitem{LiuNature20219} J.-F. Liu, et al., Nat. \textbf{575}, 618 (2019).

\bibitem{AkiyamaL12019} K. Akiyama, et al., (Event Horizon Telescope), Astrophys. J. Lett. \textbf{875}, L1 (2019).

\bibitem{AkiyamaL52019} K. Akiyama, et al., (Event Horizon Telescope), Astrophys. J. Lett. \textbf{875}, L5 (2019).

\bibitem{AkiyamaL62019} K. Akiyama, et al., (Event Horizon Telescope), Astrophys. J. Lett. \textbf{875}, L6 (2019). 

\bibitem{AkiyamaL122022} K. Akiyama, et al., (Event Horizon Telescope Collaboration), Astrophys. J. Lett. \textbf{930}, L12 (2022).

\bibitem{AkiyamaL172022} K. Akiyama, et al., (Event Horizon Telescope Collaboration), Astrophys. J. Lett. \textbf{930}, L17 (2022).

\bibitem{39} T. Do, et al., Science, {\bf 365}, 664 (2019).

\bibitem{40} R. Abuter, et al., (GRAVITY Collaboration), Astron. Astrophys. \textbf{657}, L12 (2022).

\bibitem{41} J. P. Luminet, Astron. Astrophys. \textbf{75}, 228 (1979).

\bibitem{42} H. Falcke, and F. Melia, E. Agol, Astrophys. J. Lett. \textbf{528}, L13 (2000).

\bibitem{43} R. Narayan, M. D. Johnson, and C. F. Gammie, Astrophys. J. Lett. \textbf{885}, L33 (2019).

\bibitem{44} V. Perlick, and O. Yu. Tsupko, Phys. Rep. \textbf{947}, 1 (2022).

\bibitem{45} X.-H. Feng, and H. L\"{u}, Eur. Phys. J. \textbf{C 80}, 551 (2020).

\bibitem{46} S. Hu, C. Deng, D. Li, X. Wu, and E. Liang, Eur. Phys. J. \textbf{C 82}, 885 (2022).

\bibitem{47} S. Wen, W. Hong, and J. Tao, Eur. Phys. J. \textbf{C 83}, 277 (2023).

\bibitem{TsupkoPRD2017} O. Y. Tsupko, Phys. Rev. \textbf{D 95}, 104058 (2017).

\bibitem{BroderickApJ2022} A. E. Broderick, et al., Astrophys. J. \textbf{935}, 61 (2022).

\bibitem{MizunoNatAst2018} Y. Mizuno, et al., Nat. Astron. \textbf{2}, 585 (2018).

\bibitem{AtamurotovPRD2013} F. Atamurotov, A. Abdujabbarov, and B. Ahmedov, Phys. Rev. \textbf{D 88}, 064004 (2013).

\bibitem{AbdujabbarovSS2016} A. Abdujabbarov, et al., Astrophys. Space Sci. \textbf{361}, 226 (2016).

\bibitem{AbdikamalovPRD2019} A. B. Abdikamalov, et al., Phys. Rev. \textbf{D 100}, 024014 (2019).

\bibitem{AtamurotovPRD2015} F. Atamurotov, and B. Ahmedov, Phys. Rev. \textbf{D 92}, 084005 (2015).

\bibitem{AtamurotovCPC2023} F. Atamurotov, I. Hussain, G. Mustafa, and A. Övgün, Chin. Phys. \textbf{C 47}, 025102 (2023).

\bibitem{BelhajPLB2021} A. Belhaj, et al., Phys. Lett. \textbf{B 812}, 136025 (2021).

\bibitem{BelhajCQG2021} A. Belhaj, et al., Classical Quantum Gravity \textbf{37}, 215004 (2020).

\bibitem{CunhaPLB2017} P. V. P. Cunha, et al., Phys. Lett. \textbf{B 768}, 373 (2017).

\bibitem{WeiJCAP2019} S.-W. Wei, Y.-C. Zou, Y.-X. Liu, and R.B. Mann, J. Cosmol. Astropart. Phys. \textbf{08}, 030 (2019).

\bibitem{LingPRD2021} R. Ling, H. Guo, H. Liu, X.-M. Kuang, and B. Wang, Phys. Rev. \textbf{D 104}, 104003 (2021).

\bibitem{TsukamotoPRD2018} N. Tsukamoto, Phys. Rev. \textbf{D 97}, 064021 (2018).

\bibitem{AraujoFilhoCQG2024} A. A. Ara\'{u}jo Filho, et al., Class. Quant. Grav. \textbf{41}, 055003 (2024).

\bibitem{RayimbaevPoDU2022} J. Rayimbaev, et al., Phys. Dark Universe, {\bf 35}, 100930 (2022).

\bibitem{PerlickPRD2018} V. Perlick, et al., Phys. Rev. \textbf{D 97}, 104062 (2018).

\bibitem{VagnozziCQG2023} S. Vagnozzi et al., Class. Quant. Grav. \textbf{40}, 165007 (2023).

\bibitem{KocherlakotaPRD2021} P. Kocherlakota, et al., Phys. Rev. \textbf{D 103}, 104047 (2021).

\bibitem{gy1} F. Javed, F.: Physics. Dark Universe {\bf 44}, 101450 (2024). 

\bibitem{gy2} F. Javed, and M. H. Alshehri., Results Phys. {\bf 62}, 107837 (2024).

\bibitem{q} A. Uniyal, R. C. Pantig, and A. Övgün, Phys. Dark Universe \textbf{40}, 101178 (2023).

\bibitem{qq} A. Uniyal, S. Chakrabarti, R.C. Pantig, and A. Övgün, New Astron. \textbf{111}, 102249 (2024).

\bibitem{qqq} G. Lambiase, R. C. Pantig, D. J. Gogoi, and A. Övgün, Eur. Phys. J. \textbf{C 83}, 679 (2023).

\bibitem{qqqq} M. Khodadi, and G. Lambiase, Phys. Rev. \textbf{D 106}, 104050 (2022). 

\bibitem{qqqqq} M. Khodadi, G. Lambiase, and D.F. Mota, J. Cosmol. Astropart. Phys. \textbf{09}, 028 (2021).

\bibitem{qqqqqq} G. Panotopoulos, Á. Rincón, and I. Lopes, Phys. Rev. {\bf D 103}(10), 104040 (2021).

\bibitem{qqqqqqq} B. Eslam Panah, S. Zare, and H. Hassanabadi, Eur. Phys. J. \textbf{C 84}, 259 (2024).

\bibitem{BZQ} N. Sarkar, et al., Phys. Dark Universe, {\bf 44}, 101439 (2024).

\bibitem{WuPoDU2024} S. R. Wu, B. Q. Wang, Z. W. Long, and Hao Chen, Phys. Dark Universe \textbf{44}, 101455 (2024).

\bibitem{PantigJCAP2022} R.C. Pantig, and A. Övgün, J. Cosmol. Astropart. Phys. \textbf{08}, 056 (2022).

\bibitem{CapozzielloJCAP2023} S. Capozziello, et al., J. Cosmol. Astropart. Phys. \textbf{05}, 027 (2023).

\bibitem{Capozziello2023} S. Capozziello, S. Zare, and H. Hassanabadi, arXiv:2311.12896v1 [gr-qc] (2023).

\bibitem{XuJCAP2018} Z. Xu, X. Hou, X. Gong, and J. Wang, J. Cosmol. Astropart. Phys. \textbf{09}, 038 (2018).

\bibitem{KonoplyaApJ2022} R. A. Konoplya, and A. Zhidenko, Astrophys. J. \textbf{933}, 166 (2022).

\bibitem{BG1} A. Övgün, K. Jusufi, I. Sakallı, Phys. Rev. \textbf{D  99}, 024042 (2019).

\bibitem{BG2} S. K. Jha, H. Barman, A. Rahaman, J. Cosmol. Astropart. Phys. \textbf{2021}, 036 (2021).

\bibitem{BG3} K. Jusufi, I. Sakallı, Eur. Phys. J. \textbf{C 81}, 501 (2021).

\bibitem{BG4} A. Uniyal, K. Jusufi, I. Sakallı, Eur. Phys. J. \textbf{C 83}, 668 (2023).

\bibitem{BG5} S. K. Jha, A. Rahaman, Eur. Phys. J. \textbf{C 81}, 345 (2021).

\bibitem{BG6} H. M. Wang, S. W. Wei, Eur. Phys. J. Plus \textbf{137}, 571 (2022).

\bibitem{BG7} K. M. Amarilo, M. B. F. Filho, A. A. A. Filho et al., Phys. Lett. \textbf{B 855}, 138785 (2024).

\bibitem{BG8} R. Karmakar, D. J. Gogoi, U. D. Goswami, Phys. Dark Univ. \textbf{41}, 101249 (2023).

\bibitem{BG9} Z. F. Mai, R. Xu, D. Liang, L. Shao, Phys. Rev. \textbf{D  108}, 024004 (2023).

\bibitem{BG10} Y. S. An, Phys. Dark Univ. \textbf{45}, 101520 (2024).

\bibitem{BG11} D. J. Gogoi, U. D. Goswami, J. Cosmol. Astropart. Phys. \textbf{2022}, 029 (2022).

\bibitem{BG12} X. Zhang, M. Wang, J. Jing, Sci. China Phys. Mech. Astron. \textbf{66}, 100411 (2023).

\bibitem{BG13} J. Z. Liu, W. D. Guo, S. W. Wei, Y. X. Liu, Eur. Phys. J. \textbf{C 85}, 145 (2025).

\bibitem{BG14} M. Xu,  et al., Class. Quantum Grav. \textbf{42}, 135008 (2025).

\bibitem{BG15} G. Mustafa, S. K. Maurya,  et al., Phys. Dark Univ. \textbf{47}, 101753 (2025).

\bibitem{BG16} J. de Oliveira,  et al., Class. Quantum Grav. \textbf{42}, 235018 (2025).

\bibitem{BG17} Y. P. Singh, J. Choudhury, T. I. Singh et al., Eur. Phys. J. Plus \textbf{140}, 1118 (2025).

\bibitem{MAX1} I. Bandos, K. Lechner, D. Sorokin, P.K. Townsend, Phys. Rev. \textbf{D  102}, 121703 (2020).

\bibitem{MAX2} D. Flores-Alfonso, B.A. González-Morales, R. Linares, M. Maceda, Phys. Lett. \textbf{B 812}, 136011 (2021).

\bibitem{MAX3} R.C. Pantig, L. Mastrototaro, G. Lambiase, A. Övgün, Eur. Phys. J. \textbf{C 82}, 1155 (2022).

\bibitem{MAX4} A. Banerjee, A. Mehra, Phys. Rev. \textbf{D  106}, 085005 (2022).

\bibitem{MAX5} A. Bokulić, I. Smolić, T. Jurić, Phys. Rev. \textbf{D  106}, 064020 (2022).

\bibitem{MAX6} K. Lechner, P. Marchetti, A. Sainaghi, D.P. Sorokin, Phys. Rev. \textbf{D  106}, 016009 (2022).

\bibitem{MAX7} J. Barrientos, A. Cisterna, N. Mora, A. Vásquez, L. Zapata, Phys. Lett. \textbf{B 834}, 137447 (2022).

\bibitem{MAX8} M. Ortaggio, Eur. Phys. J. \textbf{C 82}, 1056 (2022).

\bibitem{MAX9} H. Nastase, Phys. Rev. \textbf{D  105}, 105024 (2022).

\bibitem{MAX10} C. Ferko, L. Smith, V. Chandrasekaran, A. Sfondrini, Phys. Rev. Lett. \textbf{129}, 201604 (2022).

\bibitem{MAX11} D.P. Sorokin, Fortschr. Phys. \textbf{70}, 2200092 (2022).

\bibitem{MAX12} C.A. Escobar, R. Linares, B. Tlatelpa-Mascote, Int. J. Mod. Phys. \textbf{A 37}, 2250011 (2022).

\bibitem{MAX13} A. Bokulić, T. Jurić, I. Smolić, Phys. Rev. \textbf{D  105}, 024067 (2022).

\bibitem{MAX14} K. Nomura, D. Yoshida, Phys. Rev. \textbf{D  105}, 044006 (2022).

\bibitem{MAX15} M. Zhang, J. Jiang, Phys. Rev. \textbf{D  104}, 084094 (2021).

\bibitem{MAX16} S.I. Kruglov, Phys. Lett. \textbf{B 822}, 136633 (2021).

\bibitem{MAX17} I. Bandos, K. Lechner, D. Sorokin, P.K. Townsend, JHEP \textbf{10}, 031 (2021).

\bibitem{MAX18} A. Bokulić, T. Jurić, I. Smolić, Phys. Rev. \textbf{D  103}, 124059 (2021).

\bibitem{MAX19} D. Flores-Alfonso, R. Linares, M. Maceda, JHEP \textbf{09}, 174 (2021).

\bibitem{MAX20} A. Ballon Bordo, D. Kubizňák, T.R. Perche, Phys. Lett. \textbf{B 817}, 136312 (2021).

\bibitem{MAX21} H. Babaei-Aghbolagh, K.B. Velni, D.M. Yekta, H. Mohammadzadeh, Phys. Lett. \textbf{B 829}, 137079 (2022).

\bibitem{MAX22} Y. Kurmanov, et al., Phys. Dark Universe \textbf{46}, 101566 (2024).

\bibitem{MAX23} A. Ditta, G. Mustafa, G. Abbas, F. Atamurotov, K. Jusufi, JCAP \textbf{08}, 002 (2023).

\bibitem{MAX24} A. Uniyal, R.C. Pantig, A. Övgün, Phys. Dark Universe \textbf{40}, 101178 (2023).

\bibitem{MAX25} M.R. Shahzad, G. Abbas, H. Rehman, W.X. Ma, Eur. Phys. J. \textbf{C 84}, 461 (2024).


\bibitem{Maluf:2020kgf}
R. V. Maluf and J. C. S. Neves, Phys. Rev. \textbf{D 103}, 044002 (2021).

\bibitem{Liu:2024axg}
J. Z. Liu, W. D. Guo, S. W. Wei and Y. X. Liu, Eur. Phys. J. \textbf{C  85}, 145 (2025).

\bibitem{Li:2026uwx}
J. Li, Y. Zhu and B. Yang, Eur. Phys. J. \textbf{C  86}, 167 (2026).




\end{thebibliography}
\end{document}